\documentclass[twoside,12pt]{article}

\usepackage[dvips]{graphicx}
\usepackage{epsfig}
\usepackage{amsmath,amsfonts,amssymb,amsthm}
\usepackage{verbatim}
\usepackage{psfrag}
\usepackage{bm}
\usepackage{bbm}
\usepackage[square,comma,sort&compress,numbers]{natbib}
\usepackage{color}
\usepackage{slashed}
\usepackage{enumerate}
\usepackage{dsfont}
\usepackage{cancel}

\usepackage{epsf,epsfig}
\usepackage{graphics}
\usepackage{wrapfig}
\newcommand{\be}{\begin{equation}}
\newcommand{\ee}{\end{equation}}
\newcommand{\bea}{\begin{eqnarray}}
\newcommand{\eea}{\end{eqnarray}}

\newcommand{\lsim}
{\;\raisebox{-.3em}{$\stackrel{\displaystyle <}{\sim}$}\;}

\date{}

\begin{document}
\thispagestyle{empty}

\title{ \vspace{1cm} Why is there more matter than antimatter?\\
Calculational methods for leptogenesis\\and electroweak baryogenesis}

\author{Bj\"orn Garbrecht\\
\\
Physik-Department T70, James-Franck-Stra{\ss}e,\\
Technische Universit\"at M\"unchen, 85748 Garching, Germany}

\begin{flushright}
{
\small
{\tt TUM-HEP-1177-18}
}
\end{flushright}

{\let\newpage\relax\maketitle}

\begin{abstract}
We review the production of the matter--antimatter asymmetry in the early Universe, that is {\it baryogenesis}, in out-of-equilibrium
conditions induced by decays of heavy particles or by the presence of phase boundaries.
The most prominent examples are given by leptogenesis and electroweak baryogenesis, respectively.
For both cases, we derive the equations that govern the production of the asymmetries. We first use intuitive arguments based on classical fluid equations
in combination with quantum-field-theoretical effects of $CP$-violation.
As for a
more thorough approach that is well-suited for systematic improvements,
we obtain the real-time evolution of the
system of interest using the closed time-path method.
We thus provide a simple and practicable scheme to set up phenomenological fluid equations based on first principles of quantum field theory.
Necessary for baryogenesis are both, $CP$ even as well as odd phases in the amplitudes.
A possibility of generating the even phases is the coherent superposition of quantum states, i.e. mixing.
These coherence effects are essential in resonant leptogenesis as well as in
some scenarios of electroweak baryogenesis. Recent theoretical progress on
asymmetries from out-of-equilibrium decays may therefore also be applicable to baryogenesis at phase
boundaries.
\end{abstract}



\newpage

\tableofcontents

\newpage

\section{Introduction}

This article is mainly concerned with calculational methods for baryogenesis, i.e. the
hypothetical process that yields the matter--antimatter asymmetry in the early Universe.
Besides summarizing these methodical matters, we provide some minimal context
on the problem of the matter--antimatter asymmetry in this introduction. In addition,
we refer to the reviews~\cite{Riotto:1999yt,Cline:2006ts,Canetti:2012zc} that contain a general overview on baryogenesis.

\subsection{The matter--antimatter puzzle}
\label{sec:matter-antimatter}

If matter and antimatter had, apart from charges that are precisely opposite, exactly the same properties,
and we assumed the Universe to start in a big bang with symmetric initial conditions, there would be no
way for Nature to develop a preference of one above the other. Nonetheless, the evidence for an asymmetry
is overwhelming. No events where astrophysical objects composed of antimatter annihilate
with matter objects have been observed~\cite{Steigman:2008ap}. And while it could yet be conceivable that individual stars or
entire galaxies are composed of antimatter, such an hypothesis is untenable in view of the
observationally established big bang scenario, according to which the baryonic components of the
Universe have been in an almost homogeneous state at least from the time of big bang nucleosynthesis
(BBN) to the appearance of nonlinear structures.

However, it is a generic prediction of quantum field theories (QFTs) that matter and antimatter can have different properties because
certain discrete symmetries are not conserved (see Ref.~\cite{Branco:1999fs} for a comprehensive discussion): Chiral gauge theories maximally violate parity $P$ and charge $C$ conjugation,
and, crucially, mass terms or underlying Yukawa couplings can violate the
combined charge-parity symmetry $CP$. While these discrete symmetries are
observed in quantum electrodynamics,
the Standard Model (SM) of particle physics realizes the possibility of
their violation. The experiments that have discovered $P$ and $CP$ violation~\cite{Wu:1957my,Christenson:1964fg} are therefore
among the most celebrated in physics and have been pivotal for the development
of the SM. The responsible theory of electroweak symmetry breaking
and the Cabibbo-Kobayashi-Maskawa (CKM) model of fermion masses have been continuously
confirmed by present day experiments in conjunction with precision calculations.

Adding time-reversal $T$, there is another important combination: $CPT$. QFTs that are local and that observe relativistic causality predict the conservation
of this symmetry. The SM is of course an example for such a theory and no violation of
$CPT$ has been observed to the present date~\cite{Kostelecky:2008ts}. While ultimately, it could still turn out that
the more underlying theory realized in Nature violates $CPT$, we therefore apply the
working hypothesis that this is either not the case or is not of relevance in the effective theory
responsible for baryogenesis. Then, $CPT$ conservation predicts that particles and antiparticles still have precisely opposite charges, precisely the same mass and even the same lifetime, i.e. the same
inclusive decay rates. However, particular exclusive channels can have different decay rates provided
they sum up to the same total, inclusive decay rate.

Considering thus a theory respecting $CPT$, we further add the violation of baryon number $B$ or, more
appropriately in the context of the SM, some combination of $B$ and lepton number $L$. If we put
particles of that theory in a box with perfect heat insulation and wait long enough, thermal
equilibrium will be reached. This implies a vanishing baryon asymmetry because we assume that baryon number is violated and baryon and antibaryons
have exactly the same mass. Equal numbers of particles and antiparticles must therefore be present in
the equilibrium state of maximum entropy. The question of whether in the early Universe, a preexisting
asymmetry can survive the washout via baryon number violating reactions and whether {\it baryogenesis}, i.e. to produce an asymmetry dynamically, is possible therefore depends on how far the particle content of the expanding Universe
deviates from thermal equilibrium.

\begin{figure}
\begin{center}
\includegraphics[width=16cm]{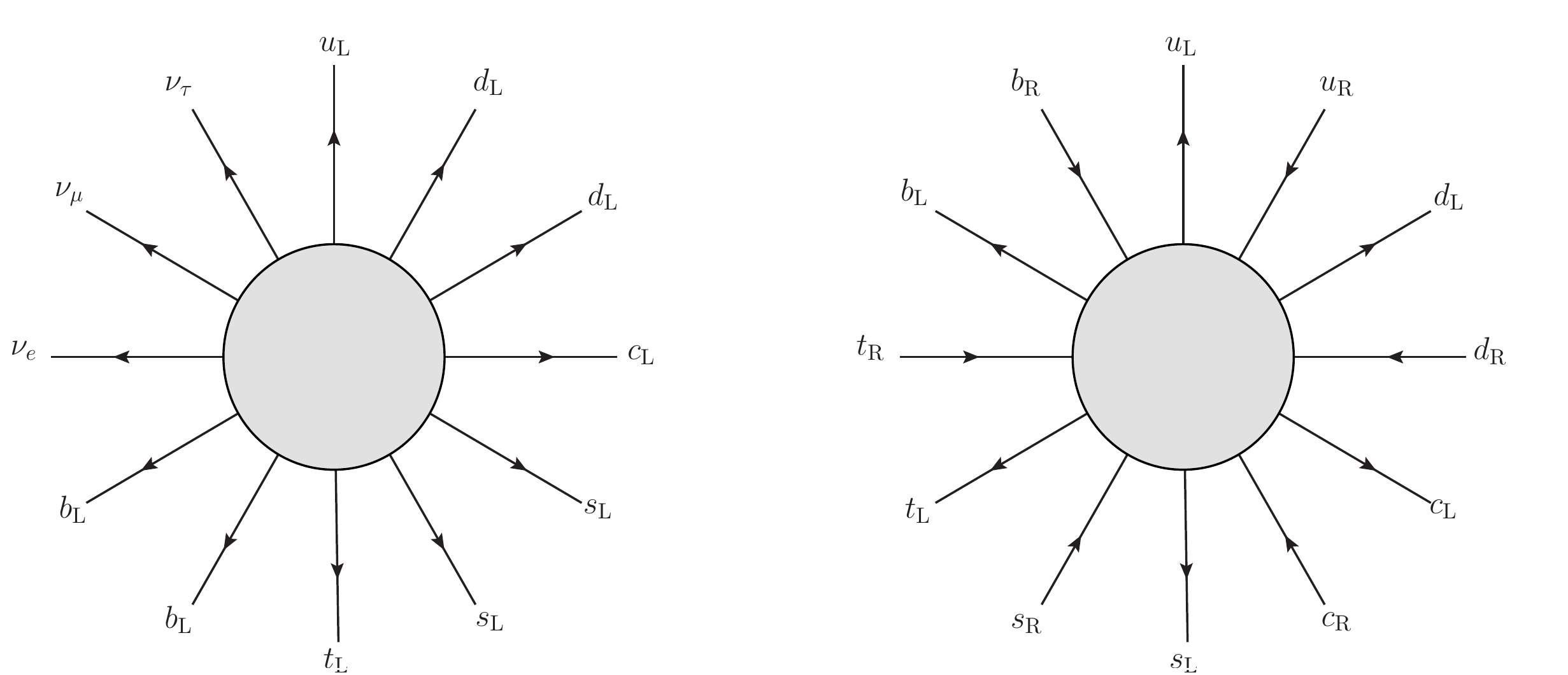}
\end{center}
\caption{\label{fig:sphalerons}
The 't Hooft vertices~\cite{tHooft:1976rip} for weak (left) and strong (right)
sphaleron interactions. For the weak sphaleron, ${\rm SU}(2)$-doublet components
may be interchanged as long as the vertex remains a gauge singlet. Note that
the weak sphaleron changes baryon-plus-lepton number $B+L$ by six units.
}
\end{figure}

By 1967, $CP$ violation~\cite{Christenson:1964fg} and the Cosmic Microwave Background (CMB) Radiation had been observed~\cite{Penzias:1965wn}. Remarkably, the
temperature of the latter turned out to be consistent with what had been predicted from BBN~\cite{AlpherHerman1,AlpherHerman2,Gamov1,Gamov2}. This pioneering work had therefore established that reactions from particle and
nuclear physics have taken place in the early Universe and that theory can make quantitative 
predictions about their turnout. The recent discoveries at that time may therefore have played a part in leading Sakharov in 1967 to propose that the matter--antimatter
asymmetry is a question that can indeed be answered by QFT. More
precisely, he has formulated the minimal requirements that successful baryogenesis poses on particle theory and its embedding in a cosmological context~\cite{Sakharov:1967dj}. These have subsequently been paraphrased in terms of the Sakharov conditions, i.e. necessary for baryogenesis are:
\begin{enumerate}
\item
the presence of baryon-number violating interactions,
\item
$C$ and $CP$ violation (the latter being necessary for asymmetries in left- and right-handed sectors not
to cancel one another),
\item
a deviation from thermal equilibrium.
\end{enumerate}
The SM turns out to qualitatively meet Sakharov's criteria:
\begin{enumerate}
\item
Baryon-plus-lepton number $B+L$ is violated through the chiral anomaly and the pertaining weak sphaleron transitions at finite temperature~\cite{tHooft:1976rip,Kuzmin:1985mm}, cf. Figure~\ref{fig:sphalerons}.
\item
$C$ is violated through the weak interactions and $CP$ through the
CKM mechanism.
\item
The expansion of the Universe leads to a deviation from thermal equilibrium and in particular, if the mass of the Higgs boson were below $70\,{\rm GeV}$, there would be a first order phase transition~\cite{Kajantie:1996mn,Rummukainen:1998as}, i.e. the coexistence of symmetric and broken electroweak phases in a certain temperature range. At the phase boundaries, which are the walls of broken phase bubbles expanding
into the sea of the symmetric phase, there would be a substantial deviation from equilibrium.
\end{enumerate}
Curiously, it is only because
of the values of its free parameters that the SM cannot solve the puzzle.
Quantitatively, it falls short of explaining the asymmetry for the following reasons: 
\begin{itemize}
\item
The $125\,{\rm GeV}$ Higgs boson is too heavy in order to support a first order phase transition, such that an electroweak crossover with continuous evolution of the expectation value of the
Higgs field has occurred instead. As a consequence, the plasma remains too close to thermal equilibrium because all of its degrees of freedom participate in gauge interactions, which are fast and therefore very effective in suppressing any deviation from
equilibrium due to the expansion of the Universe.
\item
The first $CP$ violating and rephasing invariant quantity appears at eighth order in Yukawa couplings
and involves second-generation couplings at fourth order, i.e.~\cite{Jarlskog:1985ht}
\begin{align}
{\rm Im}\left[\det[M_u M_u^\dagger,M_d M_d^\dagger]\right]\approx-2 J m_t^4 m_b^4 m_c^2 m_s^2\,,
\end{align}
what we have expressed in terms of the mass matrices $M_{u,d}$
of up and down-type quarks as well as the mass eigenvalues
$m_q$ of the quarks of flavour $q$
and where
\begin{align}
J={\rm Im}[V_{us}V_{cb}V^*_{ub}V^*_{cs}]\approx 3\times 10^{-5}
\end{align}
is constructed from the elements $V_{ij}$ of the CKM matrix.
At high temperatures, where nonperturbative effects that help to
make $CP$ violation more accessible in the laboratory are absent,
$CP$-violating effects are largely suppressed~\cite{Farrar:1993hn,Gavela:1993ts,Gavela:1994dt,Huet:1994jb},
which can be estimated through the ratio
\begin{align}
J\frac{m_t^4 m_b^4 m_c^2 m_s^2}{T^{12}}\approx3\times10^{-19}\;\textnormal{for}\;
T=100\,{\rm GeV}\,,
\end{align}
that appears too small in order to explain the oberved value~(\ref{BAU:CMB}) below.
\end{itemize}

\subsection{Asymmetry observed}

When it comes to the calculation of reaction networks in the early Universe,
it is most convenient to use entropy-normalized number or charge densities, which are
conserved unless substantial amounts of energy are injected into the approximately thermalized
plasma, e.g. through the far-from-equilibrium decay of an abundant heavy species. We thus assume that entropy
is conserved in a comoving volume element, i.e. $s a^3={\rm const.}$ where
$s$ is the entropy density and $a$ the scale factor of the Friedmann-Robertson-Walker metric.

For this reason, the calculations for BBN are formulated in terms of
entropy-normalized densities. The key cosmological parameter entering BBN is
the baryon asymmetry, which controls the density of the nucleons and the light nuclei
that are fused from these eventually. The currently reported best-fit value is~\cite{Tanabashi:2018oca}
\begin{align}
\label{BAU:BBN}
5.8\times10^{-10}\leq\frac{n_B}{n_\gamma}\leq 6.6\times10^{-10}
\end{align}
at $95\%$ confidence level,
where instead of the entropy density the number density of CMB photons $n_\gamma$ has been used for normalization.
Assuming that there are three chiral relativistic neutrino species (i.e. six degrees of freedom) that
have a temperature of $T_\nu=(4/11)^{1/3}T_{\rm CMB}$ (because photons
are heated by electron-positron annihilation after neutrinos decouple),
the entropy density today is $s=(2\pi^2/45)(2\, T_{\rm CMB}^3+6\times 7/8\, T_\nu^3)$, where
$T_{\rm CMB}=2.725 {\rm K}$ is the present temperature of the CMB. Integrating
the Bose distribution for two massless photon degrees of freedom yields
$n_\gamma=(2/\pi^2)\zeta(3)T_{\rm CMB}^3$, such that $s/n_\gamma\approx 7.04$, i.e.
$n_B/s=Y_B\approx n_B/(7.04\,n_\gamma)$.
The value~(\ref{BAU:BBN}) can therefore be interpreted such that at temperatures
above the phase transition of quantum chromodynamics (QCD), there has been roughly
one extra quark per ten billion particle-antiparticle pairs.

An entirely complementary method of determining the BAU, which has become more precise
thanks to the precision data from the WMAP~\cite{Hinshaw:2012aka} and Planck~\cite{Aghanim:2018eyx} probes, is to infer it from CMB anisotropies, in particular
from the imprint of baryon acoustic oscillations. We quote the up-to-date value as
$\Omega_B h^2=0.0224\pm 0.0001$ with $68\%$ confidence~\cite{Aghanim:2018eyx},
where $\Omega_B$ is the fraction of the baryon mass in terms of the critical
energy density $\varrho_c$ (i.e. the energy density in a spatially flat Friedmann model)
and $H_0=100 \,{\rm km}{\rm s}^{-1} {\rm Mpc^{-1}} h$ is the Hubble rate in the present Universe.
The baryon number density is then obtained by dividing the baryon mass density by
the mass of a nucleon $m_{\rm nucl}$ as
$n_B=\varrho_{\rm c}\Omega_B/m_{\rm nucl}=[3/(8\pi m_{\rm nucl})]m_{\rm Pl}^2H_0^2\Omega_b h^2$,
where $m_{\rm Pl}=1.221\times 10^{19}\,{\rm GeV}$ is the Planck mass.
Taking $n_\gamma$ as given above then yields $n_B/n_\gamma\approx 2.74\times 10^{-8}\Omega_B h^2$, such that the baryon-to-photon ratio inferred from the CMB is
\begin{align}
\label{BAU:CMB}
\frac{n_B}{n_\gamma}=6.14\pm 0.02 \times 10^{-10}\,.
\end{align}

The agreement of the results~\eqref{BAU:BBN} and~\eqref{BAU:CMB} is a key achievement of particle cosmology and impressively validates
the approach of formulating and calculating networks of QFT reactions taking place during
the early stages of the Universe.

\subsection{Outline of this article}

In the present work, we present basic aspects about calculations of the baryon asymmetry of
the Universe. To illustrate the methods on rather paradigmatic examples, we choose leptogenesis~\cite{Fukugita:1986hr}
for out-of-equilibrium decays and electroweak baryogenesis~\cite{Kuzmin:1985mm,Shaposhnikov:1986jp,Shaposhnikov:1987tw,Cohen:1994ss} for $CP$ violation on phase boundaries.

The generation of the baryon asymmetry is typically described through fluid equations. These can be derived in classical mechanics using conservation laws, i.e. energy, momentum and current conservation.
Microscopically, they can also be derived from Boltzmann equations. In the given context of particle physics, reaction rates then have to be added from scattering theory, assuming that individual scattering
events are well separated in spacetime, which is not always justifiable. One may therefore prefer to combine the
derivation of the reaction rates with the kinematic evolution of the system by evolving the QFT in real time. An approach that is based on first principles
that proves efficient in specifying initial conditions,
evolving the system and evaluating observables is given by the closed-time path (CTP) formalism~\cite{Schwinger:1960qe,Keldysh:1964ud,Calzetta:1986cq}. We present and
compare these different approaches in Section~\ref{sec:fluid}.

In Section~\ref{sec:CPV}, we discuss aspects of the calculation of $CP$-violating decay rates
of heavy particles. In view of comparing with reaction rates obtained in the CTP
framework, we explicitly evaluate these rates in Section~\ref{sec:decayasymmetry} using the optical theorem. We also discuss how the decay asymmetries for Dirac fermions  can be reduced to
results for Majorana fermions, thus covering and relating both of these scenarios
for out-of-equilibrium decays. In Section~\ref{sec:mixing:variants}, we review several approaches to
computing the decay rates of mixing right-handed neutrinos (RHNs), the differences of which
become relevant for very mass-degenerate mixing systems.

The results for the decay asymmetry are then used in Section~\ref{sec:classLG} in order
to derive fluid equations governing leptogenesis in the early Universe. An important technical point
is the subtraction of real intermediate states (RIS), that is necessary in order to unitarize the
system such that no asymmetry can be generated or persist in equilibrium. This procedure is owed
to the fact that scattering processes with intermediate RHNs are not
clearly distinguishable from decay and inverse decay processes. Also the expansion of the Universe is accounted for.
Individual reaction rates appear in the fluid equations when integrated over phase space. These are
closely related to expressions that we encounter in calculations using the CTP approach and
are therefore presented in Section~\ref{sec:integralform} in integral form for comparison.

In Section~\ref{sec:kineq:firstprinciples}, we then derive the fluid equations
from first principles of QFT using the CTP formalism. In integral form, the results
will be compared with those from Section~\ref{sec:integralform}. This should
give some concrete insights into the relation between standard methods
using Boltzmann equations with scattering rates and the CTP approach. Notably,
in the CTP approach no subtraction of RIS is necessary because the correct
counting is implemented by construction.

To give an example for the benefit of using the CTP method, we discuss
leptogenesis in the resonant regime in Section~\ref{sec:reslg}, with a
particular focus on the extremely mass-degenerate case where the masses
of the RHNs are only separated within their decay width.
It turns out that the off-diagonal correlations of the RHNs,
that the decay asymmetry depends on, can be computed by
solving the Schwinger--Dyson equations on the CTP, what effectively
amounts to the correct resummation of the one-loop absorptive
self-energy corrections. This method therefore resolves the questions
brought up in Section~\ref{sec:mixing:variants}.

In Section~\ref{sec:pedestrian}, we then briefly review how to solve the
fluid equations for leptogenesis in the strong washout regime and arrive
at an analytic approximation in a form
that can be compared with the observed value~(\ref{BAU:CMB}) for the asymmetry.
To give a flavour
of more advanced applications, we outline in Section~\ref{sec:beyond:sw} how the CTP methods have recently been used
to include radiative effects that become important beyond the scenario of minimal strong
washout.

Similar to resonant leptogenesis, mixing and oscillating systems may
play a role in electroweak baryogenesis. In Section~\ref{sec:ewbg}, we review
the computation of $CP$-violating
source terms using the CTP approach and gradient expansion.
As an introduction, we first discuss force terms in classical kinetic equations.
Applying the CTP approach to
a fermionic system, we next discuss the semiclassical force that is present
independent of mixing and then identify a resonantly enhanced force term
for mixing systems. We eventually give some directions how to complete the calculation
in order to arrive at a prediction for the asymmetry that is to be compared with the
observed value~(\ref{BAU:CMB}).

This outline also sets the scope of the present article which are calculational methods
for baryogenesis. Given the large number of possible scenarios for baryogenesis
and variants thereof a comprehensive phenomenological survey of the field
would require a far more extensive treatment than is offered in this work. We note
nonetheless that recently, a series of review articles on leptogenesis has appeared that
provides some comprehensive, yet detailed, review of the state of the art~\cite{Dev:2017trv,Drewes:2017zyw,Dev:2017wwc,Biondini:2017rpb,Chun:2017spz,Hagedorn:2017wjy}.
The present article presents basic elements necessary in order to set up some minimal
fluid equations for baryogenesis, and it additionally justifies these equations
in a more detailed way based on the CTP methods. It is therefore aimed to be complementary
to the more detailed but also more specific and less general calculations presented in the typical
research literature.

We further remark that while discussing leptogenesis in the strong washout regime and
electroweak baryogenesis,
we do not cover the
many other aspects where CTP or finite-temperature techniques are relevant for
baryogenesis calculations, apart from the qualitative discussion of Section~\ref{sec:beyond:sw}. Among these are leptogenesis involving relativistic RHNs~\cite{Akhmedov:1998qx,Asaka:2005pn,Abada:2015rta,Hernandez:2015wna,Drewes:2016gmt,Antusch:2017pkq,Ghiglieri:2018wbs,Drewes:2017zyw},
flavoured leptogenesis~\cite{Endoh:2003mz,Pilaftsis:2005rv,Abada:2006fw,Nardi:2006fx,Beneke:2010dz,Blanchet:2011xq} or thermal and other radiative corrections to
leptogenesis in general~\cite{Anisimov:2010gy,Laine:2011pq,Besak:2012qm,Garbrecht:2013gd,Garbrecht:2013urw,Laine:2013lka,Biondini:2013xua}. While noting that many of these works use methods from equilibrium field
theory or the time evolution in the canonically quantized formalism, we emphasize that the CTP approach is of particular appeal because it uses the exact time evolution of the system as derived from first principles of QFT as a platform and nonetheless
allows for a very efficient representation in terms of Feynman diagrams. This time evolution then has to be approximated systematically, which to some large extent is the work lying ahead.

\section{Theoretical foundation of kinetic and fluid equations for baryogenesis}
\label{sec:fluid}

Predictions for the BAU are usually obtained from a solution to fluid equations
describing the processes that generate the asymmetry in the early Universe. These are
cast as ordinary, linear differential equations that can easily be solved numerically
or even using analytic approximations~(cf. Section~\ref{sec:pedestrian} for leptogenesis
in the strong washout regime). To some large extent, the fluid equations can be formulated in terms of averaged interaction rates for charge and number densities.
Balancing these rates for the creation and annihilation of particles is a
very efficient way of setting up these equations, cf. the discussion in Section~\ref{sec:setup:fluid:eq} on leptogenesis. This procedure can be justified
on a more basic level from the Boltzmann kinetic equations, i.e. from \emph{classical statistical
mechanics}, combined with $S$-matrix elements from QFT. Alternatively, we
can derive fluid equations based on QFT altogether, where the \emph{closed time-path
approach} turns out to be very suitable and powerful. In this section, we give an
overview about both methods.

\subsection{Boltzmann equations from classical statistical mechanics}
\label{sec:stat:mech}

\paragraph{Kinetic equations}
In many cases, reaction networks can be effectively described in terms of
Boltzmann equations for distribution functions $f(\mathbf{x},\mathbf{p},t)$ of particles
with mass $m$, i.e. in the form
\begin{align}
\label{Eq:Boltzmann}
\frac{df}{dt}=
\frac{p\cdot\partial}{p^0}f=\frac{\partial f}{\partial t}+\frac{\partial f}{\partial \mathbf x}\cdot{\mathbf v}
+\frac{\partial f}{\partial \mathbf{p}}\cdot \frac{d \mathbf p}{dt}
={\cal C}\,,
\end{align}
where $\tau$ is the proper time of the particles with momentum $\mathbf p$, partial
derivatives with respect to $\mathbf x$ and $\mathbf p$ are understood to be the
spatial and momentum gradients, $\mathbf v=d\mathbf x/dt=\mathbf p/[\gamma(\mathbf v) m]$, $p^0=\sqrt{\mathbf{p}^2+m^2}$ and $\gamma(\mathbf v)$ is the Lorentz factor.
The left hand side of this equation is referred to as the convection term.
In the convection term, we can also identify a term that is relevant in presence
of a force field $d\mathbf p/d\tau=\mathbf F$.
On the right-hand side, there is the collision term ${\cal C}$.

When we let $X$ label the particle that the distribution $f_X(p_X)$ refers to, the collision term takes the form
\begin{align}
\label{coll:term}
{\cal C}=&\frac{1}{2p^0}\int\prod\limits_i\frac{d^3 p_i}{(2\pi)^32p_i^0}(2\pi)^4\delta^4(p+p_{A1}+\cdots-p_{B1}-\cdots)
\\\notag
\times&
\big\{
(1\pm f)(1\pm f_{A1})\cdots f_{B1}\cdots|{\cal M}_{B_1 B_2\cdots\to X A_1 A_2\cdots}|^2
\\\notag&
-
f f_{A1}\cdots (1\pm f_{B1})\cdots|{\cal M}_{X A_1 A_2\cdots\to B_1 B_2\cdots}|^2
\big\}\,,
\end{align}
where
$p^0=\sqrt{\mathbf p^2 +m^2}$, $p^0_i=\sqrt{\mathbf p_i^2 +m_Y^i}$, the index
$i$ runs through the particles labeled by $B_1, B_2,\ldots$
and $A_1, A_2,\ldots$, and ${\cal M}$ is the invariant matrix element of the reaction indicated
in the subscript. We have included here the Bose enhancement and Pauli suppression terms
$\pm f_X$ such that this collision term also accounts for quantum statistics.

A key assumption here is that the range of the force between two particles is much shorter than
their average distance. Famously, this does not apply to the Coulomb force, which is a long-range interaction, leading to divergences in the collision integral. In classical electrodynamics, long range interactions can be included through Vlasov equations from  which one can obtain important plasma
phenomena such as Landau damping and Debye screening. Another approximation made is that
we assume that the system can be described by a set of distribution functions for
the single particle species rather than one joint phase-space distribution function
accounting for the trajectory of each single particle, i.e. we truncate the
Bogoliubov-Born-Green-Kirkwood-Yvon (BBGKY) hierarchy at its lowest order.

\paragraph{Fluid equations}
In cosmological calculations one is mostly interested in the evolution of
charge and number densities that are conserved under the long-range gauge interactions (e.g. baryon or lepton number or the number of dark matter particles) such that
the pertaining issues not necessarily need to be dealt with. The role
of the gauge interactions is then mainly to maintain kinetic equilibrium, i.e. to establish the Fermi-Dirac
or Bose-Einstein distributions
\begin{align}
f(p,x)=\frac{1}{{\rm e}^{\frac{p_\nu u_{{\rm b}}^{\nu}-\mu}{T}}\pm 1}\,.
\end{align}
Rather than on the distribution function accounting for an infinite number of degrees
of freedom at each spacetime point, the problem now depends on the temperature field $T(x)$,
the field of bulk velocities $u_{\rm b}^\nu(x)$
and the  field of chemical potentials $\mu(x)$. These fields
can be extracted from the distributions that appear in the kinetic equations
by taking moments
\begin{subequations}
\label{eq:densities}
\begin{align}
\varrho(x)=&\int\frac{d^3 p}{(2 \pi)^3} f(p,x)\,,\\
n^\mu(x)=&\int\frac{d^3 p}{(2 \pi)^3} \frac{p^\mu}{m} f(p,x)\,,\\
&\ldots\notag\,,
\end{align}
\end{subequations}
where the dots indicate that
these generalize to a tower of quantities. In many cases---as in the discussions
of the present article---it is sufficient to consider only the number density
$\varrho$ or in addition the fluid-density current $n^{\mu}$.
A charge density $q$ and a current are obtained from taking the difference for particles
and antiparticles. In kinetic equilibrium, the chemical potentials
are opposite such that e.g.
\begin{align}
\label{q:mu}
q=\varrho-\left(\varrho|_{\mu\to-\mu}\right)
=\left\{\begin{array}{l}\frac{T^2}{6} \mu+\cdots\quad\textnormal{(massless chiral fermions)}\\\frac{T^2}{3}\mu+\cdots\quad\textnormal{(massless bosons)} \end{array}\right.
\,,
\end{align}
where it has been assumed that $\mu/T\ll 1$ and the dots indicate higher order terms
in this parameter.

It is often also the case that the bulk velocity can be nonrelativistically
approximated as $u_{\rm b}^\mu=(1,\mathbf{v}_{\rm b})$, i.e. $|\mathbf v_{\rm b}|\ll 1$, such that
we can extract it through
\begin{align}
\int \frac{d^3 p}{(2\pi)^3} \mathbf{p} f(p)=\left\{\begin{array}{l}\frac{7}{90} \pi^2 \mathbf{v_{\rm b}} T^4+\cdots\quad\textnormal{(massless chiral fermions)}\\\frac{4}{45}\mathbf{v_{\rm b}} T^4+\cdots\quad\textnormal{(massless bosons)}\end{array}\right.\,.
\end{align}

In terms of these variables, we may therefore obtain from the Boltzmann equations a
simplified network of fluid equations. A very typical form of these is (cf. Section~\ref{sec:CPconservingprocesses})
\begin{subequations}
\label{fluid:equations}
\begin{align}
\label{fluid:equation:q}
\dot q+\nabla\cdot \mathbf j=& -\Gamma q\,,\\
\mathbf{j}=\mathbf{v_{\rm b}} q=& - D \mathbf\nabla q
\,,
\end{align}
\end{subequations}
where $\mathbf j$ is the current density of the charge density $q$.
The coefficients $\Gamma$, which may be referred to as decay rates, and
$D$, the diffusion constants, have to be extracted by taking the corresponding
moments of the collision terms in the Boltzmann equations.
While the fluid equations~(\ref{fluid:equations}) form a closed network
of four equations for the four variables which are $q$ and the components of $\mathbf v_{\rm b}$,
this closure is typically an idealization and relies on the approximate validity of
discarding terms involving higher moments of the distribution functions.

Fluid equations in the form of Eqs.~(\ref{fluid:equations}) are a typical starting point for
setting up calculations of particle reactions in the early Universe. In the absence
of quantum coherence, $q$ and $\mathbf{j}$ are promoted to carry a species index and
$\Gamma$ and $D$ will then in general be matrix valued. It is also possible to account
for quantum coherence when promoting $q$ and $\mathbf{j}$ to matrices in the space of
particle species. In that case, additional commutator and anticommutator structures
appear in the equations that will play an important part in the remainder of this review,
cf. Eqs~(\ref{sterile:oscillations}) and~(\ref{eq:transport:gradients}).

\subsection{First principles of QFT: closed time-path approach}
\label{sec:CTP}

As an alternative to introducing distribution functions within classical kinetic theory
in order to obtain results for e.g. current densities and their time evolution,
we may also note that current densities can be defined as expectation values
of two-point functions, i.e.
\begin{align}
\label{currents}
j^\mu(x)=\left\langle[{\rm i}
\partial^\mu\phi^*(x)] \phi(x)-\phi^*(x)[{\rm i}\partial^\mu \phi(x)]\right\rangle\,,
\quad
j^\mu(x)=\langle\bar\psi(x)\gamma^\mu\psi(x)\rangle={\rm tr}\langle\psi(x) \gamma^\mu \bar\psi(x)\rangle
\end{align}
for scalar and fermion fields, respectively. For fermion fields, one may in addition
compute e.g. an axial current or currents associated with states of definite helicity
or spin, as we will get back to in more detail in Section~\ref{sec:ewbg}
[cf. Eqs.~(\ref{currents:densities:speven}) and~(\ref{currents:densities:spodd})].

Given a certain state (that in general can be a statistically mixed quantum state),
we aim to evaluate expectation values of the above type. In equilibrium field theory, the
state is given by a canonical or grand canonical ensemble and is time-translation invariant.
In general however, the system evolves with time, and we specify the state through some initial condition at one time and evaluate the expectation values at another. We therefore need the time evolution, and a particularly powerful method
of computing it results from
combining the Schwinger--Keldysh CTP
method~\cite{Schwinger:1960qe,Keldysh:1964ud} with Schwinger--Dyson equations that are derived from a two-particle irreducible (2PI) effective action~\cite{Calzetta:1986cq,Prokopec:2003pj,Prokopec:2004ic,Berges:2004yj}.

\begin{figure}[t]
\includegraphics[width=\textwidth]{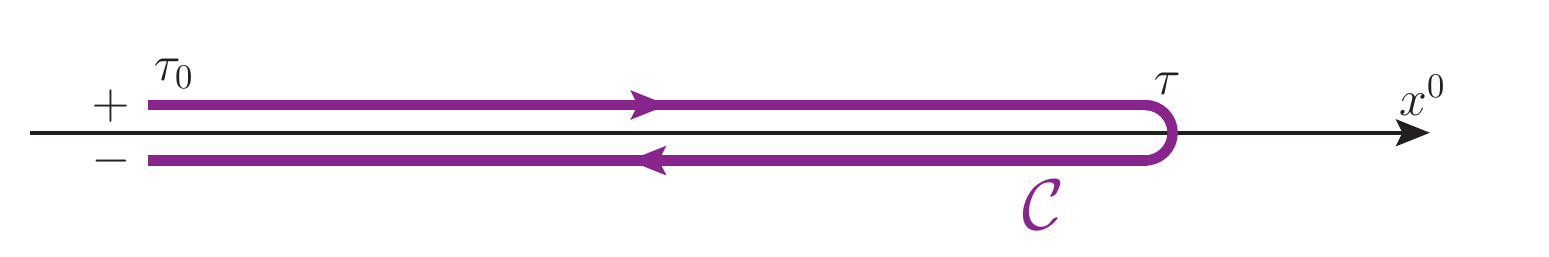}
\vskip-.4cm
\caption{\label{fig:CTP}The closed time-path (CTP).}
\end{figure}

In order to discuss this approach, we begin with a real scalar field $\phi$
for simplicity. For this,
we define the \emph{in-in generating functional}:
\begin{align}
\label{Z:in:in}
Z[J_+,J_-]=&\int\cal{D \phi(\tau)}{\cal D}\phi^-_{\rm in}{\cal D}\phi_{\rm in}^+\langle \phi^-_{\rm in}|\phi(\tau)\rangle
\langle\phi(\tau)|\phi_{\rm in}^+\rangle \langle\phi^-_{\rm in}|\varrho|\phi_{\rm in}^+\rangle
\\\notag
=&\int{\cal D}\phi^-{\cal D}\phi^+
{\rm e}^{{\rm i}\int d^4 x\{{\cal L}(\phi^+)-{\cal L}(\phi^-)+J_+\phi^+-J_-\phi^-\}}\langle\phi^-_{\rm in}|\varrho|\phi_{\rm in}^+\rangle
\,.
\end{align}
In the first step (cf. Figure~\ref{fig:CTP}), we integrate over a set of in states $|\phi_{\rm in}^\pm\rangle$ that
can be weighted statistically by a density matrix $\varrho$. Then, we evolve $|\phi_{\rm in}^+\rangle$ specified at the time $\tau_0$
to some finite time $\tau$, where we insert a complete set of states $|\phi(\tau)\rangle\langle\phi(\tau)|$, and eventually we evolve back to $\langle\phi_{\rm in}^-|$
at $\tau_0$.
For both steps of this time evolution, we introduce Lagrangian terms $J_\pm(x)\phi^\pm(x)$ for
variational purposes. When writing down the corresponding path-integral representation
in the second step, we note that
this procedure generates two branches of integration that we denote by $+$ and $-$. The
factor of minus one from the backward time integration along the $-$ branch is then attributed to the
integrand, such that both branches appear as integrands in an integral over $d^4x$, where each of the
Lagrangian terms pertaining to the minus branch attains a factor of minus one.

Taking variational derivatives, we arrive at path-ordered Green functions:\\
\begin{align}
{\rm i}\Delta^{ab}(x,y)=-\frac{\delta^2}{\delta J_a(x) \delta J_b(y)}\log Z[J_+,J_-]\Big|_{J_\pm=0}
=\langle{\cal C}[\phi^a(x)\phi^b(y)]\rangle\,,
\end{align}
where the subscript ${\cal C}$ indicates that operators are arranged from
the right to the left according to the sequence in which their time arguments appear
on the closed time-path shown in Figure~\ref{fig:CTP}, and higher correlation
functions are obtained in an analogous way. Observables such as the currents~(\ref{currents}) can then be calculated straightforwardly from these
Green functions.

From the path integral representation in Eq.~(\ref{Z:in:in}), we infer that
the standard Feynman rules hold, with the generalization that there are $+$
and $-$ vertices. These vertices are connected through the Green functions ${\rm i}\Delta^{ab}$ with the matching CTP superscripts, and vertices of the $-$ type receive
an extra factor of minus one beause of the negative sign in front of the $-$ Lagrangian
in the exponent in Eq.~(\ref{Z:in:in}).

It is further customary to introduce specific superscripts for the various types
of two-point functions on the CTP that are more or less connected to their
mathematical and physical significance. For a  two-point function $G^{ab}$,
we write
\begin{align}
\label{CTP:indices}
G^>(x,y)=G^{-+}(x,y)\,,\quad G^<(x,y)=G^{+-}(x,y)\,,\quad
G^T(x,y)=G^{++}(x,y)\,,\quad G^{\bar T}(x,y)=G^{--}(x,y)\,,
\end{align}
where $G^T$ ($G^{\bar T}$) is the time (anti-time) ordered two-point function
and $G^{<,>}$ are \emph{Wightman functions}. Of practical importance
are the causal, retarded and advanced two-point functions
\begin{align}
G^{\rm r}=G^T-G^<=G^>-G^{\bar T}\,,\quad G^{\rm a}=G^T-G^>=G^<-G^{\bar T}
\end{align}
as well as the spectral function
\begin{align}
G^{\cal A}=\frac{\rm i}{2}(G^>-G^<)=\frac{1}{2{\rm i}}(G^{\rm a}-G^{\rm r})
\end{align}
and the Hermitian function
\begin{align}
G^{\rm H}=\frac{1}{2}(G^{\rm a}+G^{\rm r})\,.
\end{align}
These relations also imply the identity
\begin{align}
\label{T:Tbar:gr:le}
G^T+G^{\bar T}=G^>+G^<\,.
\end{align}

To progress further towards models of phenomenological interest, we
now proceed with complex scalar fields $\phi$ and four-component spinors
$\psi$, i.e.
\begin{align}
{\rm i}\Delta(x,y)=\langle{\cal C}[\phi(x)\phi^*(y)]\rangle\,,\quad
{\rm i}S(x,y)=\langle{\cal C}[\psi(x) \bar\psi(y)]\rangle\,.
\end{align}
Note that the spinor indices are not contracted here, i.e. that
$S$ is endowed with  $4\times 4$ structure in spinor space.
We have suppressed here the CTP indices as well as possible flavour indices.

Throughout this article, we are concerned with complex fermion masses or mass matrices, that
are in general Nonhermitian. For a complex mass, we use the notation of a lower case $m$
that is decomposed into its real and imaginary part as
\begin{align}
m^{\rm R}={\rm Re}[m]\,,\quad m^{\rm I}={\rm Im}[m]\,.
\end{align}
We may also attach subscripts indexing the particle species to $m$ and $m^{\rm R,I}$. It is further
useful to define the term pertaining to this mass that appears in spinor products,
\begin{align}
\widehat m= m^{\rm R}+{\rm i} \gamma^5 m^{\rm I}\,,\quad
\widehat m^*= m^{\rm R}-{\rm i} \gamma^5 m^{\rm I}\,.
\end{align}
When dealing with mass matrices, we use capital letters (aside from the exception of masses for RHNs, where customarily capital letters are used for the mass terms). A matrix $M$ of general form can be decomposed into Hermitian
and Antihermitian parts as
\begin{align}
\label{M:decomp}
M^{\rm H}=\frac12\left(M+M^\dagger\right)\,,\quad M^{\rm A}=\frac{1}{2{\rm i}}\left(M-M^\dagger\right)\,,
\end{align}
which appear in spinor products as
\begin{align}
\label{M:spinor}
\widehat M= M^{\rm H}+ {\rm i}\gamma^5 M^{\rm A}\,.
\end{align}
General expressions for mass matrices can be reduced to the case of a single fermion species
by taking $M^{\rm H}\to m^{\rm R}$ and $M^{\rm A}\to m^{\rm I}$.

In order to derive kinetic equations, it turns out useful to note the
Hermiticity properties of the Wightman functions
\begin{subequations}
\begin{align}
\left[{\rm i}\Delta^{<,>}(x,y)\right]^\dagger=&{\rm i}\Delta(y,x)\,,\\
\left[{\rm i}\gamma^0 S^{<,>}(x,y)\right]^\dagger=&{\rm i}\gamma^0 S(y,x)\,.
\end{align}
\end{subequations}
Here, we have explicitly replaced the spinor matrix $A$
defined in Eq.~(\ref{A:C}) with $\gamma^0$, as it is
appropriate for all common representations of Dirac matrices
such as the Weyl and the Dirac representations.

Green functions on the CTP can be expanded using straightforward perturbation theory.
However, a more powerful computation method arises from using Schwinger--Dyson equations and deriving the latter from the two-particle-irreducible (2PI) effective
action. The main advantage is that the Schwinger--Dyson equations readily express the
time evolution of the system in an integro-differential form similar to kinetic equations.
The 2PI effective action can be expressed as
\begin{align}
\label{Gamma2PI}
\Gamma[\Delta,S]=B+{\rm i}\,{\rm tr}[{\Delta^{(0)}}^{-1}\Delta]-{\rm i}\,{\rm tr}[{S^{(0)}}^{-1}S]
+{\rm i}\,{\rm tr}\log\Delta^{-1}-{\rm i}\,{\rm tr}\log S^{-1}+\Gamma_2[\Delta,S]\,,
\end{align}
where $B$ is the classical action, ${\Delta^{(0)}}^{-1}$ the Klein-Gordon,
${S^{(0)}}^{-1}$ the Dirac operator,
\begin{align}
\label{Gamma2}
\Gamma_2[\Delta,S]\equiv -{\rm i}\times \textnormal{the sum of 2PI vacuum graphs}\,,
\end{align}
and the relative signs between the terms concerning fermions and bosons arise from
the contributions of the quadratic fluctuations about the classical field configuration
in the path integral (i.e. the one-loop determinants).

Defining the self energies
\begin{subequations}
\begin{align}
\Pi^{ab}(x,y)=&{\rm i}ab\frac{\delta\Gamma_2[\Delta,S]}{\delta\Delta^{ba}(y,x)}\,,\\
\slashed\Sigma^{ab}(x,y)=&-{\rm i}ab\frac{\delta\Gamma_2[\Delta,S]}{\delta S^{ba}(y,x)}\,.
\label{selferg:fermi}
\end{align}
\end{subequations}
and taking functional derivatives, we obtain
\begin{subequations}
\begin{align}
\frac{\delta\Gamma[\Delta,S]}{\delta\Delta(y,x)}=&0
\Leftrightarrow
{\rm i}{\Delta^{(0)}}^{-1}(x,y)
-{\rm i}\Delta^{-1}(x,y)-{\rm i}\Pi(x,y)=0\,,\\
\frac{\delta\Gamma[\Delta,S]}{\delta S(y,x)}=&0
\Leftrightarrow
-{\rm i}{S^{(0)}}^{-1}(x,y)
+{\rm i}S^{-1}(x,y)+{\rm i}\slashed \Sigma(x,y)=0\,.
\end{align}
\end{subequations}
The convolution from the right with the full propagators yields the Schwinger--Dyson equations on the CTP
\begin{subequations}
\begin{align}
\label{SDE:scalar}
\left[-\partial^2-M^2\right]{\rm i}\Delta^{ab}(x,y)=&a\delta_{ab}{\rm i}\delta^4(x-y) + 
\sum\limits_c c \int d^4z\Pi^{ac}(x,z){\rm i}\Delta^{cb}(z,y)\,,\\
\label{SDE:fermi}
\left[{\rm i}\slashed\partial - M^{\rm H} -{\rm i}\gamma^5 M^{\rm A}\right]{\rm i}S^{ab}(x,y)=&a\delta_{ab}{\rm i}\delta^4(x-y) + 
\sum\limits_c c \int d^4z\slashed\Sigma^{ac}(x,z){\rm i} S^{cb}(z,y)\,.
\end{align}
\end{subequations}
To this end, we have presented the expressions for scalar fields and Dirac fermions, in order
to highlight agreements and differences in their treatment. As the examples in this article are mainly concerned with the evolution of fermions, we proceed with these and leave the scalar particles aside
for now. The theory for the latter, along with fermions, is developed in detail in Refs.~\cite{Prokopec:2003pj,Prokopec:2004ic}.

Since for the Green functions on the CTP, there are two linearly independent combinations,
the same holds true for the Schwinger--Dyson equations as well. It proves useful to
set up equations for the retarded or advanced Green functions
\begin{align}
\left[{\rm i}\slashed\partial-M^{\rm H}-{\rm i}\gamma^5 M^{\rm A}\right]S^{\rm r,a}(x,y)-\int d^4z \slashed\Sigma^{\rm r,a}(x,z) S^{\rm r,a}(z,y)=\delta^4(x-y)\,,
\end{align}
and for the Wightman functions
\begin{align}
&\left[{\rm i}\slashed\partial-M^{\rm H}-{\rm i}\gamma^5 M^{\rm A}\right]S^{<,>}(x,y)-
\int d^4z \left[\slashed\Sigma^{\rm H}(x,z) S^{<,>}(z,y)+\slashed\Sigma^{<,>}(x,z) S^{\rm H}(z,y) \right]
\notag\\
=&\frac12\int d^4z \left[\slashed\Sigma^>(x,z)S^<(z,x)-\slashed\Sigma^<(x,z)S^>(z,x)\right]\,,
\end{align}
where this latter equation is called the Kadanoff--Baym equation.

To proceed with deriving kinetic equations that are suitable for a reduction to the Boltzmann
or fluid form, we perform a Wigner transformation of the two-point functions,
\begin{align}
G(k,x)=\int d^4r\, {\rm e}^{{\rm i} k\cdot r} G\left(x+\frac r2,x-\frac r2 \right)\,,
\end{align}
where $G$ stands here either for a propagator or a self energy. When applying this transformation
to the Schwinger--Dyson equations, we encounter convolutions that transform into~\cite{Groenewold:1946kp,Moyal:1949sk,Greiner:1998vd}
\begin{align}
\int d^4r\, {\rm e}^{{\rm i} k\cdot r} \int d^4 z G\left(x+\frac r2,z\right)F\left(z,x-\frac r2 \right)
={\rm e}^{-{\rm i}\diamond} \left\{G(k,x)\right\}\left\{F(k,x)\right\}\,,
\end{align}
where
\begin{align}
\diamond \left\{G(k,x)\right\}\left\{F(k,x)\right\}=\frac 12 \left(\frac{\partial G(k,x)}{\partial x^\mu} \frac{\partial F(k,x)}{\partial k_\mu}-\frac{\partial G(k,x)}{\partial k^\mu} \frac{\partial F(k,x)}{\partial x_\mu}\right)\,.
\end{align}
In Wigner space, the Kadanoff--Baym equation thus reads
\begin{align}
&\left[\slashed k +\frac{\rm i}{2}\slashed\partial-M^{\rm H}{\rm e}^{-\frac{\rm i}{2}\overleftarrow\partial\cdot\overrightarrow\partial_{\!k}}-{\rm i}\gamma^5 M^{\rm A}{\rm e}^{-\frac{\rm i}{2}\overleftarrow\partial\cdot\overrightarrow\partial_{\!k}} \right]S^{<,>}-{\rm e}^{-{\rm i}\diamond}\{\slashed\Sigma^{\rm H}\}\{S^{<,>}\}-{\rm e}^{-{\rm i}\diamond}\{\slashed\Sigma^{<,>}\}\{S^{\rm H}\}\notag\\
=&\frac12 {\rm e}^{-{\rm i}\diamond}\left(\{\slashed\Sigma^>\}\{S^<\}-\{\slashed\Sigma^<\}\{S^>\}\right)\,,
\label{KBE:Wigner:Fermions}
\end{align}
and, similarly, the retarded and advanced propagators obey
\begin{align}
&\left[\slashed k +\frac{\rm i}{2}\slashed\partial-M^{\rm H}{\rm e}^{-\frac{\rm i}{2}\overleftarrow\partial\cdot\overrightarrow\partial_{\!k}}-{\rm i}\gamma^5 M^{\rm A}{\rm e}^{-\frac{\rm i}{2}\overleftarrow\partial\cdot\overrightarrow\partial_{\!k}} \right]S^{\rm r,a}-{\rm e}^{-{\rm i}\diamond}\{\slashed\Sigma^{\rm r,a}\}\{S^{\rm r,a}\}
=1\,,
\label{retav:Wigner:Fermions}
\end{align}
where we recall that we have suppressed the spinor and flavour indices, such that the right-hand side
is to be understood as a an identity operator in tensor space. The arrows over the partial derivatives
are indicating onto which side these are acting, and the subscript $k$ indicates a partial
derivative with respect to the four-momentum $k^\mu$.

It turns out that the right-hand side of the Kadanoff--Baym equation~(\ref{KBE:Wigner:Fermions}) can be physically interpreted
as the QFT analogue of the collision term in the classical Boltzmann equations. It should therefore
vanish in thermal equilibrium which can be readily seen from the Kubo-Martin-Schwinger (KMS) relation~\cite{Kubo:1957mj,Martin:1959jp}:
Any two-point function $G$ (that here may be a propagator or a self energy) in equilibrium
at a temperature $T$ satisfies
\begin{align}
\label{eq:KMS}
G^{>}(k,x)=\pm{\rm e}^{k^0/T}G^{<}(k,x)\,,
\end{align}
where the plus sign holds for bosons, the minus sign for fermions. Since the collision term vanishes in
equilibrium, this automatically implies that no matter--antimatter asymmetry can then be generated
in compliance with Sakharov's conditions.

The Kadanoff--Baym equations can then be decomposed into kinetic and constraint equations, where
the former can be further reduced to Boltzmann and then to fluid equations, while the latter
yield the necessary input on the spectral properties. These matters are best illustrated
on concrete examples, as we present in
Section~\ref{sec:kineq:firstprinciples} for leptogenesis in
the strong washout regime and in Section~\ref{sec:ewbg} for baryogenesis at
phase boundaries, where the most important scenario is electroweak baryogenesis. We
therefore outline the remaining steps in formulating kinetic equations by pointing to
the relevant results in the subsequent sections.

At tree level, when neglecting self energies,
the solutions to Eqs.~(\ref{KBE:Wigner:Fermions}) and~(\ref{retav:Wigner:Fermions})
are shown in Appendix~\ref{App:treeprop}.
While depending on the particular
problem, these solutions may or may not be a suitable starting point for a perturbative
expansion, their form is useful in order to understand the physical meaning
of the particular terms that appear in the Schwinger--Dyson equations on the CTP.
To see already to this end how to proceed toward fluid equations, we note
that the Kadanoff--Baym equations can be decomposed in kinetic and so-called constraint
equations. Kinetic equations for the Wightman
functions of nonrelativistic fermions are given by Eq.~(\ref{kineq:fermi:simple})
which are then reduced to kinetic equations~(\ref{kin:eq:N}) for the distribution functions. The general (fully relativistic) theory for fermionic Wightman functions turns out to be more involved and is
discussed for spatially varying mixing mass matrices in Section~\ref{sec:ewbg} on electroweak baryogenesis.

\subsection{Comparing remarks concerning Boltzmann and the closed time-path methods}
\label{sec:compare:methods}

When substituting $S$-matrix elements into the Boltzmann equations, we implicitly
make use of the full machinery of scattering theory: amplitudes are computed in time-ordered
perturbation theory and related to matrix elements using the Lehmann-Symanzik-Zimmermann (LSZ)
reduction formula. This constitutes a conceptual detour because for baryogenesis,
we are interested
in the time-evolution of observables such as current densities that can be expressed as expectation
values of operators. Moreover, in a finite density system, it is often not possible to
identify free asymptotic states as scattering theory requires. As we discuss in this article,
in the context of baryogenesis, this leads to problems concerning the counting of real intermediate
states or the treatment of quasi-degenerate mixing states.

The method of directly computing the time evolution of the correlation functions of interest, as
it is done in the CTP approach, is therefore more direct. The evolution of the system
is expressed entirely in terms of correlation functions. These can also be used to specify initial conditions as well as to derive the observables of interest. In this functional approach, there is no reference to operators nor we have to rely on the computation of amplitudes that need to be related
to observables via the LSZ machinery. The price to pay for this simplification is to give up Lorentz
symmetry because it is explicitly broken by the finite density background that implies a preferred plasma frame, such
that we have to give up the simple form in which the Green functions behave under Wick rotations that
is enjoyed in vacuum field theory. A combination of both methods (and in addition with methods of
equilibrium field theory) may therefore often prove as the most efficient means of achieving the particular
calculational goal.

\section{$CP$ violation}
\label{sec:CPV}

The violation of $CP$ symmetry is a hallmark of quantum physics that requires the
interference of amplitudes with $CP$-even and $CP$-odd phases. For mesons of the SM, where $CP$-violation has actually been observed, obtaining the $CP$-even phases requires experimental or numerical input.
In contrast, in many scenarios of baryogenesis, the $CP$-even phase emerges in a comparably
simple manner and can be computed in terms of on-shell cuts of loop amplitudes. We review
the basics of such calculations as well as the interplay with the $CP$-odd phases in Section~\ref{sec:decayasymmetry}.
We first discuss the calculation of the decay asymmetry in a model with heavy Dirac fermions
and then show how this can be reduced to the case of Majorana neutrinos that is relevant
for leptogenesis~\cite{Fukugita:1986hr}. It should therefore be clear that baryogenesis from out-of-equilibrium decays
does not necessarily rely on Majorana fermions. For the calculation of the asymmetry
produced in the early Universe, we nonetheless focus on leptogenesis as the
most plausible scenario.

For the most time since its inception, there has been some debate on wave-function
or mixing contributions to the $CP$ asymmetry in leptogenesis, a point that
is most relevant in the resonant regime, where there is a pronounced mass degeneracy
for the RHN states. We review some
approaches concerning this matter in Section~\ref{sec:mixing:variants}, while
in Section~\ref{sec:reslg} we show that the decay asymmetry in the limit of
strong mass degeneracy can strongly depend on the dynamical circumstances,
i.e. on the oscillation time
and washout strength of the system of mixing RHNs in the early Universe.

\subsection{Decay asymmetries, odd and even phases under $CP$ conjugation}

\label{sec:decayasymmetry}

Some salient features of $CP$-violating decays can be explained on
a model given by the Lagrangian
\begin{align}
\label{L:simplemodel}
{\cal L}=&\bar F\left({\rm i}\slashed\partial-m_F^{\rm R}-{\rm i}\gamma^5 m_F^{\rm I}\right)F+\bar G\left({\rm i}\slashed\partial-m_G^{\rm R}-{\rm i}\gamma^5 m_G^{\rm I}\right)G
+\phi^*\partial^2\phi+\bar f {\rm i} \slashed\partial f\notag\\
-&\sum_{X=F,G}\left(y_X^* \bar f \phi P_{\rm R} X+\tilde y_X^*\overline{f^C}\phi^* P_{\rm L} X +{\rm h.c.}\right)\,,
\end{align}
where $F,G$ are Dirac spinors, $f$ a left-chiral fermion, $\phi$ a massless complex scalar field and h.c. denotes Hermitian conjugation. As indicated in the sum, $X$ is a Dirac-fermion
field that either stands for $F$ or $G$, a notation that is also used in the subsequent discussion.
The left and right chiral projectors are given by
\begin{align}
P_{\rm L,R}=\frac{1\mp\gamma^5}{2}\,.
\end{align}
Given this Lagrangian, we discuss here how the decay of the massive particles $F$ and $G$ 
and their antiparticles can lead to an asymmetry in $f$ and $\phi$.

We have written the Yukawa interactions in a form such that it is manifest that $F$ and $G$
can decay into both, $f$ and $f^C$ but note that alternatively, we could express the
Yukawa interaction terms as
\begin{align}
\overline{f^C}\phi^* P_{\rm L} X=\overline{X^C} \phi^* P_{\rm L} f\,.
\end{align}
From the transformation properties stated in Appendix~\ref{app:discrete},
in particular Eqs.~(\ref{CP:scalar}) and~(\ref{CP:pseudoscalar}),
we see that $CP$ conjugation takes the effect $m_X\to m_X^*$,
$y_X\to y_X^*$. We therefore refer to the
arguments of these complex parameters as $CP$-odd phases.
In general, there are also extra arbitrary phases (i.e. $\alpha^{CP}$ in Appendix~\ref{app:discrete})
from the definition of the action of $CP$ conjugation on the fields,
that we choose to be zero for convenience in the present discussion.
However, due to the freedom of field redefinitions, individual terms in the Lagrangian
that violate $CP$ do in general (and typically) not lead to $CP$ violation. Rather,
physical $CP$ indiscretion relies on rephasing invariants, i.e. combinations that are $CP$ odd and that
are invariant under field redefinitions by complex phases. We will shortly come back to this point on
the present example.

For definiteness, we focus on the decays of the massive fermion $F$.
Working in the rest frame
of the decaying particle $F$, where $p^\mu=(|m_F|,\mathbf 0)$,
up to one loop order, the decay rate into fermions and antiscalars
can be written as
\begin{align}
\label{decay:rate}
\Gamma_{F\to f\phi^*}
=\frac{1}{2 |m_F|}\int\frac{d^3 q}{(2\pi)^3 2|\mathbf q|}\frac{d^3 q^\prime}{(2\pi)^3 2|\mathbf q^\prime|}(2\pi)^4\delta^4(p-q-q^\prime)\overline{\sum_{\rm pol}}\left|{\rm i}{\cal M}^{\rm LO}_{F\to f\phi^*} +{\rm i}{\cal M}^{\rm vert}_{F\to f\phi^*}
+{\rm i}{\cal M}^{\rm wv}_{F\to f\phi^*} \right|^2\,,
\end{align}
where $q$ and $q^\prime$ are the momenta of $f$ and $\phi^*$,
and the $CP$-conjugate rate can be obtained accordingly.
Here ${\rm i}{\cal M}$ are the invariant matrix elements, ${\rm LO}$ indicates the tree level, leading order contribution and ${\rm vert}$
and ${\rm wv}$ the one-loop vertex and wave-function type corrections. Under the sum sign, ``${\rm pol}$''
indicates the polarization sum over the particles $f$ and the bar above the average over an unpolarized sample of $F$.
The one-loop terms decompose into absorptive and dispersive contributions,
\begin{align}
{\rm i}{\cal M}^{\rm vert}_{F\to f\phi^*}
+{\rm i}{\cal M}^{\rm wv}_{F\to f\phi^*}
={\rm i}{\cal M}^{\rm abs}_{F\to f\phi^*}
+{\rm i}{\cal M}^{\rm dis}_{F\to f\phi^*}\,,
\end{align}
and correspondingly for the additional decays implied by the Lagrangian~(\ref{L:simplemodel}).
The absorptive contributions can be isolated by the use of the optical theorem, or, equivalently
by extracting the discontinuous part of the loop integrals, which appears when
it is possible to cut the loop such that all cut particles are kinematically allowed to simultaneously be on shell. It then gives rise to an
extra $CP$-even phase of ${\rm i}={\rm e}^{{\rm i}\frac{\pi}{2}}$. (In general, it is not
necessary to attribute the $CP$ even phase to purely absorptive contributions to the
amplitude, i.e. we could add absorptive and dispersive contributions at
one loop together such that the phase in general is different from $\pi/2$.)
For the $CP$-conjugate rates, this implies the respective relations
\begin{subequations}
\begin{align}
{\rm i}{\cal M}^{\rm LO}_{F\to f\phi^*}=&\left({\rm i}{\cal M}^{\rm LO}_{F^{CP}\to f^{CP}\phi}\right)^*\,,\\
\label{conj:abs}
{\rm i}{\cal M}^{\rm abs}_{F\to f\phi^*}=&-\left({\rm i}{\cal M}^{\rm abs}_{F^{CP}\to f^{CP}\phi}\right)^*\,,\\
{\rm i}{\cal M}^{\rm dis}_{F\to f\phi^*}=&\left({\rm i}{\cal M}^{\rm dis}_{F^{CP}\to f^{CP}\phi}\right)^*\,.
\end{align}
\end{subequations}
We note that if we were not setting the explicit phases in the definition of
$CP$ conjugation to zero, there would
be additional overall phases multiplying each of the three amplitudes on the right-hand side.
These are spurious phases that turn out to be immaterial when it comes to physical observables,
i.e. the decay rate.

In our setup, we can therefore explicitly attribute the complex conjugation in Eq.~(\ref{conj:abs})
to a $CP$-odd, the factor of minus one to the square of a $CP$-even phase. As a consequence,
\begin{align}
\left|{\rm i}{\cal M}^{\rm LO}_{F\to f\phi^*} +{\rm i}{\cal M}^{\rm abs}_{F\to f\phi^*}\right|^2
\not=
\left|{\rm i}{\cal M}^{\rm LO}_{F^{CP}\to f^{CP}\phi} +{\rm i}{\cal M}^{\rm abs}_{F^{CP}\to f^{CP}\phi}\right|^2\,,
\end{align}
and therefore
the decay rates of particles and antiparticles as per Eq.~(\ref{decay:rate}) are different.
Crucially, the difference in the rates is due to the interference of amplitudes that
have both, different $CP$-odd and $CP$-even phases. Since physical $CP$ violation thus
relies on interference it is a hallmark phenomenon of quantum physics.

In view of the comparison with the results derived in the CTP framework,
it is interesting to explicitly quote those crucial interference terms
that~\cite{Yanagida:1980gf,Masiero:1981vp,Fukugita:1986hr,Covi:1996wh} are proportional to
both, the $CP$-even and $CP$-odd factors.

The leading contribution to the decay rate~(\ref{decay:rate}) is proportional to
\begin{align}
\overline{\sum_{\rm pol}}\left|{\rm i}{\cal M}_{F\to f\phi^*}^{\rm LO}\right|^2
=\overline{\sum_{\rm pol}}\left|{\rm i}{\cal M}_{F\to f^{CP}\phi}^{\rm LO}\right|^2
=\frac{|y_F|^2}{2} |m_F|^2\,.
\end{align}
In the following, we refer to the momenta of the fermions $f$ and bosons $\phi$ that
appear as decay products by $q$ and $q^\prime$, when they appear in internal cuts
by $k$ and $k^\prime$  (as well as for the antiparticles).
Accounting for the extra factor of $1/(2|m_F|)$ in Eq.~(\ref{decay:rate})
and integrating over the phase space $\{d^3 q,d^3 q^\prime\}$, what gives a factor of
$1/(8\pi)$, the decay rate at leading order is
\begin{align}
\Gamma^{\rm LO}_{F\to f \phi^*}=\Gamma^{\rm LO}_{F\to f^{CP} \phi}=\frac{|y_F|^2|m_F|}{32\pi}\,.
\end{align}

\begin{figure}[t]
\begin{align*}
&\raisebox{-.8cm}{\includegraphics[scale=0.5]{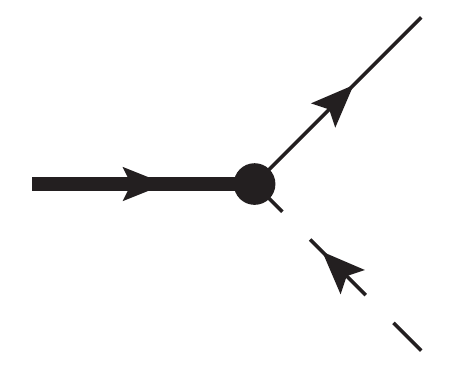}}\times\left(\raisebox{-.9cm}{\includegraphics[scale=0.5]{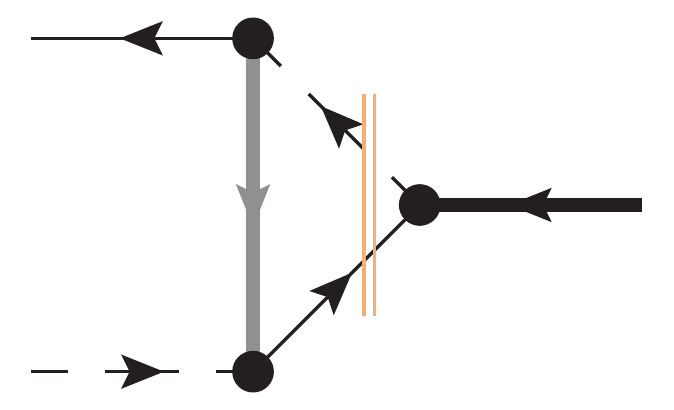}}+\raisebox{-.9cm}{\includegraphics[scale=0.5]{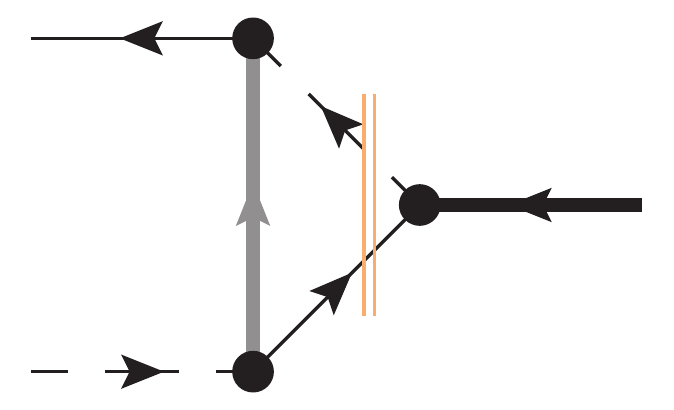}}\right)\\
=&\frac{1}{2}\raisebox{-.8cm}{\includegraphics[scale=0.5]{TreeFstar.pdf}}\times
\mbox{\Huge$\int$}\left(\raisebox{-.9cm}{\includegraphics[scale=0.5]{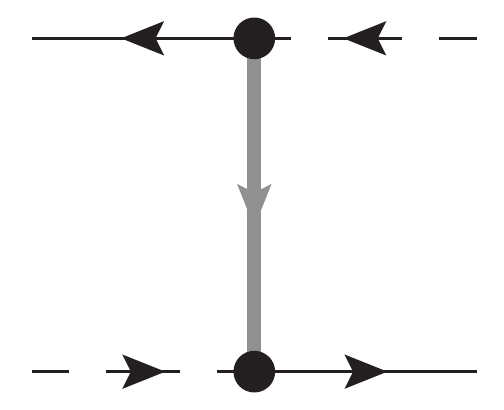}}d\Gamma\raisebox{-.8cm}{\includegraphics[scale=0.5]{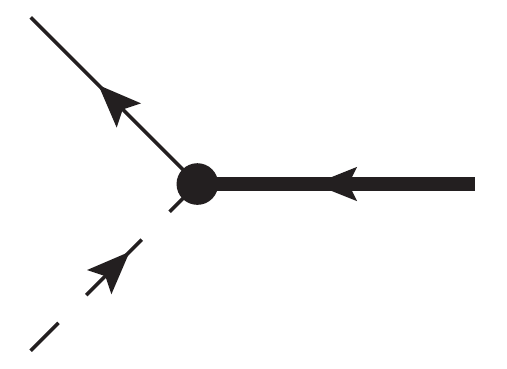}}+
\raisebox{-.9cm}{\includegraphics[scale=0.5]{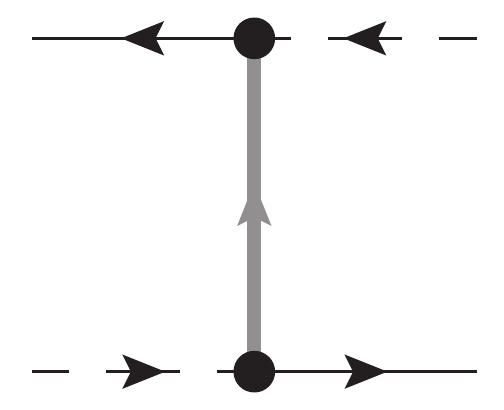}}d\Gamma\raisebox{-.8cm}{\includegraphics[scale=0.5]{TreeF.pdf}}
\right)
\end{align*}

\begin{align*}
&\raisebox{-.8cm}{\includegraphics[scale=0.5]{TreeFstar.pdf}}\times\left(\raisebox{-.9cm}{\includegraphics[scale=0.5]{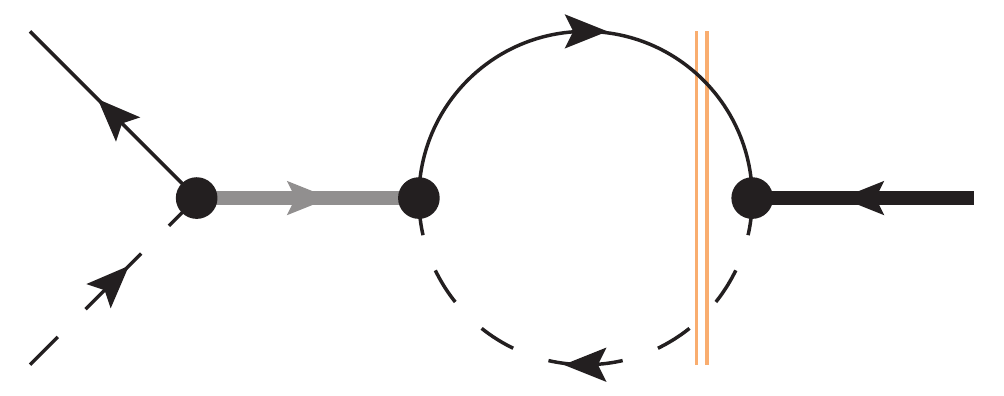}}+\raisebox{-.9cm}{\includegraphics[scale=0.5]{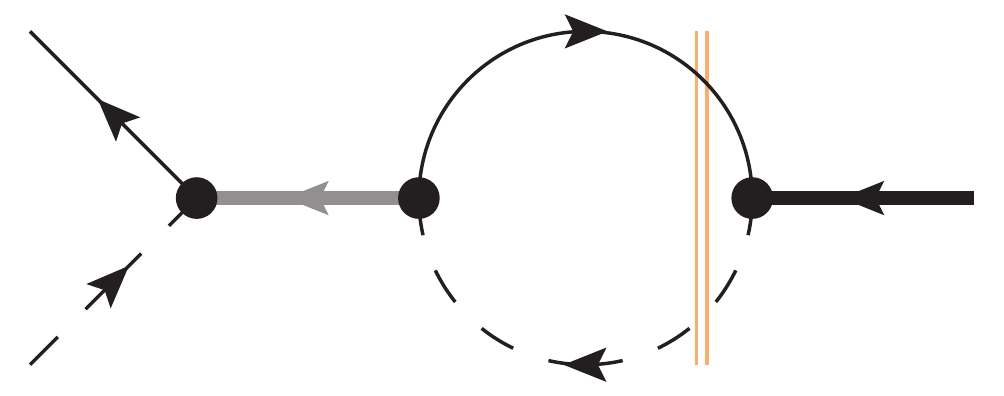}}\right)\\
=&\frac{1}{2}\raisebox{-.8cm}{\includegraphics[scale=0.5]{TreeFstar.pdf}}\times
\mbox{\Huge$\int$}\left(\raisebox{-.9cm}{\includegraphics[scale=0.5]{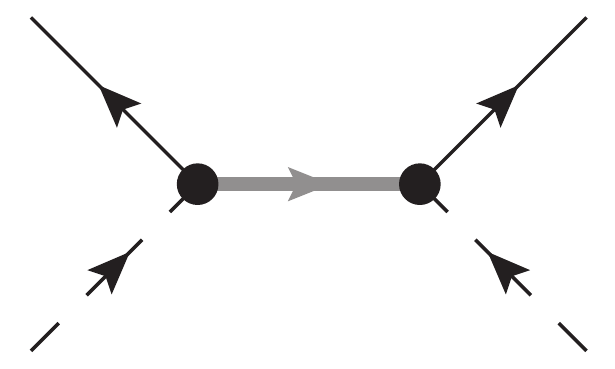}}d\Gamma\raisebox{-.8cm}{\includegraphics[scale=0.5]{TreeF.pdf}}+
\raisebox{-.9cm}{\includegraphics[scale=0.5]{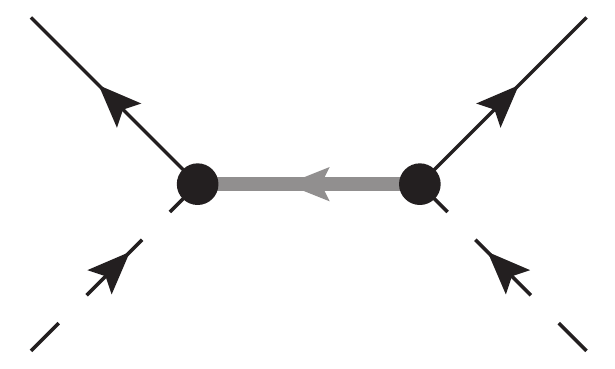}}d\Gamma\raisebox{-.8cm}{\includegraphics[scale=0.5]{TreeF.pdf}}
\right)
\end{align*}
\caption{\label{fig:optical}Diagrammatic representation of the application of the
optical theorem in Eqs.~(\ref{optical:vertex}) and~(\ref{msquared:wv}). 
Individual graphs correspond to amplitudes ${\rm i}{\cal M}$ when the decay products
appear on the left,  ${\rm i}{\cal M}^*$ when they appear on the right.
The black bold solid lines
stand for propagators of the fermion $F$, grey bold solid lines for the fermion $G$, thin solid
lines for the fermion $f$ and dashed lines for the scalar field $\phi$. The integrals are over
the phase space of the on-shell cuts indicated by the placement of $d\Gamma$.
We refer to the cut momenta appearing in the integrals $d\Gamma$ as
$\mathbf k$ and $\mathbf k^\prime$ for fermions $f$ and bosons $\phi$, respectively. The decay products carry the momenta $\mathbf q$ and $\mathbf q ^\prime$. All cut and all external
particles are imposed to be on shell.
}
\end{figure}

Using the optical theorem as
illustrated in Figure~\ref{fig:optical}, i.e. for on-shell cut momenta, the vertex contribution is obtained as
\begin{align}
\overline{\sum_{\rm pol}}\left({\rm i}{\cal M}_{F\to f\phi^*}^{\rm LO}\right)^* {\rm i}{\cal M}_{F\to f\phi^*}^{\rm abs,vert}
=&-\frac12 y_F \tilde y_F^* \tilde y_G^* y_G \int\frac{d^3 k d^3 k^\prime}{(2\pi)^6 2 |k^0| 2 |{k^\prime}^0|}
(2\pi)^4\delta^4(p-k-k^\prime)\notag\\
\times&
\frac{{\rm i}\,{\rm tr}\left[(\slashed p +\widehat m_F^*)P_{\rm L}\slashed q P_{\rm R} (\slashed q - \slashed k^\prime +\widehat m_G^*) P_{\rm R} \slashed k P_{\rm L}\right]}{(q-k^\prime)^2-|m_G|^2}\notag\\
=&
-\frac12 y_F \tilde y_F^* \tilde y_G^* y_G
\frac{m_F m_G^*}{32\pi}\int\limits_{-1}^1 d\cos\vartheta \frac{{\rm i} |m_F|^2 (1-\cos\vartheta)}{\frac12 |m_F|^2(1+\cos\vartheta)+|m_G|^2}
\,.
\label{optical:vertex}
\end{align}
Note the appropriate prefactors due to the averaging over the spin states of the
decaying particle and due to the application of the optical theorem.
Further,
$\mathbf q$ is the  momentum of the outgoing lepton, $\mathbf k$ of the lepton running in
the loop, and $\vartheta$ is the angle between these.
This expression is represented by the right-hand side of the first equation
in Figure~\ref{fig:optical}. In particular, the integration over $d^3k d^3 k^\prime$
corresponds to the phase-space integral $d\Gamma$ over the on-shell cuts in the graphical expression.

Similarly, the wave-function contribution is obtained as
\begin{align}
&\overline{\sum_{\rm pol}}\left({\rm i}{\cal M}_{F\to f\phi^*}^{\rm LO}\right)^* {\rm i}{\cal M}_{F\to f\phi^*}^{\rm abs,wv}\notag\\
=&-\frac12 y_F \tilde y_F^* \tilde y_G^* y_G \int\frac{d^3 k d^3 k^\prime}{(2\pi)^6 2 |k^0| 2 |{k^\prime}^0|}(2\pi)^4\delta^4(p-k-k^\prime)\notag\\
&\hspace{2.5cm}
\times
{\rm tr}\left[(\slashed p +\widehat m_F^*)P_{\rm L}\slashed q P_{\rm R} (\slashed p +\widehat m_G^*) P_{\rm R} \slashed k P_{\rm L}\right]
\frac{{\rm  i}}{|m_F|^2- |m_G|^2}
\notag\\
=&-\frac12 y_F^* \tilde y_F \tilde y_G y_G^*\frac{{\rm  i}m_F m_G^*}{|M_F|^2- |M_G|^2}4 q_\mu \frac{p^\mu}{32\pi}
=-2{\rm i} y_F \tilde y_F^* \tilde y_G^* y_G\frac{|m_F|^2}{64\pi}\frac{ m_F m_G^*}{|m_F|^2 - |m_G|^2}\,,
\label{msquared:wv}
\end{align}
which corresponds to the right-hand side of the second equation
in Figure~\ref{fig:optical}.
In the second step, we have isolated the loop function, that again is
essentially the result of a phase-space integral,
\begin{align}
\label{Sigma:hat}
\hat\Sigma^\mu=p^\mu/(32\pi)\,,
\end{align}
which we will use to compare with corresponding quantities appearing in the CTP approach.
Note that by virtue of the optical theorem,
the computation of the $CP$-violating effects has been reduced to
deriving tree-level amplitudes and their interference terms and eventually
to performing a phase-space integral.
The limit $|m_F|^2 - |m_G|^2\to 0$ is referred to as the resonant regime. Clearly, as the mass difference approaches zero, the validity of the result~(\ref{msquared:wv})
will break down. We discuss this further in Section~\ref{sec:mixing:variants} as
well as Section~\ref{sec:reslg}, where we present the resolution relevant for
scenarios of out-of-equilibrium decay in the early Universe.

Using this result and carrying out the
integration in Eq.~(\ref{optical:vertex}), we define
\begin{subequations}
\begin{align}
{\cal I}^{\rm vert}(m_F,m_G)=&-\frac{1}{16 \pi}\left[1-\left(1+\frac{|m_G|^2}{|m_F|^2}\right)\log\left(1+\frac{|m_F|^2}{|m_G|^2}\right)\right]\,,\\
{\cal I}^{\rm wv}(m_F,m_G)=&-\frac{1}{32 \pi}\frac{|m_F|^2}{|m_F|^2-|m_G|^2}\,.
\end{align}
\end{subequations}
In terms of these, the crucial interference terms are given by
\begin{align}
\label{intereference:terms}
\overline{\sum_{\rm pol}}\left({\rm i}{\cal M}^{\rm LO}_{F\to f\phi^*}\right)^*
{\rm i}{\cal M}^{\rm abs}_{F\to f\phi^*}
= {\rm i}\left({\cal I}^{\rm vert}+{\cal I}^{\rm wv}\right)y_F \tilde y_F^* y_G^* \tilde y_G m_F m_G^*
\,.
\end{align}

It is next convenient to parametrize the decay rate as
\begin{align}
\Gamma_{F\to f\phi^*}=\Gamma^{\rm LO}_{F\to f\phi^*}\left(1+\varepsilon\right)\,.
\end{align}
The effect of $CP$ conjugation on the interference term~(\ref{intereference:terms}) is the complex conjugation
of the $CP$-odd phase, i.e. the conjugation of all explicit masses and couplings, while the
$CP$-even phase
remains unaffected. It therefore follows that
\begin{align}
\Gamma_{F^{CP}\to f^{CP}\phi}=\Gamma^{\rm LO}_{F\to f\phi^*}\left(1-\varepsilon\right)\,.
\end{align}
At the present level of accuracy,
the parameter $\varepsilon$ can therefore be identified with the decay asymmetry
\begin{align}
\varepsilon=&\frac{\Gamma_{F\to f\phi^*}-\Gamma_{F^{CP}\to f^{CP}\phi}}{\Gamma_{F\to f\phi^*}+\Gamma_{F^{CP}\to f^{CP}\phi}}
=\frac{\overline{\sum}_{\rm pol} \left[\left({\rm i}{\cal M}^{\rm LO}_{F\to f\phi^*}\right)^*
{\rm i}{\cal M}^{\rm abs}_{F\to f\phi^*}-\left({\rm i}{\cal M}^{\rm LO}_{F^{CP}\to f^{CP}\phi}\right)^*
{\rm i}{\cal M}^{\rm abs}_{F^{CP}\to f^{CP}\phi}\right]+{\rm c.c.}}{2\overline{\sum}_{\rm pol}\left|{\rm i} {\cal M}^{\rm LO}_{F\to f\phi^*}\right|^2}\notag\\
=&\frac{4 \, {\rm Im}[y_F^* \tilde y_F y_G \tilde y_G^* m_F^* m_G]}{|y_F|^2|M_F|^2}
\left({\cal I}^{\rm vert}+{\cal I}^{\rm wv}\right)\,,
\label{epsilon:simplemodel}
\end{align}
where c.c. stands for complex conjugation.

We can now check explicitly that this result is invariant under field redefinitions
through rephasings
\begin{align}
F\to {\rm e}^{{\rm i}\varphi^F \gamma^5} F\,,\quad G\to {\rm e}^{{\rm i}\varphi^G \gamma^5} G\,,\quad
\phi\to{\rm e}^{{\rm i}\varphi^\phi}\phi\,,\quad f\to {\rm e}^{-{\rm i}\varphi^f} f\,,
\end{align}
upon which the terms in the Lagrangian~(\ref{L:simplemodel}) transform as
\begin{align}
m_{F,G}\to m_{F,G} {\rm e}^{-2 {\rm i} \varphi^{F,G}}\,,\quad
y_{F,G} \to y_{F,G} {\rm e}^{{\rm i} \varphi^{F,G}+{\rm i} \varphi^\phi +{\rm i}\varphi^f}\,,\quad
\tilde y_{F,G} \to \tilde y_{F,G} {\rm e}^{- {\rm i} \varphi^{F,G}-{\rm i} \varphi^\phi -{\rm i}\varphi^f}\,.
\end{align}
Indeed, this leaves the decay asymmetry~(\ref{epsilon:simplemodel}) unaffected.
The combination $\arg[y_F^* \tilde y_F y_G \tilde y_G^* m_F^* m_G]$ is therefore
a physical $CP$-odd phase that leads in conjunction
with interference effects involving a $CP$-even, absorptive phase
to $CP$-violating decays.

A viable scenario of baryogenesis can emerge when attributing baryon number e.g. to the
field $f$. As the Universe expands and cools, equal populations of heavy particles
$F$ and $F^{CP}$ can then decay and leave behind a baryon asymmetry. We discuss such a
scenario on the simpler example of leptogenesis, that is of great
phenomenological relevance, however.

To obtain the decay asymmetry for leptogenesis from the above results we introduce the spinor fields
$N_1=(1/\sqrt2)\left(F+ F^C\right)$,
$N^\prime_1=({\rm i}/\sqrt2)\left(F- F^C\right)$,
as well as
$N_2=(1/\sqrt2)\left(G+ G^C\right)$
$N^\prime_2=({\rm i}/\sqrt2)\left(G-G^C\right)$.
By construction, these are Majorana spinors, i.e. $N_{1,2}^C=N_{1,2}$ and $N^{\prime C}_{1,2}=N^\prime_{1,2}$.
Imposing further $\tilde y_X^*=y_X$ on the Yukawa couplings, $N^\prime_{1,2}$
decouple such that we are effectively left with the Lagrangian
\begin{align}
{\cal L}=\sum_{i=1,2}\left[\frac12\bar N_i\left({\rm i}\slashed \partial -M_i{\rm e}^{{\rm i}\alpha_i\gamma^5}\right)N_i
-Y^*_i \bar{f} \phi P_{\rm R} N_i - Y_i \bar N_i \phi^* P_{\rm L} f
\right]\,,
\end{align}
where $M_{1,2}=m_{F,G}$, $\alpha_{1,2}=\arg(m_{F,G})$, $Y_{1,2}=\sqrt{2} y_{F,G}$. Following
widely used notation, we
denote here the masses of the RHNs with a capital letter (not being entirely consequent about
the notation declared above, where we have reserved this for mass
matrices). When dealing with
mixing and oscillations of RHNs in Sections~\ref{sec:mixing:variants}
and~\ref{sec:kineq:firstprinciples}, it is convenient to take $M_{1,2}$
as entries of a mass matrix for these particles that will be defined in due course.

Further, we identify $f$ with the lepton weak isodoublet $\ell$ of the SM, and  the
scalar field is related to the Higgs doublet $H$ as $\phi=(\epsilon H)^\dagger$
(where $\varepsilon$ is the totally antisymmetric rank-two tensor and we suppress the notation of all ${\rm SU}(2)$ contractions), what leads to
an extra factor of $g_w=2$ in the wave-function contributions relative to the vertex ones.
From the above results, we then obtain the decay asymmetry
\begin{align}
\label{epsilon:N1}
\varepsilon=\frac{\Gamma_{N\to \ell\phi^*}-\Gamma_{N\to \ell^{CP}\phi}}{\Gamma_{N\to \ell\phi^*}+\Gamma_{N\to \ell^{CP}\phi}}
=\frac{2\,{\rm Im}[{Y_1^*}^2 {Y_2}^2 M_1^* M_2]}{|Y_1|^2 |M_1|^2}\left({\cal I}^{\rm vert}(M_1,M_2)+g_w {\cal I}^{\rm wv}(M_1,M_2)\right)\,.
\end{align}

\subsection{Variants of computing $CP$ violation from mixing and oscillations}
\label{sec:mixing:variants}

So far, we have assumed an external state for the RHN that is
purely $N_1$ without an admixture of $N_2$. However,  the lighter mass eigenstate, for 
which we have been computing the decay asymmetry
contains an admixture of $N_2$ that is generated by the Yukawa couplings.
This admixture is present even when attributing the renormalized dispersive loop corrections
to the mass terms, resulting in a redefined diagonal mass matrix, because the absorptive effects
cannot be removed this way.
In the previous section, we have apparently accounted for this mixing
through the absorptive part of the wave-function correction for the RHN. Another point of view one can take calculationally
is to attribute the mixing to the external RHN states of the $S$ matrix. One may do so by generalizing the LSZ reduction formula to account
for the mixing induced by absorptive corrections~\cite{Pilaftsis:1997jf,Pilaftsis:2003gt}.

To work this out in more detail, in Ref.~\cite{Pilaftsis:1997jf} the one-loop improved
Dirac operator
\begin{align}
S^{-1}=\left(
\begin{array}{cc}
\slashed p- M_1 -\slashed\Sigma_{N 11} & -\slashed\Sigma_{N 12}\\
-\slashed \Sigma_{N 21} & \slashed p -M_2 - \slashed\Sigma_{N 22}
\end{array}
\right)
\end{align}
is inverted as
\begin{subequations}
\label{prop:improved}
\begin{align}
S_{11}=&\left(\slashed p -M_1 -\slashed \Sigma_{N 11} -\slashed \Sigma_{N 12}\frac{1}{\slashed p- M_2 -\slashed\Sigma_{N 22}}\slashed\Sigma_{21}\right)^{-1}\,,\\
S_{12}=&S_{11}\slashed\Sigma_{N 12}\frac{1}{\slashed p- M_2 -\slashed\Sigma_{N 22}}
=\frac{1}{\slashed p- M_1 -\slashed\Sigma_{N 11}}\slashed\Sigma_{N 12}S_{22}\,,
\end{align}
\end{subequations}
where the remaining entries follow from replacing $1\leftrightarrow 2$.
The loop functions are obtained from the one defined in Eq.~(\ref{Sigma:hat}) as
\begin{align}
\hat{\slashed\Sigma}=\gamma_\mu\hat\Sigma^\mu\,,\quad
\slashed\Sigma^{\cal A}_{N ij}=g_w Y_i^* Y_j \hat{\slashed \Sigma} P_{\rm R}
+g_w Y_i Y_j^* \hat{\slashed \Sigma} P_{\rm L}\,,\quad
\slashed\Sigma_{N ij}=\slashed\Sigma^{\rm disp}_{N ij}-{\rm i}\slashed\Sigma^{\cal A}_{N ij}\,.
\end{align}
We identify here the absorptive part of $\slashed\Sigma_N$ with the {\it spectral} self energy
$\slashed\Sigma^{\cal A}_N$, where the sign is dictated by imposing the time-ordered boundary conditions
on the resummed propagator.
The factor $g_w=2$ 
serves also as a reminder that the two fundamental ${\rm SU}(2)_{\rm L}$ degrees of freedom are understood to run in the loop.

Next, consider the three-point Green function pertaining to the decay of the lighter
RHN state to $\ell$ and $\phi^*$. The leg for the RHN is then given by the resummed external propagator $S$. In order to arrive at an amplitude, it is amputated
by multiplying with the diagonal components $S_{11}^{-1}$. As a result, the
squared amplitude reads
\begin{align}
\label{squamp:LSZ}
\overline{\sum_{\rm pol}} \left({\rm i}{\cal M}_{N_1\to \ell\phi^*}\right)^*{\rm i}{\cal M}_{N_1\to \ell\phi^*}
=&\frac12 {\rm tr}\left[ (S^{-1}_{11})^\dagger \left(Y_1 S_{11} + Y_2 S_{21}\right)^\dagger \slashed q \left(Y_1 S_{11} + Y_2 S_{21}\right)S^{-1}_{11} P_{\rm R} (\slashed p +M_1) \right]\notag\\
\supset& -{\rm i}\frac{g_w}{64 \pi} Y_1^2 {Y_2^*}^2\frac{M_1 M_2^*}{|M_1|^2-|M_2|^2+2\slashed p\, {\rm i}\slashed\Sigma^{\cal A}_{N 22}} +{\rm c.c.}\,,
\end{align}
where we can immediately verify
the agreement with Eq.~(\ref{msquared:wv}). In addition, there
appears an extra Antihermitian term in the denominator which we 
recognize as the decay width of the off-shell particle $N_2$:
$2\slashed p \slashed\Sigma^{\cal A}_{N 22}=\mathbbm{1}_4\, \frac12 {\rm tr}[\slashed p \slashed\Sigma^{\cal A}_{N 22}]=\mathbbm{1}_4\,p^0\Gamma_{22}$.

Yet another perspective arises from describing the time evolution through
a Hamiltonian. For the mixing of light neutrinos,
such an approach has first been used in Ref.~\cite{Sigl:1992fn}. For nonrelativistic RHNs in the context of leptogenesis, the kinetic energy can be neglected
and the different helicity states evolve in the
same way such that one can use the effective Hamiltonian~\cite{Flanz:1994yx,Flanz:1996fb}
\begin{align}
H=\left(\begin{array}{cc}M_1 -\frac{\rm i}2\Gamma_{11} & -\frac{\rm i}2\Gamma_{12}\\ -\frac{\rm i}2\Gamma_{12}^* & M_2-\frac{\rm i}2\Gamma_{22}\end{array}\right)\,,
\end{align}
where
\begin{align}
\label{Gamma:ij}
\Gamma_{ij}=\frac{1}{2 p^0} {\rm tr}[\slashed p \slashed\Sigma^{\cal A}_{ij}]
=\frac{g_w}{32\pi}\left(Y_i Y_j^*+Y_i^* Y_j\right) p^2\,,
\end{align}
such that $\Gamma_{ii}$ is the total decay rate of $N_i$.
This system has two eigenstates of mass and lifetime. If 
the lifetimes of the two states are different enough, only one of these
is relevant at late times. Also, if the mass difference and hence the
oscillation time is much shorter than the lifetime, one will produce
states that perform fast oscillations about the eigenstates, such that
the phases pertaining to these oscillations average out and can be neglected~\cite{Garbrecht:2011aw}. Under
these circumstances, only the relative phase that appears when constructing
these eigenstates from $N_{1,2}$ is of relevance for $CP$ violation from mixing.
Provided $|\Gamma_{ij}|\ll|M_1-M_2|$ (An exact expression can be easily found but its form is less compact and illuminating.), these eigenstates are
\begin{align}
\label{Hamiltonian:mixing:states}
v_1\approx\left(\begin{array}{c}1\\0\end{array}\right)+\left(\begin{array}{c}0\\1\end{array}\right)\delta v_1\,,\quad
v_2\approx\left(\begin{array}{c}0\\1\end{array}\right)+\left(\begin{array}{c}1\\0\end{array}\right)\delta v_2\,,
\end{align}
where
\begin{align}
\delta v_1=\frac{\frac{\rm i}{2}\Gamma_{12}^*}{M_1-M_2+\frac{\rm i}{2}\Gamma_{11}-\frac{\rm i}{2}\Gamma_{22}}\,,\quad
\delta v_2=-\frac{\frac{\rm i}{2}\Gamma_{12}}{M_1-M_2+\frac{\rm i}{2}\Gamma_{11}-\frac{\rm i}{2}\Gamma_{22}}\,,
\end{align}
and, for simplicity, we take here $M_{1,2}$ to be real.
Assuming in addition that we are close to the resonance, $|M_1-M_2|\ll \bar M=(M_1+M_2)/2$, we
can express the admixture to $N_1$ as
\begin{align}
\label{admixture}
\delta v_1\approx\frac{{\rm i}\bar M\Gamma_{12}^*}{M_1^2-M_2^2+{\rm i}\bar M\left(\Gamma_{11}-\Gamma_{22}\right)}\,.
\end{align}
Thus, we can once more compute the wave-function contribution to the decay asymmetry
when taking for the RHN $N_1$ a mixing state according to $v_1$ in Eq.~(\ref{Hamiltonian:mixing:states}), with the result
\begin{align}
\label{reslg:Hamiltonian}
&\overline\sum_{\rm pol} \left({\rm i}{\cal M}_{N_1\to \ell\phi^*}\right)^*{\rm i}{\cal M}_{N_1\to \ell\phi^*}\notag\\
=&\frac12 {\rm tr}\left[\slashed k P_{\rm R} (\slashed p +M_1)\right]
\left(Y_1+Y_2\frac{{\rm i}\left(Y_1Y_2^*+Y_1^* Y_2\right)\frac{g_w}{32\pi}\bar M^2}{M_1^2-M_2^2+{\rm i}\bar M(\Gamma_{11}-\Gamma_{22})}\right)
\left(Y_1^*-Y_2^*\frac{{\rm i}\left(Y_1^*Y_2+Y_1 Y_2^*\right)\frac{g_w}{32\pi}\bar M^2}{M_1^2-M_2^2-{\rm i}\bar M(\Gamma_{11}-\Gamma_{22})}\right)
\notag\\
\supset& -\frac{{\rm i} g_w Y_1^2 {Y_2^*}^2}{64 \pi}\frac{\bar M^2}{|M_1|^2-|M_2|^2+{\rm i}\bar M(\Gamma_{11}-\Gamma_{22})}+{\rm c.c.}
\end{align}
As stated above, we see that the $CP$ phase is generated by the interference
of the tree-level contributions with the admixture~(\ref{admixture}).
Again, when $|M_1-M_2|\gg |\Gamma_{11}-\Gamma_{22}|$ we recognize agreement
with the result~(\ref{msquared:wv}). However, now the corrections due to the
finite width of the RHNs are different from what is stated in Eq.~(\ref{squamp:LSZ}).

Regarding the denominator term involving
the RHN width, the expression~(\ref{reslg:Hamiltonian}) agrees with what is found
in Ref.~\cite{Anisimov:2005hr}. However, one needs to be aware of the
fact that the admixtures only build up on the oscillation time-scale
that is for nonrelativistic neutrinos given by $|M_1-M_2|^{-1}$.
Given that the lifetime of the decaying RHN is $\Gamma_{11}^{-1}$, as noted in Ref.~\cite{Garny:2011hg},
the results~(\ref{admixture}) and~(\ref{reslg:Hamiltonian}) cannot be used
for leptogenesis calculations
in the interesting regime where $|M_1-M_2|$ is smaller or even much smaller
than $\Gamma_{11}$.

The apparent resolution to the discrepancies in the contributions
from the finite width of the RHNs to the denominator terms is found when appreciating
that the amount of mixing depends on the full real-time dynamics, i.e. on the way the system
deviates from equilibrium as well as on the background evolution and initial conditions.
In particular, it is necessary to solve systematically
and in general settings for the correlation between $N_1$ and $N_2$, that 
yields the crucial $CP$-even phase due to quantum mechanical interference~\cite{Garbrecht:2011aw,Iso:2013lba}.
While the result based on the Hamiltonian evolution in this section assumes
RHN states that decay in vacuum, in Ref.~\cite{Garny:2011hg} an approach of a
vanishing initial distribution of RHNs toward equilibrium in a finite temperature
background is considered. Neither setup realistically models the dynamics in
the expanding Universe.
In Section~\ref{sec:reslg}, we
work out this dynamics in the context of the strong-washout regime of leptogenesis
in the early Universe, using CTP methods~\cite{Iso:2014afa,Garbrecht:2014aga,Dev:2014laa}. Again, the result differs from Eqs.~(\ref{squamp:LSZ}) and~(\ref{reslg:Hamiltonian}) in the way the finite width of the RHNs affects the denominator
of the term describing the resonant enhancement.

We finally note that when carefully specifying the initial conditions
as well as decomposing into helicity eigenstates,
the method based on the Hamiltonian evolution can well be
applied to cosmological calculations, as it has been carried out for neutrino
oscillations~\cite{Sigl:1992fn} or for leptogenesis from oscillations of RHNs with ${\rm GeV}$-scale masses~\cite{Akhmedov:1998qx,Asaka:2005pn,Abada:2015rta,Hernandez:2015wna}. We nonetheless
pursue here the CTP approach because it leads to a straightforward expansion in terms
of Feynman diagrams as explained in Section~\ref{sec:CTP} and because we do not
need to specify a basis of quantum states for the interacting system since the Schwinger--Dyson equations are formulated in terms of Green functions.

\section{Baryogenesis from out-of-equilibrium decays and inverse decays---Classical fluid equations
with QFT cross sections}
\label{sec:classLG}

In order to explain the calculation for baryogenesis based on classical kinetic theory and, in contrast, based on
first principles of QFT, we pick leptogenesis as the simplest and phenomenologically
most relevant scenario from out-of-equilibrium decays. In addition, we choose a parametrically simple situation where
$M_1\ll M_2$ and $\Gamma_{11}\gg H|_{T=M_1}=(4\pi^3 g_\star/45)^{1/2} M_1^2/{m_{\rm Pl}}$, where $g_\star$ is the effective number of relativistic degrees of freedom and $H$ is the Hubble rate. Due to the first relation,
we may assume that during the times relevant for leptogenesis, the abundance of $N_2$ is strongly Maxwell suppressed and can be neglected. The second condition characterizes the {\it strong-washout}
regime. It is of great importance because any preexisting asymmetry at higher temperatures
will be erased by the lepton-number violating interactions involving $N_1$ as the Universe cools (unless the asymmetry is partly preserved in so-called
spectator fields as discussed in Section~\ref{sec:beyond:sw} or the asymmetry
from possible decays of $N_2$ is stored in a different flavour combination
of the doublet leptons that is not aligned with the one coupling to $N_1$~\cite{DiBari:2005st,Engelhard:2006yg}).

\subsection{Setting up the fluid equations}
\label{sec:setup:fluid:eq}

Under the assumptions stated above, we may readily proceed to formulate kinetic or
fluid equations for leptogenesis. Given that the RHNs are nonrelativistic
and that there is kinetic equilibrium
for the charged particles, it is sufficient to only track charge and number densities
rather than the distribution
functions of the species involved. This is because for  $\ell$ and $\phi$, kinetic equilibrium maintains the Fermi-Dirac and Bose-Einstein form of the distributions whereas for the nonrelativistic RHNs, the details of the distribution function
are irrelevant in order to obtain leading order accurate results.
It is therefore the most commonly used and quickest approach to
sidestep the setup of kinetic equations and to directly go for fluid equations in the first place. After having resolved some important matter
regarding real intermediate states that occur in matrix elements
necessary for $CP$ violation in Section~\ref{sec:RIS}, we return in Section~\ref{sec:integralform} to the
systematic derivation of the fluid equations from Boltzmann kinetic equations
along the general lines discussed in Section~\ref{sec:stat:mech}.

For the simple case present, there are only
three densities that we need to track: $n_{N1}$, i.e. the number density of $N_1$ (counting
both helicity degrees of freedom),
and the charge densities of leptons and Higgs bosons $q_\ell$ and $q_\phi$, respectively.
We define the charge densities such that they only count the contribution from
one component of the weak isodoublet. Due to gauge invariance, the charge densities
for the different components in a multiplet are equal.
When normalized to the entropy density
\begin{align}
s=\frac{2\pi^2}{45}g_\star T^3\,,
\end{align}
these quantities are referred to as yields,
$Y_{X}=n_{X}/s$, $Y_{\Delta X}=q_X/s=(n_X-n_{X^{CP}})/s$, where $n_X$ is the number density of the
particle $X$ and $q_X$ is the charge density when summed over particles and
antiparticles.
Further, since we are working in the limit of nonrelativistic RHNs, where
$M_1\gg T$, we approximate the averaged decay rate
\begin{align}
\label{gamma:av}
\gamma=
\Gamma_{N_1 \to \ell \phi^*,\ell^{CP} \phi}
\frac
{
\int\frac{d^3 p}{(2\pi)^3}
\frac{M_1}{\sqrt{\mathbf p^2+M_1^2}}
{\rm e}^{-\sqrt{\mathbf p^2+M_1^2}/T}
}
{
\int\frac{d^3 p}{(2\pi)^3}
{\rm e}^{-\sqrt{\mathbf p^2+M_1^2}/T}
}
=
\frac{K_1(z)}{K_2(z)}
\Gamma_{N_1 \to \ell \phi^*,\ell^{CP} \phi}
\,,
\end{align}
where
$z=M_1/T$ and $\Gamma_{N\to \ell \phi^*,\ell^{CP} \phi}\equiv \Gamma_{11}$,
as defined in Eq.~(\ref{Gamma:ij}), is the
total vacuum decay rate of a singlet neutrino $N_1$ into particles and
antiparticles.
The factor $M_1/\sqrt{\mathbf p^2+M_1^2}$ accounts for time dilation
and can, of course, be verified when evaluating the decay rates for $\mathbf p\not=\mathbf 0$.
Note that
\begin{align}
\frac{K_1(z)}{K_2(z)}=1-\frac{3}{2 z}+\frac{15}{8z^2}+\cdots\,,
\end{align}
such that we explicitly see that this factor accounts for relativistic corrections. Further,
we have approximated the distribution of RHNs by Maxwell statistics as it is appropriate
for $M_1\gg T$. Being sterile particles, the RHNs are not maintained in kinetic equilibrium by
gauge interactions. Deviations of their distribution from the equilibrium form should however
only have a subdominant impact on the final result in the present nonrelativistic limit.

The fluid equations can now easily be written down by balancing
the number densities of the individual particles with their rate of change in the particular reactions.
For the simple model of leptogenesis under consideration, we thus obtain
\begin{subequations}
\label{fluid:RIS}
\begin{align}
\frac{dY_{N1}}{dt}=&-\gamma\left(Y_{N1}-Y_{N1}^{\rm eq}\right)
\,,\\
\frac{dY_{\Delta\ell}}{dt}=&
Y_{N1}\frac{1+\varepsilon}{2}\frac{\gamma}{g_w}
-Y_{N1}\frac{1-\varepsilon}{2}\frac{\gamma}{g_w}
+\left(1-\frac{\mu_\ell}{T}\right)\frac{1+\varepsilon}{2}\frac{\gamma Y_{N1}^{\rm eq}}{g_w}
-\left(1+\frac{\mu_\ell}{T}\right)\frac{1-\varepsilon}{2}\frac{\gamma Y_{N1}^{\rm eq}}{g_w}
\notag\\
-&2\gamma_{\ell \phi^*\to \ell^{CP} \phi}
+2\gamma_{\ell^{CP} \phi \to \ell \phi^*}
+2\gamma^{\rm RIS}_{\ell \phi^*\to  \ell^{CP} \phi}
-2\gamma^{\rm RIS}_{\ell^{CP} \phi \to \ell \phi^*}\,,
\label{BE:ql}
\end{align}
\end{subequations}
where
\begin{align}
\label{YN1:eq}
Y_{N1}^{\rm eq}=\frac{45}{2\pi^4 g_\star}z^2 K_2(z)\approx\frac{45 z^\frac32}{2^\frac32\pi^\frac72g_\star}{\rm e}^{-z}
\end{align}
is the value that $Y_{N1}$ takes in thermal equilibrium (for Maxwell statistics)
and the approximation holds for $z\gg1$.

The right hand side of
Eq.~(\ref{BE:ql}) can be derived from the collision term~(\ref{coll:term}),
where the integrations lead to terms involving the averaged decay rates
$\gamma$ in Eq.~(\ref{gamma:av}) multiplied with the
charge or number densities (i.e. the yields when normalized
to entropy density). We show in more detail how these terms arise
in the derivation from the Boltzmann kinetic equations in Section~\ref{sec:integralform} and from first principles of QFT in Section~\ref{sec:kineq:firstprinciples}.
To this end, we make some remarks on important features of the individual collision terms, i.e. why they take their particular form and the reactions
that they describe.

The first two of the collision terms in Eq.~(\ref{BE:ql})
account for decays of $N$ into $\ell$ and $\ell^{CP}$ and the pertaining decay asymmetry, and
the third and fourth term describe inverse decays.
In Eq.~(\ref{epsilon:N1}), we have obtained the asymmetry $\varepsilon$ for the processes
$N_1\to \ell \phi^*,\ell^{CP} \phi$ in vacuum, which is
applicable to the nonrelativistic regime, where $M_1\gg T$, because finite-temperature
effects are only of subleading importance.
 Together with the averaged decay rate~(\ref{gamma:av}),
this leads directly to a contribution to the rate of change in lepton asymmetry.
In the denominators, explicit factors of
two are present because the individual decay rates into
$\ell \phi^*$ and $\ell^{CP} \phi$ are equal in the absence of $CP$ violation, while $\gamma$
accounts for the total decay rate into both of these final states. Explicit factors of $1/g_w$ arise
because the decay rate $\gamma$ accounts for both weak isodoublet final states, while
$q_\ell$ and $q_\phi$ only account for an individual component.
As for the third and the fourth term, the asymmetries for the inverse processes follow
directly from the $CPT$ theorem.
In order to understand the overall coefficient of these terms,
note that the product $\gamma Y_{N1}^{\rm eq}$ is the total rate for the one-to-two decay as well
inverse decay processes in equilibrium. The contribution of 
the nonvanishing charge density of the doublet leptons $\ell$
to the deviation from
equilibrium can be easily accounted for when realizing that in the
nonrelativistic regime, where Pauli-blocking of the RHNs can be neglected, the rate of the inverse decay processes is proportional
to the distribution functions $f_\ell(\mathbf q)$ and
$f_{\ell^{CP}}(\mathbf q)$ of doublet leptons and antileptons, respectively, where
the momenta $|\mathbf q|\approx M_1/2\gg T$, i.e. they are within the Boltzmann
tail of the Fermi-Dirac distribution.
Hence, compared to the case of equilibrium distributions for
$f^{\rm eq}_{\ell}$ of doublet leptons and antileptons, the rates need to be rescaled
with the factors
\begin{align}
\label{factors:inversedecays}
\left.\frac{f_\ell(\mathbf q)}{f^{\rm eq}_\ell(\mathbf q)}\right|_{|\mathbf q|\gg T}
\approx\left(1+\frac{\mu_\ell}{T}\right)\,,
\quad
\left.\frac{f_{\ell^{CP}}(\mathbf q)}{f^{\rm eq}_\ell(\mathbf q)}\right|_{|\mathbf q|\gg T}
\approx\left(1-\frac{\mu_\ell}{T}\right)\,.
\end{align}
The chemical potential can again be expressed using Eq.~(\ref{q:mu}) through
$Y_{\Delta \ell}$ via $q_\ell$. We note that this procedure leads to a slight numerical discrepancy with the one used e.g. in Ref.~\cite{Buchmuller:2004nz}. In that work,
the factors~(\ref{factors:inversedecays}) are replaced in favour of the
ratios of yields ${Y_{\ell}}/{Y_\ell^{\rm eq}}$ and ${Y_{\ell^{CP}}}/{Y_\ell^{\rm eq}}$.
This however disregards the fact that the rates do not directly scale
with the lepton number-densities
but rather with their distribution functions evaluated at $|\mathbf q|\gg T$.
Applying a Maxwell-Boltzmann approximation for the number densities as in Ref.~\cite{Buchmuller:2004nz}, the coefficients of the collison terms proportional
to $Y_{\Delta\ell}$ are larger here by a factor of $12/\pi^2$ compared to Ref.~\cite{Buchmuller:2004nz} (cf. also Ref.~\cite{Garbrecht:2019zaa}).
Finally, the fifth to eighth of the collision terms in Eq.~(\ref{BE:ql})
account for two-by-two lepton-number violating scattering
processes which entail important subtleties that we discuss in the following subsection.

\subsection{Real intermediate states, $CP$ violation and deviation from equilibrium}
\label{sec:RIS}

Now, when taking $\varepsilon>0$ for definiteness,
we immediately observe from the first four of the collision terms in Eq.~(\ref{BE:ql}) that
leptons would be preferred over antileptons even in equilibrium.
However, this cannot be the full picture because it would imply that an asymmetry can
be present even in thermal equilibrium, in contradiction with Sakharov's
nonequilibrium condition based on the $CPT$ theorem along the reasoning in
Section~\ref{sec:matter-antimatter}.

\begin{figure}[t]
\begin{center}
\includegraphics[width=4cm]{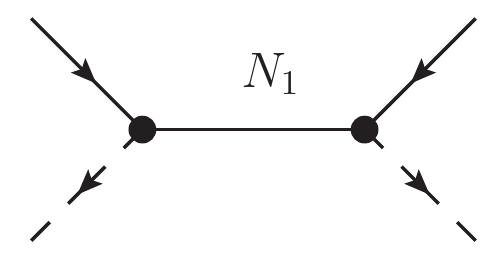}
\end{center}
\vskip-.4cm
\caption{\label{fig:DeltaL2} The tree-level contribution to the two-by-two process changing lepton number by two units.
Solid lines with arrows denote doublet leptons $\ell$, dashed lines with arrows Higgs bosons $\phi$ and
solid lines without arrows RHNs $N_i$.}
\end{figure}

\begin{figure}[t]
\begin{center}
\includegraphics[width=\textwidth]{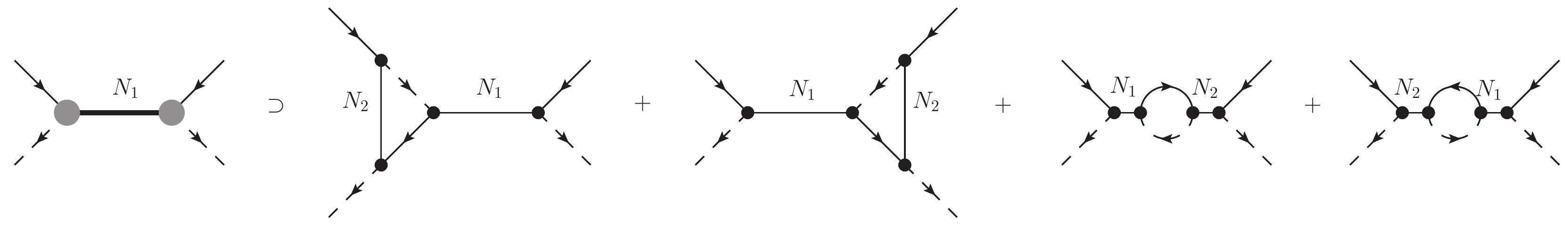}
\end{center}
\vskip-.4cm
\caption{\label{fig:DeltaL2Full} In these vertex and wave-function corrections to the
lepton-number changing two-by-two process, the $CP$-odd phases cancel. Including all contributions, processes of this type that change lepton
number by two units therefore do not lead to $CP$ violation. The bold line
denotes the full propagator of $N_1$
and the grey blobs the full vertices, such that the diagram on the left represents the full amplitude for this two-by-two reaction.}
\end{figure}

A way of resolving this issue appears when adding the last four collisional
terms describing two-by-two scatterings~\cite{Kolb:1979qa} shown in Figure~\ref{fig:DeltaL2}, where we
have extracted explicit factors of two because these processes change lepton number
by two units. The rates $\gamma_{\ell \phi^*\to \ell^{CP} \phi}$ and
$\gamma_{\ell^{CP} \phi \to \ell \phi^*}$ denote the full two-by-two rates,
and one can easily see (cf. the diagrammatic representation in Figure~\ref{fig:DeltaL2Full}) that these are $CP$ even because each Feynman diagram
has a counterpart with complex conjugated Yukawa couplings, such that the
$CP$-odd phases cancel.
Since we assume that $M_1\ll M_2$, we can concentrate on contributions mediated by
the exchange of $N_1$. Provided $N_1$ is off shell, at tree level,
the two-by-two rates are $CP$ even and of order $Y^4$.
(We omit absolute values and indices on the Yukawa coyplings $Y$ in these power counting arguments.)
They are therefore subdominant when compared with the $CP$-even one-to-two rates that
are of order $Y^2$.
However, when $N_1$ is exchanged in the $s$ channel, the tree-level two-by-two rates
lead to contributions of order $Y^2$ from the portion of the phase-space integral where
the internal $N_1$ propagator is on shell, i.e. where there is
a so-called real intermediate state (RIS).
This on-shell enhancement from order $Y^4$ to $Y^2$ can be seen when substituting Eqs.~\eqref{prop:improved}
for the full RHN propagator in Figure~\ref{fig:DeltaL2Full}.
However, the production of on-shell $N_1$
is already accounted for by the explicit one-to-two processes, such that these portions
need to be subtracted from the two-by-two rates.

These RIS contributions to the two-by-two rates
can be written in the following suggestive way:
\begin{subequations}
\begin{align}
\gamma^{\rm RIS}_{\ell \phi^*\to \bar \ell^{CP} \phi}
=&\left(1+\frac{\mu_\ell}{T}\right)\times\frac{1-\varepsilon}{2}\frac{\gamma Y_N^{\rm eq}}{g_w}
\times\frac{1-\varepsilon}{2}
\approx \left(1+\frac{\mu_\ell}{T}\right)\frac{\gamma Y_N^{\rm eq}}{g_w}\frac{1-2\varepsilon}{4}\,,\\
\gamma^{\rm RIS}_{\ell^{CP} \phi \to \ell \phi^*}
=&\left(1-\frac{\mu_\ell}{T}\right)\times\frac{1+\varepsilon}{2}\frac{\gamma Y_N^{\rm eq}}{g_w}
\times\frac{1+\varepsilon}{2}
\approx\left(1-\frac{\mu_\ell}{T}\right)\frac{\gamma Y_N^{\rm eq}}{g_w}\frac{1+2\varepsilon}{4}\,.
\end{align}
\end{subequations}
In each of these equations,
the first factor involving $\varepsilon$ corresponds to the $CP$-violating
inverse decay rate of $N_1$, while the second factor involving $\varepsilon$ is the
branching ratio of the $CP$-violating decays. Substituting these equations
into Eq.~(\ref{BE:ql}), a number of cancellations leads to
\begin{align}
\frac{Y_{\Delta\ell}}{dt}=&
\varepsilon\frac{\gamma}{g_w}\left(Y_{N1}-Y_{N1}^{\rm eq}\right)
-2\gamma_{\ell \phi^*\to \ell^{CP} \phi}
+2\gamma_{\ell^{CP} \phi \to \ell \phi^*}\,.
\end{align}
Notice that, according to our above remark, the two-by-two rates are $CP$ conserving
such that all remaining effects of $CP$ violation are proportional to the deviation of the RHN $N_1$ from
equilibrium, in agreement with Sakharov's conditions. Further, we now refer to the first term on the right-hand side as a
source term and the second one as a washout term. The washout processes are $CP$ conserving.

Next, we recall that two-by-two washout processes contain still those mediated by RIS. In
a wide range of parameter space, where washout mediated by off-shell RHNs may be neglected,
it is a good approximation to only account for the RIS contributions. In that case, the fluid equation
for the lepton asymmetry simplifies to
\begin{align}
\frac{d Y_{\Delta\ell}}{dt}=&
\varepsilon\frac{\gamma}{g_w}\left(Y_{N1}-Y_{N1}^{\rm eq}\right)
-2\frac{\mu_\ell}{T}\frac{\gamma Y_{N1}^{\rm eq}}{2 g_w}
=
\varepsilon\frac{\gamma}{g_w}\left(Y_{N1}-Y_{N1}^{\rm eq}\right)
-6 \frac{q_\ell}{T^3}\gamma \frac{Y_{N1}^{\rm eq}}{g_w}
=
\varepsilon\frac{\gamma}{g_w}\left(Y_{N1}-Y_{N1}^{\rm eq}\right)
-W Y_{\Delta_\ell}
\,,
\end{align}
where the terms proportional to $\mu_\ell$ (or $q_\ell$ and $Y_{\Delta\ell}$, respectively) describe the washout, and where we note that
this equation is of the form advertised in Eq.~(\ref{fluid:equation:q}).
We have substituted here the chemical potential $\mu_\ell$ in favour of
the charge density $q_\ell$ using the relation~\eqref{q:mu} and have eventually
defined the washout rate as
\begin{align}
\label{Wnobar}
W=\gamma s \frac{6}{T^3} \frac{1}{g_w}Y^{\rm eq}_{N 1}\,.
\end{align}

When the RIS are subtracted, we can denote the remaining contribution
from the two-by-two washout processes by $\Delta W$.
In most regions of parameter
space, $\Delta W\ll W$ for $T\ll M_1$, such that, as stated above, $\Delta W$ can be neglected.
Nonetheless, this relation may be saturated when the Yukawa couplings are large compared
to the values generically predicted by the type-I seesaw mechanism in conjunction with
the observed light neutrino masses, provided these are hierarchical.
In that case,
the contribution of $\Delta W$ to washout gives important upper bounds on the mass of the lightest RHN
as well as on the masses of the observed active neutrinos~\cite{Buchmuller:2002rq,Buchmuller:2002jk,Buchmuller:2003gz}. Since all RHNs are exchanged
in these processes that change lepton number by two units, in the type-I seesaw model and the limit $T\ll M_i$, the rates are just proportional to the Majorana masses of
the light neutrinos themselves, i.e. they lead to an extra contribution to the washout
rate of the form
$\Delta W\sim \bar m^2 T^3/v^4$,
where $\bar m^2=m_1^2+m_2^2+m_3^2$, $m_i$ are the mass eigenvalues of
the light neutrinos and $v$ is the vacuum expectation value of the Higgs field. For minimal, unflavoured, nonresonant leptogenesis, this leads to the bound $\bar m<0.2\,{\rm eV}$, which is of interest in view of the current and future search for the absolute neutrino mass scale in neutrinoless double beta decay and in large-scale cosmological structure.
Scenarios where
the two-by-two rates are of quantitative importance for washout without leading to
large values of $\bar m$ are studied
e.g. in Refs.~\cite{Pilaftsis:2005rv,Blanchet:2009kk}.


To conclude this discussion, the first four collision terms in Eq.~(\ref{BE:ql}) by themselves
fail to comply with the $CPT$ theorem (that relies on unitarity of the $S$ matrix)
and consequently also with Sakharov's nonequilibrium condition.
Operating with $S$-matrix elements for unstable states, here for $N_1$, leads apparently to
a nonunitary time-evolution, which is fixed trough the subtraction of certain parts of the RIS contributions.
Whether or not the procedure presented here corresponds to a satisfactory solution
of the problem, it has motivated studies of leptogenesis in the CTP framework~\cite{Buchmuller:2000nd,DeSimone:2007gkc,Garny:2009rv,Garny:2009qn,Anisimov:2010aq,Garny:2010nj,Beneke:2010wd,Beneke:2010dz,Garny:2010nz,Anisimov:2010dk}, where
instead of using $S$-matrix elements, one operates with the real-time evolution of
quantum mechanical correlation functions in first place. We review these
matters in Section~\ref{sec:kineq:firstprinciples}.

\subsection{Final adjustments and expansion of the Universe}
\label{sec:expansionoftheUniverse}

There are two more simple, yet substantial adjustments that we are going to apply to the fluid equations: First, we take account of the  additional bias that the charge density in Higgs bosons implies for the washout term and second, we include the expansion of the Universe, as this creates the crucial nonequilibrium conditions for baryogenesis in first place.


The decay of a RHN produces either a lepton and an anti-Higgs boson or the corresponding antiparticles. When ignoring additional processes that
change the charge densities, this amounts to the relation $q_\ell=-q_\phi$.
In order to include the bias of the charge density
of Higgs bosons $q_\phi$ next to the lepton charge density $q_\ell$ in the washout term, we once again make use of the relation~(\ref{q:mu})
between charge and chemical potentials for fermions and bosons and follow
the arguments of Section~\ref{sec:setup:fluid:eq}. 
In the washout term, this amounts to the replacement
$W Y_{\Delta\ell}\to W (Y_{\Delta\ell}-1/2\,Y_{\Delta\phi})=3/2\,W Y_{\Delta\ell}$,
i.e. the multiplication of
the washout rate
by a factor of $3/2$.
In the SM, the charges in the Higgs bosons and doublet leptons are further redistributed
by the Yukawa interactions and by strong and weak sphaleron processes (cf. Figure~\ref{fig:sphalerons}) to the remaining SM particles~\cite{Barbieri:1999ma,Buchmuller:2001sr}. However, due to the large number of degrees of freedom, this additional
correction is comparably small, and we neglect it here for simplicity. It is technically easy though to incorporate such fully equilibrated \emph{spectator effects} such that
these should be nonetheless included in phenomenological studies. Partially equilibrated spectator
fields may lead to an effective protection of the asymmetry from washout, which
is of importance in some regions of the parameter space discussed in Refs.~\cite{Garbrecht:2019zaa,Garbrecht:2014kda}.

Crucially for baryogenesis, we yet need to account for the expansion
of the Universe that creates the necessary out-of-equilibrium conditions.
At the present stage, where the derivations to this end
have been made in Minkowski background, this is most easily
implemented when describing the Friedmann-Robertson-Walker Universe
through conformal coordinates and the pertaining metric tensor
\begin{align}
\label{metric:tensor}
g_{\mu\nu}=a^2(\eta){\rm diag}(1,-1,-1,-1)\,,
\end{align}
where $\eta$ is conformal time and $a(\eta)$ the scale factor.

In standard scenarios for leptogenesis, the Universe is dominated by relativistic radiation at
the time when the asymmetry is generated. The expansion during radiation domination
is given by
\begin{align}
a(\eta)=a_{\rm R}\eta\,.
\end{align}
When we define a comoving temperature $T_{\rm com}=a T$,
where $T$ is the physical temperature, we note that
the choice $a_{\rm R}=T_{\rm com}$ is particularly convenient, as this implies that
$T=1/\eta$. Relating the Hubble rate
to the energy density of the plasma through the Friedmann equation, one obtains
\begin{align}
H^2=\left(\frac{1}{a^2}\frac{d}{d\eta} a\right)^2=\frac{T^4}{a_{\rm R}^2}=\frac{8\pi}{3}\frac{\frac{\pi^2}{30}g_\star T^4}{m_{\rm Pl}^2}
\;\Leftrightarrow\;
a_{\rm R}=\frac{m_{\rm Pl}}{2}\sqrt{\frac{45}{\pi^3 g_\star}}\,,
\end{align}
where $m_{\rm Pl}=1.22\times 10^{19}{\rm GeV}$ is the Planck mass.

Given the metric tensor~(\ref{metric:tensor}), we can view the kinetic equations
for the charge and number densities (i.e. not the entropy-normalized versions)
derived to this end as expressed in terms of comoving momenta, i.e. $k \to k_{\rm com}=a(\eta) k_{\rm ph}$, and the comoving temperature, i.e. $T\to T_{\rm com}$ provided we replace the masses $M_i \to a(\eta) M_i$ (up to potential effects from the coupling of the scalar fields
to the background curvature that are negligible in the present context) and the derivatives $d/dt\to d/d\eta$. When integrating over $d^3 k_{\rm com}$, we then
obtain equations for the comoving number densities.
In the absence of collisions, these are conserved as the Universe expands and
directly proportional to the entropy-normalized yields.
Next, in order to have the fluid equations temporally depend on an order-one parameter, we
define $z=M_1/T=\eta M_1$, which implies that
\begin{align}
\frac{d}{d\eta}=M_1\frac{d}{d z}\,,\quad a(\eta)=z\frac{T_{\rm com}}{M_1}\,.
\end{align}
Carrying out these rescalings, the fluid equations derived in Minkowski space are recast
to a form suitable for the radiation-dominated Universe as
\begin{align}
\label{replace:rule:expansion}
\frac{d n_i}{dt}=-\Gamma_{ij}\left(\left\{M_k\right\},T\right) n_j
\;\;\to\;\;
\frac{d Y_i}{dz}=-\frac{1}{M_1} \Gamma_{ij}\left(\left\{M_k/M_1 \times z T_{\rm com}\right\},T_{\rm com}\right) Y_j\,.
\end{align}

In summary, the apparently simplest scenario of leptogenesis is described
by the following set of coupled differential equations:
\begin{subequations}
\label{fluid:eq:leptogenesis}
\begin{align}
\label{fluid:eq:leptogenesis:N1}
\frac{dY_{N1}}{dz}=&-\bar\gamma\left(Y_{N1}-Y_{N1}^{\rm eq}\right)\,,\\\label{fluid:eq:leptogenesis:ell}
\frac{d Y_{\Delta\ell}}{dz}=&
\varepsilon\frac{\bar\gamma}{g_w}\left(Y_{N1}-Y_{N1}^{\rm eq}\right)
-\bar W\frac32 Y_{\Delta\ell}\,,
\end{align}
\end{subequations}
where, applying the replacement rule~(\ref{replace:rule:expansion})
to the expressions~(\ref{gamma:av}) and~(\ref{Wnobar}),
\begin{subequations}
\label{rates:exp:bkg}
\begin{align}
\label{gammabar}
\gamma=&
\frac{K_1(z)}{K_2(z)}|Y_1|^2\frac{M_1}{8\pi}
\;\;\to\;\;
\bar\gamma=\frac{K_1(z)}{K_2(z)}|Y_1|^2\frac{z T_{\rm com}}{8\pi M_1}
\,,
\\
\label{Wbar}
W=& \frac{3 |Y_1|^2 M_1^3}{8\pi^3 T^2} K_1\left(\frac{M_1}{T}\right)
\;\;\to\;\;
\bar W=\frac{3 |Y_1|^2 z^3 T_{\rm com}}{8\pi^3 M_1} K_1(z)
=
\frac{3 |Y_1|^2 z^\frac52 T_{\rm com}}{2^\frac72 \pi^\frac52 M_1}{\rm e}^{-z}
\times\left(1+{\cal O}(1/z)\right)\,.
\end{align}
\end{subequations}
Numerically, these can be solved easily. An approximate analytic solution,
that yields some insight into the general mechanics of baryogenesis from out-of-equilibrium
decays is also available and will be reviewed in Section~\ref{sec:pedestrian}.

\subsection{Integral expressions for the rates in the fluid equations}
\label{sec:integralform}

\begin{figure}[t]
\raisebox{-.1cm}{
\parbox{4.3cm}{
\center
\includegraphics[scale=0.5]{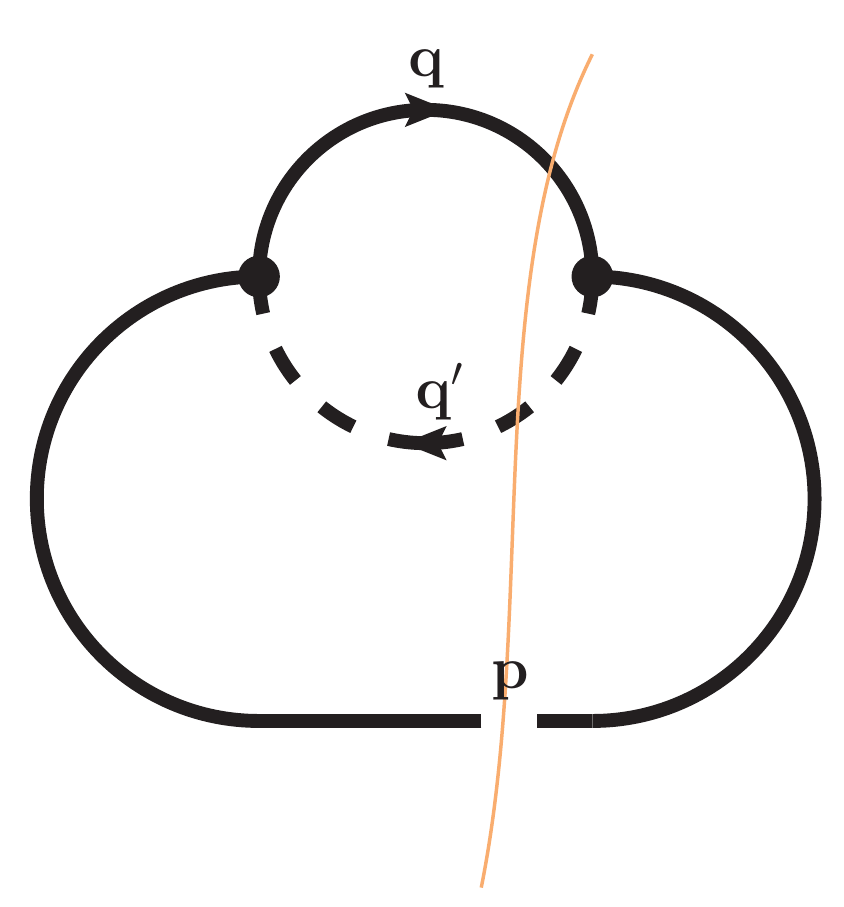}\\
\vskip-1.cm
$\cal D$
}
}
\raisebox{-.1cm}{
\parbox{4.3cm}{
\center
\includegraphics[scale=0.5]{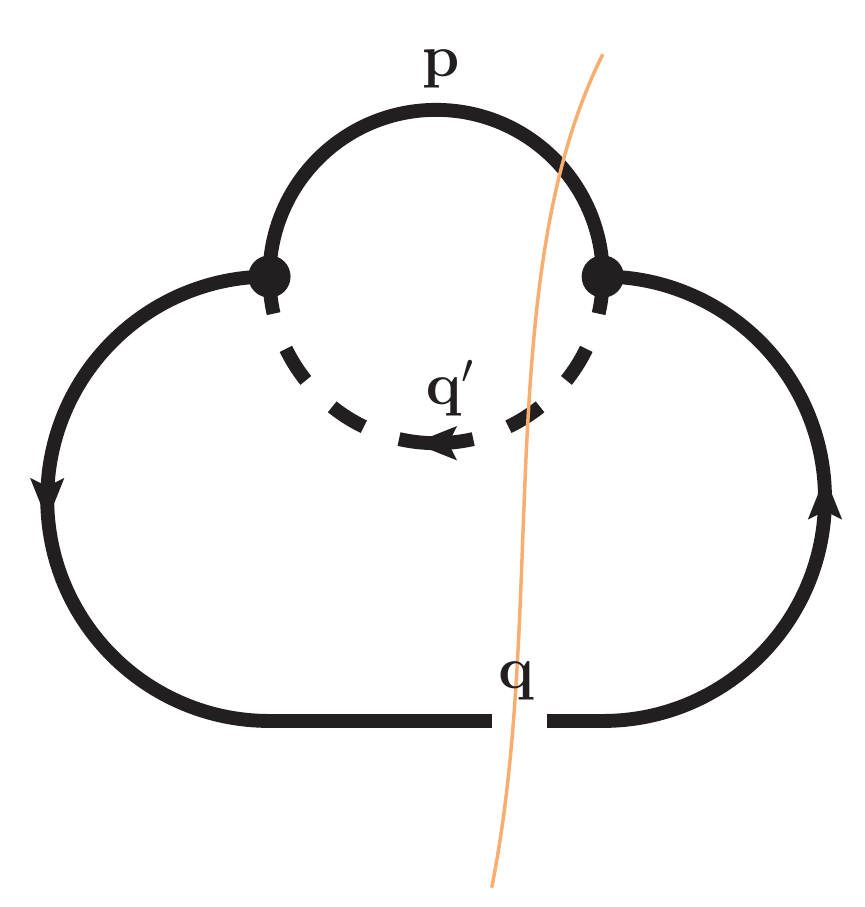}
\vskip-1.cm
${\cal W}$
}
}
\parbox{4.3cm}{
\center
\includegraphics[scale=0.5]{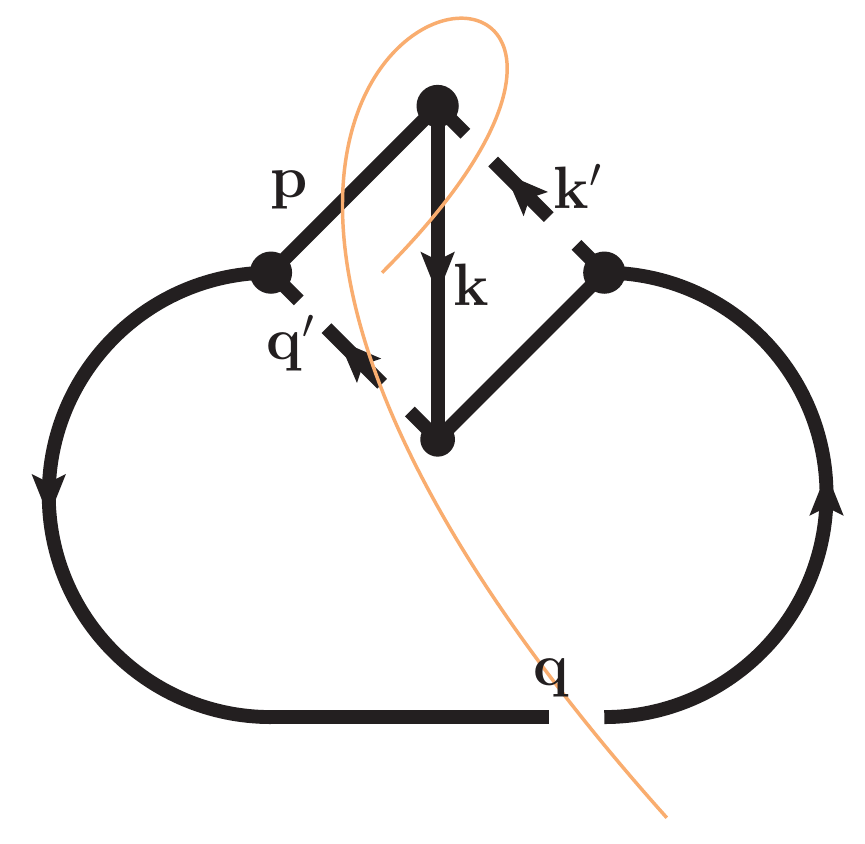}
\vskip-.7cm
${\cal S}^{\rm vert}$
}
\raisebox{-.2cm}{\parbox{4.3cm}{
\center
\includegraphics[scale=0.5]{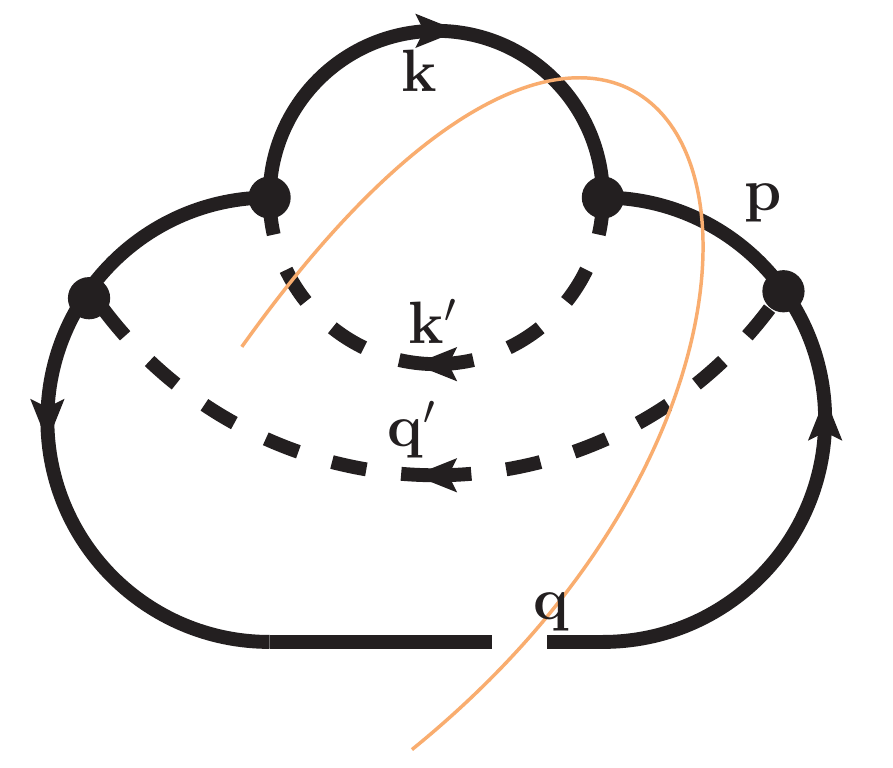}
\vskip-.7cm
${\cal S}^{\rm wv}\phantom{XX}$
}
}
\vskip.3cm
\caption{
\label{fig:integralexpr}
For the various collision terms
in the fluid equations,
these diagrams show the correspondence of the derivation using classical
kinetic theory combined with QFT cross sections (Section~\ref{sec:integralform})
with the derivation based on the effective action in the CTP approach (Section~\ref{sec:kineq:firstprinciples}).
Bold solid lines without arrows stand for RHNs, with arrow for leptons
and bold dashed lines with arrows stand for Higgs bosons. The on-shell cuts are indicated by thin orange lines.
A gap at the cut through the RHN or lepton lines indicates that the pertaining integral originates from the zeroth moment of the associated kinetic equation.
In the limit $M_1\ll M_2$, the leading
contributions to ${\cal S}^{\rm vert}$ and ${\cal S}^{\rm wv}$ are those where the
cut goes through an $N_1$ propagator whereas $N_2$ remains off shell.
The expression~\eqref{integral:decay} for $\cal{D}$ is to be compared with Eq.~\eqref{D:CTP} in Section~\ref{sec:integralform},
expression~\eqref{integral:washout} for $\cal{W}$ with Eq.~\eqref{coll:tree}.
In Eq.~\eqref{integral:decayasymmetry} the contribution ${\cal S}^{\rm vert}$ is to be compared with Eq.~\eqref{coll:vert}
and ${\cal S}^{\rm wv}$ with Eq.~\eqref{source:wv}.
We also
note that ${\cal D}$ and ${\cal W}$ can be recovered from the two-loop diagrams
and ${\cal S}^{\rm vert}$ from the three-loop diagram that contribute to the effective
action $\Gamma_2$, Eq.~\eqref{Gamma2:diagrammatic}. The contribution ${\cal S}^{\rm wv}$
descends from the two-loop diagram in $\Gamma_2$ when inserting a one-loop correction
into the RHN propgagator. (Recall that $\Gamma_2$ is understood to be expressed
in terms of full propagators, such that this correction is implicit.)
}
\end{figure}

To this end, we have inferred the fluid equations for leptogenesis by
balancing the rates of change in the charge and number densities of 
leptons $\ell$ and RHNs $N_i$. Nonetheless, as advertised in Section~\ref{sec:stat:mech}, these can also be derived from the Boltzmann equations by integration over three-momentum, i.e. the fluid equations~(\ref{fluid:eq:leptogenesis}) can be expressed as
\begin{align}
\label{Beq:Kineq}
\frac{d q_\ell}{dt}=\frac{d}{dt}\int\frac{d^3 q}{(2\pi)^3}\left(f_{\ell}(\mathbf{q})-\bar f_{\ell}(\mathbf{q})\right)={\cal S}-{\cal W}\,,\quad \frac{d n_{N1}}{dt}=\frac{d}{dt}\int\frac{d^3 p}{(2\pi)^3}2 f_{N1}(\mathbf p)=-{\cal D}\,,
\end{align}
where the decay term ${\cal D}$, the washout term ${\cal W}$ and the
source term ${\cal S}$ can be expressed as momentum integrals over
the collision term~(\ref{coll:term}). The bar over the function
$\bar f_X$ indicates that this is the distribution of the antiparticle of $X$.
In the integral for the RHNs, there is an explicit factor of two accounting for
the two helicity states.
For simplicity, we do not
include here the expansion of the Universe and the normalization to entropy that
can be reintroduced as explained in Section~\ref{sec:expansionoftheUniverse}.
These integral expressions will also be useful for comparison with the corresponding
results derived in the CTP approach in Section~\ref{sec:kineq:firstprinciples}.
In order to condense the notation of phase-space and four-momentum integrals,
we use the shorthand expressions
\begin{subequations}
\begin{align}
\int\limits_{p_1\cdots p_n}=&\int\frac{d^4 p_1}{(2\pi)^4}\cdots\frac{d^4 p_n}{(2 \pi)^4}\,,\\
\int\limits_{\mathbf p_1 \cdots \mathbf p_n}=&\int\frac{d^3p_1}{(2\pi)^3 2\sqrt{\mathbf p_1^2+m_1^2}} \cdots\frac{d^3p_n}{(2\pi)^3 2\sqrt{\mathbf p_n^2+m_n^2}}\,,\\
\delta_p=&(2\pi)^4\delta^4(p)\,,
\end{align}
\end{subequations}
where $m_i$ is the mass of the particle with momentum $p_i$.


We proceed by neglecting quantum-statistical factors (i.e. by replacing $(1\pm f)\to 1$) in the collision term, as it is appropriate
in the strong-washout regime where the RHNs are nonrelativistic.
The decay term can then be expressed as
\begin{align}
\label{integral:decay}
{\cal D}=\gamma\left(n_{N1}-n_{N1}^{\rm eq}\right)
=2 g_w \int\limits_{\mathbf{p}\mathbf{q}\mathbf{q^\prime}}\delta_{p-q-q^\prime}
\sum_{\rm pol} \left|{\rm i}{\cal M}^{\rm LO}_{N_1\to \ell\phi^*}\right|^2 \delta f_{N1}
=g_w\int\limits_{\mathbf{p}\mathbf{q}\mathbf{q^\prime}}\delta_{p-q-q^\prime}
 4 p\cdot q \delta f_{N1}(\mathbf p)\,,
\end{align}
where the explicit factor of  two accounts for the decay channels into particles and antiparticles.
The distribution $\delta f_{N1}=f_{N1}-f_{N1}^{\rm eq}$ is the deviation from the
equilibrium distribution $f_{N1}^{\rm eq}$ for $N_1$.
Note also that there appears a polarization sum rather than
the average over the RHN polarizations because we account here for both of their helicity states,
i.e. ``pol'' under the sum now refers to the polarizations of $N_1$ as well as $\ell$.
The integral form of the washout rate is
\begin{align}
\label{integral:washout}
{\cal W}= W Y_{\Delta\ell}
=&\int\limits_{\mathbf p\mathbf q\mathbf q^\prime} \delta_{p-q-q^\prime} 
\sum_{\rm pol} \left|{\rm i}{\cal M}^{\rm LO}_{N_1\to l\phi^*}\right|^2 f_\phi(q^\prime)\left[ f_\ell(\mathbf q)-\bar f_\ell(\mathbf q)\right]\notag\\
=&\int\limits_{\mathbf p\mathbf q\mathbf q^\prime} \delta_{p-q-q^\prime} {\rm tr}[\slashed p \slashed q]
f_\phi(q^\prime)
\left[ f_\ell(\mathbf q)-\bar f_\ell(\mathbf q)\right]
\,,
\end{align}
and for the $CP$-violating source, we write
\begin{align}
{\cal S}=&{\cal S}^{\rm vert}+{\cal S}^{\rm wv}
=\varepsilon\frac{\gamma}{g_w}\left(n_{N1}-n_{N1}^{\rm eq}\right)\notag\\
=&\int\limits_{\mathbf{p}\mathbf{q}\mathbf{q^\prime}}
\sum_{\rm pol}\left[\left({\rm i}{\cal M}^{\rm LO}_{N_1\to \ell\phi^*}\right)^*{\rm i}{\cal M}_{N_1\to \ell\phi^*}-\left({\rm i}{\cal M}^{\rm LO}_{N_1\to \ell^{CP}\phi}\right)^*{\rm i}{\cal M}_{N_1\to \ell^{CP}\phi}+{\rm c.c.}\right]\delta_{p-q-q^\prime} \delta f_{N1}(\mathbf p)\notag\\
=&
-\left(Y_1^2 {Y_2^*}^2-{Y_1^*}^2 Y_2^2\right)
\int\limits_{\mathbf{p}\mathbf{q}\mathbf{q^\prime} \mathbf{k}\mathbf{k^\prime}}
\delta_{p-k-k^\prime}\delta_{p-q-q^\prime}
\bigg\{
\frac{{\rm i}\,{\rm tr}\left[\left(\slashed p +\widehat M_1^*\right)P_{\rm L}\slashed q P_{\rm R}\left(\slashed q -\slashed k^\prime +\widehat M_2^*\right)P_{\rm R}\slashed k P_{\rm L}\right]}{(q-{k^\prime})^2-|M_2|^2}\notag\\
+&g_w\frac{{\rm i}\,{\rm tr}\left[\left(\slashed p +\widehat M_1^*\right)P_{\rm L}\slashed q P_{\rm R}\left(\slashed p +\widehat M_2^*\right)P_{\rm R}\slashed k P_{\rm L}\right]}{|M_1|^2-|M_2|^2}
\bigg\}\delta f_{N1}(\mathbf p)\,,
\label{integral:decayasymmetry}
\end{align}
where we have substituted Eqs.~(\ref{optical:vertex}) and~(\ref{msquared:wv})
for the interference terms. [Note the appropriate rescaling of the Yukawa coupling and that this result~(\ref{integral:decayasymmetry})
accounts for the sum of both polarization states of $N_1$ whereas Eq.~(\ref{optical:vertex}) does so for the average.]

One can check that these integral expressions
agree with the definitions of $\gamma$ in Eq.~(\ref{gamma:av}),
$W$ in Eq.~(\ref{Wbar}) and $\varepsilon$ in Eq.~(\ref{epsilon:N1}) when further
relating the distribution functions to the densities as in Eq.~(\ref{eq:densities}).

The rates appearing in the fluid equations can also be represented
diagrammatically as shown in Figure~\ref{fig:integralexpr}. Note that the diagrams
for the $CP$-violating rates are obtained from those in Figure~\ref{fig:optical}
by sewing together the external lines.
The resulting diagrams take the form of contributions
to an effective action, i.e. of ``vacuum graphs'' (cf. Eq.~\eqref{Gamma2:diagrammatic} below). Within the CTP approach discussed in the following Section~\ref{sec:kineq:firstprinciples}, we see that such an interpretation
is indeed meaningful.

We also note that in the source terms ${\cal S}$, depending on whether we attribute the cut
to $N_1$ or to $\ell$ and $\phi$, we either obtain an interference between a tree-level
and a one-loop, one-two-two amplitude or an interference between two tree-level, two-by-two scatterings,
one mediated by an on-shell $N_1$ and one by an off-shell $N_2$. This ambiguity is
related to the correct counting of the reaction rates that has been implemented in this
section by subtracting the RIS. The CTP approach presented in the following section
automatically takes care of the right counting.

\section{Baryogenesis from out-of-equilibrium decays and inverse decays based on first principles
in the closed time-path approach}
\label{sec:kineq:firstprinciples}

When reviewing the standard approach to leptogenesis based on classical kinetic theory
and QFT cross sections and decay rates, we have encountered two somewhat unsatisfactory
arguments:
\begin{itemize}
\item
In order to meet the requirement that no $CP$ asymmetry may be present or be generated
in thermodynamic equilibrium, certain contributions from RHNs propagating as RIS have to
be subtracted from the two-by-two scattering rates. While this procedure leads to correct
results, the argument faces the complications due to unstable
particles as external states in scattering theory. It appears convoluted and calls for a simpler and more direct treatment.
The discussion in Section~\ref{sec:compare:methods} suggests that the CTP approach offers
such a method.
\item
The results from the wave-function or mixing contribution to leptogenesis
are, as discussed in Section~\ref{sec:mixing:variants}, inconclusive
in the resonant limit where $|M_1^2-M_2^2|\gg M_1 \Gamma_{ij}$ does not hold.
\end{itemize}
Both of these issues have to do with the appearance of long-lived states in the
reactions that drive the kinetic equations. In the first case, it is  the RHN $N_1$
whose lifetime by definition is of the same order as the time between
decays and inverse decays, in the second case, $N_2$ can be produced from $N_1$
without external radiation within the bounds of energy uncertainty. Therefore, in both situations,
a region where the particles evolve as free in and out states outside of
a region in spacetime where interactions may not be neglected, cannot be clearly identified. The presence
of regions of freely propagating states is however a prerequisite for defining
the matrix elements that are substituted into the Boltzmann equations. In the case of
leptogenesis, the one-to-two rate cannot be clearly separated from the two-to-two
rates involving RIS (cf. Section~\ref{sec:RIS}) or external states may not be
clearly identifiable because they correspond to unstable, mixing particles (cf. Section~\ref{sec:mixing:variants}). In the Boltzmann approach, it therefore proves complicated
to construct kinetic equations that respect unitarity and the $CPT$ theorem.

Due to these problems associated with the formulation of matrix elements, the computation
of the real-time evolution of the quantum mechanical correlation functions is
a more straightforward approach (save for many practitioners
being more familiar with scattering theory). In fact, all observables of interest, in particular
the lepton asymmetry, can be calculated from the correlation functions.
The time evolution of these is found by solving
a closed system of Schwinger--Dyson equations for the causal (i.e. retarded and advanced)
propagators and the Wightman functions, as reviewed in Section~\ref{sec:CTP}.

Self-consistent solutions to these equations can be obtained based on systematic approximations (e.g.
perturbative expansions or resummed variants of these, numerical methods or combinations of both).
The correct account of finite-width effects in Wigner space is explained in Ref.~\cite{Garbrecht:2011xw} but these
can be safely ignored if the quasi-particles can be approximated to occupy a sharp mass shell.
Further, in general kinematic regimes, the spinor structure leads to technical  complications
that can be dealt with using the methods developed in Refs.~\cite{Joyce:1999uf,Kainulainen:2001cn,Prokopec:2003pj,Prokopec:2004ic} that we review in the context
of electroweak baryogenesis in Section~\ref{sec:ewbg}. Compared to
that, in the present context, which is leptogenesis in the
strong washout regime, the nonrelativistic approximation for the RHNs leads to a considerably simplified treatment of the fermion fields.

\subsection{Kinetic equations for leptogenesis in the strong-washout regime}

We start writing down the loop contributions~(\ref{Gamma2}) to the 2PI effective action
\begin{align}
\Gamma_2=
-{\rm i}&\raisebox{-1.4cm}{\includegraphics[scale=0.6]{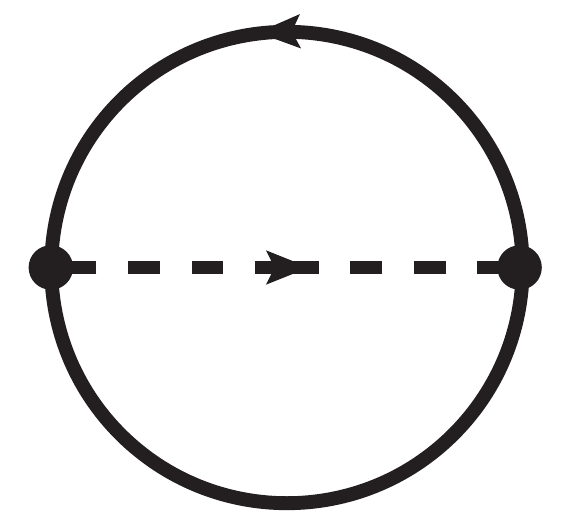}}-{\rm i}\raisebox{-1.4cm}{\includegraphics[scale=0.6]{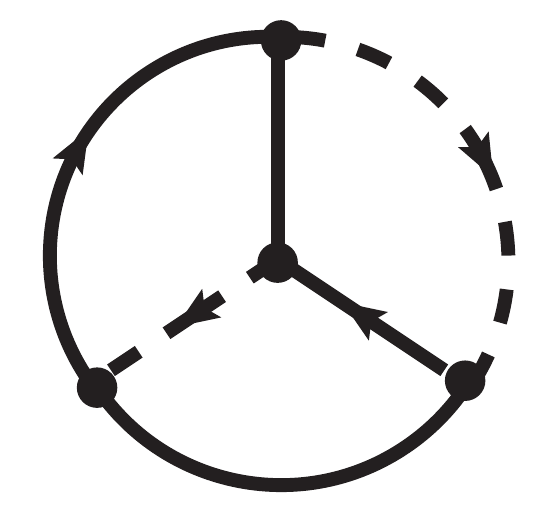}}
+\cdots\,.
\label{Gamma2:diagrammatic}
\end{align}
The bold lines that appear here represent full propagators, and when these are solid,
they stand for fermions, and when dashed,
for  Higgs bosons. Fermion lines without arrow are for RHNs and with arrow
for doublet leptons.
Taking functional derivatives with respect to the propagators of doublet leptons and RHNs
according to Eq.~(\ref{selferg:fermi}) yields the self energies that appear
in the Schwinger--Dyson equations~(\ref{SDE:fermi}). For the present case, these take
the diagrammatic form
\begin{subequations}
\begin{align}
\raisebox{.0cm}{\includegraphics[scale=.6]{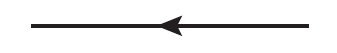}}^{-1}\;
\raisebox{-.1cm}{\includegraphics[scale=.6]{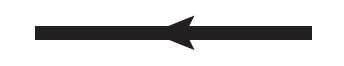}}
=&\delta+\raisebox{-1.1cm}{\includegraphics[scale=.6]{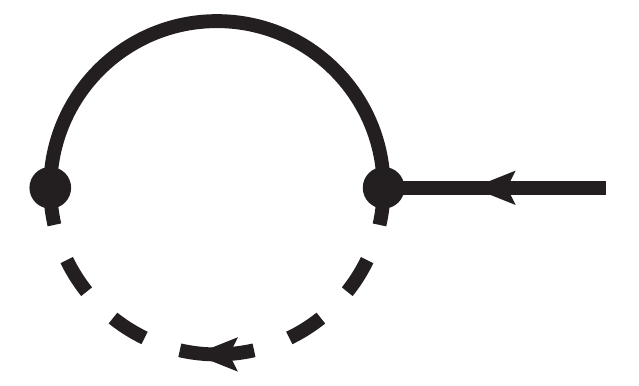}}
+\raisebox{-1.1cm}{\includegraphics[scale=.6]{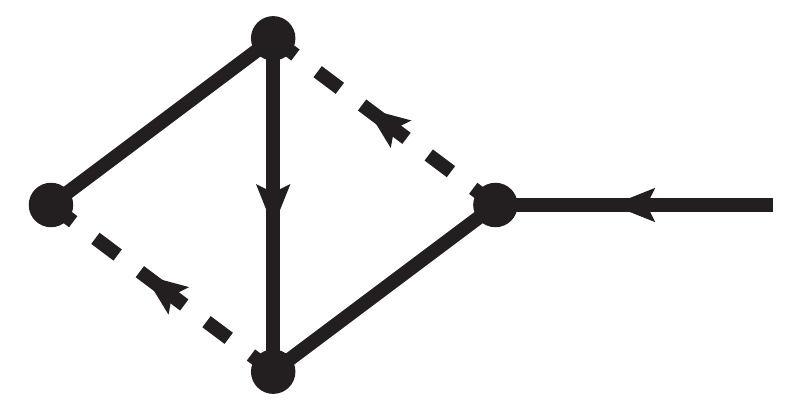}}\,,\label{SDE:lepton:diagrammatic}\\
\raisebox{.0cm}{\includegraphics[scale=.6]{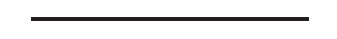}}^{-1}\;
\raisebox{-.1cm}{\includegraphics[scale=.6]{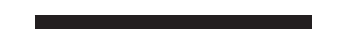}}
=&\delta+\raisebox{-1.13cm}{\includegraphics[scale=.6]{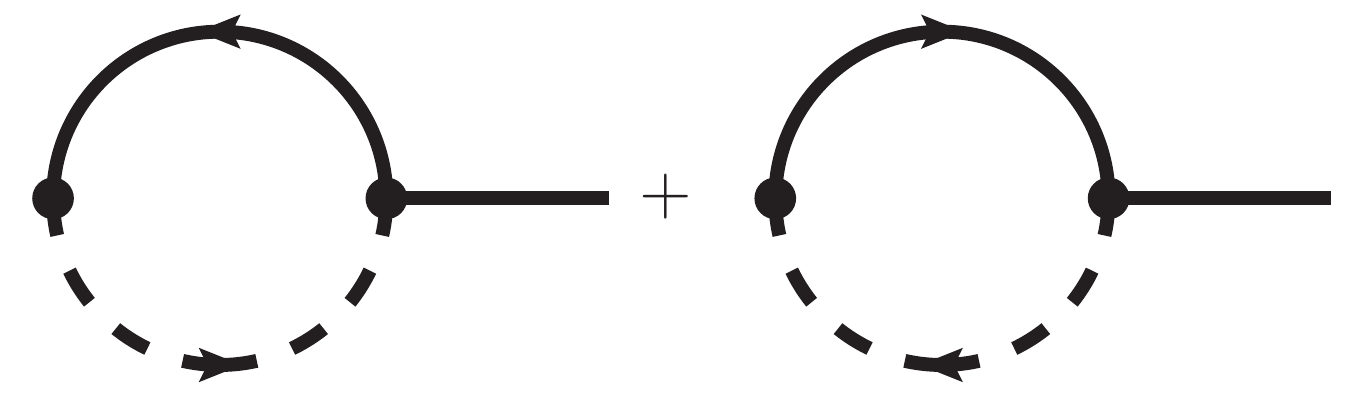}}
\,.
\label{SDE:RHN:diagrammatic}
\end{align}
\end{subequations}
Thin lines stand for tree propagators and the inverse of these amounts to
Dirac operators. The term $\delta$ stands for the first term on the RHS of Eq.~(\ref{SDE:fermi}).
As described in Section~\ref{sec:CTP}, the Schwinger--Dyson equations then lead to
Kadanoff--Baym equations~(\ref{KBE:Wigner:Fermions}) in Wigner space.

For the present problem of leptogenesis in the strong washout regime,
we apply the following simplifications:
\begin{itemize}
\item
We discard the terms involving $\slashed \Sigma^{\rm H}$ that amount to a correction in the dispersion relation, e.g. a thermal mass. For the RHN, the squared thermal mass is
of order $Y^2 T^2$
and can be neglected compared to the masses $M_{1,2}\gg T$ as well as the term associated with the decay
width $\sim Y^2 M^2_{1,2}$ appearing in the denominator of the propagator for the RHN~\cite{Garbrecht:2014aga}. Note that the term with $\slashed \Sigma^{\rm H}$ may nonetheless be of importance in scenarios
where the asymmetry is generated when the RHNs are relativistic and when at the same time their mass splitting is very small~\cite{Drewes:2016gmt}. For the doublet leptons,
the correction to the dispersion relation of order $g T$ is small compared
to the average energy of order $M_1$, where $g$ schematically stands for the ${\rm U}(1)_Y$ and ${\rm SU}(2)_{\rm L}$ gauge couplings.
\item
We further drop the term involving $S^{\rm H}$ that constitutes an inhomogeneous contribution
to the differential equation for $S^{<,>}$. It gives rise to the equilibrium contribution to the Green function in a form that also resolves the finite width~\cite{Garbrecht:2011xw}.
In many cases of phenomenological interest (such as the present one), it is sufficient to approximate spectral distributions of finite width
by Dirac-$\delta$ functions. Even when dropping this term, the equilibrium form of the RHN distribution
follows from the KMS relation when imposing that the collision term should vanish, such that the solution is independent of time.
\item
We furthermore neglect gradient
effects that are in the present case controlled by the parameter $H/M_1$ (the temporal rate of change
given by the Hubble rate divided by the typical momentum scale given by the mass of the decaying RHN), i.e. we truncate $\exp({-{\rm i}\diamond})$
at zeroth order.
\item
We assume spatial isotropy, such that $\partial_i S^{<,>}(x)=0$ for $i=1,2,3$.
\end{itemize}
We are thus left with
\begin{align}
&\left[-{\rm i}\slashed k\gamma^0 +\partial_t+{\rm i}M\gamma^0\right]{\rm i}\gamma^0S^{<,>}
=-\frac12 \left({\rm i}\slashed\Sigma^>\gamma^0{\rm i}\gamma^0S^<-{\rm i}\slashed\Sigma^<\gamma^0{\rm i\gamma^0}S^>\right)\,,
\end{align}
where we have inserted factors of ${\rm i}\gamma^0$ such that we can readily
take the Hermitian and Antihermitian parts that are
\begin{subequations}
\label{kinetic:constraint:fermi:simple}
\begin{align}
\label{kineq:fermi:simple}
\frac{d}{dt}\gamma^0{\rm i}S^{<,>}-\frac{\rm i}{2}\left[M \gamma^0, {\rm i}\gamma^0 S^{<,>}\right]=&-\frac12\left\{{\rm i}\slashed\Sigma^>\gamma^0,{\rm i}\gamma^0 S^<\right\}
+\frac12\left\{{\rm i}\slashed\Sigma^<\gamma^0,{\rm i}\gamma^0 S^>\right\}\,,\\
\label{constrainteq:fermi:simple}
\frac12 \left\{\slashed k  - M, S^{<,>}\right\}
=&-\frac12\left[{\rm i}\slashed\Sigma^>\gamma^0,{\rm i}\gamma^0 S^<\right]
+\frac12\left[{\rm i}\slashed\Sigma^<\gamma^0,{\rm i}\gamma^0 S^>\right]\,.
\end{align}
\end{subequations}

We refer to the Antihermitian part~(\ref{constrainteq:fermi:simple}) as the constraint equation.
Neglecting the loop terms on the right-hand side, the solutions
are given in terms of the tree-level propagators~(\ref{prop:N:expl}). In addition,
for a system with several flavours, the solutions support flavour-off-diagonal correlations.
For now, we will leave these correlations aside and perform the calculation
in a perturbative expansion based on the tree-level propagators.
In Section~\ref{sec:reslg}, we show that this is equivalent to solving
Eqs.~(\ref{kinetic:constraint:fermi:simple}) also for the off-diagonal correlations in the RHNs
through most of the parameter space and that this also resolves the question of
how to correctly treat the degenerate regime where $|M_i-M_j|\gg\Gamma_{ii}$
is not satisfied.

The Hermitian part~(\ref{kineq:fermi:simple}) is called kinetic equation. To appreciate this,
we note that we can extract the distribution functions as
\begin{subequations}
\begin{align}
\label{tr:int:n_ell}
{\rm tr}\int\limits_{-\infty}^\infty\frac{d q^0}{2\pi}{\rm i}\gamma^0 S_\ell^{<,>}(q)=\left[\bar f_\ell(\mathbf q)-f_\ell(\mathbf q)\right]\,,\\
\label{tr:int:n_N}
{\rm tr}\int\limits_{-\infty}^\infty\frac{d p^0}{2\pi}{\rm sign}(p^0) S_{Ni}^{<,>}(p)
=-4 f_{Ni}(\mathbf p)\,,
\end{align}
\end{subequations}
where we have made use of the tree-level solutions~(\ref{prop:N:expl}) and
where the first equation corresponds to the zero-component of the
fermionic current, cf. Eq.~(\ref{currents}). By inspection of Eqs.~(\ref{prop:N:expl}),
one may also notice that the contributions for $\mathbf q^0=\pm |\mathbf q|$ in
Eq.~\eqref{tr:int:n_ell},
as imposed by the $\delta$-functions, account for leptons and antileptons, respectively.
For the RHNs, which are Majorana fermions, the distribution functions on the
positive and negative shell $p^0=\pm\sqrt{\mathbf p^2+M_i^2}$ must be equal due
to the Majorana condition. Since the respective distributions appear within $S^{<,>}_{Ni}$ with opposite sign, the purpose of the sign function in Eq.~\eqref{tr:int:n_N} is
to project on the particle distribution rather than the charge distribution as opposed to
Eq.~\eqref{tr:int:n_ell}.
Taking the trace of Eq.~(\ref{kineq:fermi:simple}) and integration over the zero component of the four momentum then leads to the following form of the
kinetic equations for RHNs and doublet leptons:
\begin{subequations}
\label{kin:eq}
\begin{align}
\label{kin:eq:N}
\frac d{dt}f_{Ni}(\mathbf p)&=
{\cal C}_N(\mathbf p)=
\frac 14\int\frac{dp^0}{2\pi}{\rm sign}(p^0)
{\rm tr}
\left[
{\rm i}\slashed\Sigma^>_{Ni}(p){\rm i}S_{Ni}^<(p)
-{\rm i}\slashed\Sigma_{Ni}^<(p){\rm i}S^>_{Ni}(p)
\right]\,,
\\
\frac d{dt}
\left(
f_\ell(\mathbf q)-\bar f_\ell(\mathbf q)
\right)
&={\cal C}_\ell(\mathbf q)
=\int\frac{dq^0}{2\pi}
{\rm tr}
\left[
{\rm i}\slashed\Sigma^>_\ell(q){\rm i}S_\ell^<(q)
-{\rm i}\slashed\Sigma_\ell^<(q){\rm i}S^>_\ell(q)
\right]\,.
\label{kin:eq:leptons}
\end{align}
\end{subequations}

Consequently, we obtain equations for the  number and charge densities when using Eq.~(\ref{Beq:Kineq}), and we calculate the particular collision terms in the following subsection.

\subsection{Decays, inverse decays and washout}
In order to derive the washout rate for the lepton asymmetry, we note
that the leading-order self-energy on the CTP is
\begin{align}
{\rm i}\slashed\Sigma^{{\rm LO} ab}_\ell(q)=&P_{\rm R} Y_i^* Y_i \int\limits_{p q^\prime} \delta_{p-q-q^\prime}
P_{\rm R}{\rm i}S_{Ni}^{ab}(p)P_{\rm L}{\rm i}\Delta^{ba}_\phi(q^\prime)\,,
\end{align}
which corresponds to the amputated one-loop diagram in Eq.~(\ref{SDE:lepton:diagrammatic}).
Assuming that $f_\phi=\bar f_\phi$, we obtain
\begin{align}
-{\cal W}=&\int\limits_q {\rm tr}
\left[
{\rm i}\slashed\Sigma^{{\rm LO}>}_\ell(q){\rm i}S_\ell^<(q)
-{\rm i}\slashed\Sigma_\ell^{{\rm LO}<}(q){\rm i}S^>_\ell(q)
\right]\notag\\
=&-|Y_i|^2\int\limits_{\mathbf p \mathbf q \mathbf q^\prime}
\delta_{p-q-q^\prime}
{\rm tr}\left[P_{\rm R} \slashed p \slashed k\right]
\left[f_{N1}(\mathbf p) + f_\phi(\mathbf q^\prime)\right]\times
\left[f_\ell(\mathbf q)-\bar f_\ell(\mathbf q)\right]\,,
\label{coll:tree}
\end{align}
which is proportional to the lepton asymmetry.
This agrees with the result~(\ref{integral:washout}), up to an extra term involving
$f_{Ni}(\mathbf p)$. It can also be reproduced in the Boltzmann approach when
accounting for the full quantum statistics. However, it is negligible in the
strong-washout regime, where the RHN distribution is exponentially suppressed compared
to the one of Higgs bosons. When including spectator effects, one should also expand in terms
of the charge asymmetry in Higgs bosons in addition to the one in leptons.

Next, for the rate of decays and inverse decays that drives the RHNs toward thermal equilibrium,
we use the leading-order self-energy of the RHNs,
\begin{align}
\label{Sigma:N:ij}
{\rm i}\slashed\Sigma_{Nij}^{ab}(p)=
g_w \int\limits_{k k^\prime}\delta_{p-k-k^\prime}
\left\{
Y_i Y_j^* P_{\rm L}{\rm i}S_\ell^{ab}(k)
\left[{\rm i}\Delta_\phi^{ab}(k^\prime)\right]^*
+
Y_i^* Y_j P_{\rm R}{\rm i}S_\ell^{Cab}(k)
{\rm i}\Delta_\phi^{ab}(k^\prime)
\right\}\,,
\end{align}
i.e. the amputated one-loop diagrams in Eq.~(\ref{SDE:RHN:diagrammatic}).

Using the approximations $f_\ell=\bar f_\ell$ and $f_\phi=\bar f_\phi$,
this leads to the decay term
\begin{align}
{\cal D}=&-\frac12 \int\limits_p{\rm tr}\; {\rm sign}(p^0) \left[
{\rm i}{\slashed\Sigma}_{Nii}^{>}(p) {\rm i}S_{Ni}^{<}(p)
-{\rm i}{\slashed\Sigma}_{Nii}^{<}(p) {\rm i}S_{Ni}^{>}(p)
\right]
=-g_w |Y_i|^2 \int\limits_p{\rm tr}\; {\rm sign}(p^0) \hat{\slashed{\Sigma}}_N^{\cal A} {\rm i}\delta S_{Nii}
\notag\\
=& g_w |Y_i|^2 \int_{\mathbf p \mathbf q \mathbf q^\prime}\delta_{p-q-q^\prime}
{\rm tr}\left[\slashed k \slashed p\right]
\left(1-f_\ell(\mathbf q)+f_\phi(\mathbf q^\prime)\right)
\delta f_{Ni}(\mathbf p)\,,
\label{D:CTP}
\end{align}
which is in agreement with Eq.~(\ref{integral:decay}), again up to quantum statistical
corrections. For comparison with the results in Section~\ref{sec:reslg}, we also define here
\begin{align}
\label{Sigma:hat:CTP}
\hat{\slashed\Sigma}^{\cal A}(p)=\gamma_\mu\hat\Sigma^{{\cal A}\mu}(p)
=\frac12\int\limits_{\mathbf q\mathbf q^\prime} \delta_{p-q-q^\prime} \slashed q \left(1-f_\ell({\mathbf q})+f_\phi(\mathbf q^\prime)\right)\,,
\end{align}
such that we can decompose
\begin{align}
\label{Sigma:spec:chiral}
\slashed\Sigma_{Nij}^{\cal A}=g_w\left(Y_i Y_j^* P_{\rm L}+Y_i^* Y_j P_{\rm R}\right)\gamma_\mu\hat\Sigma^{{\cal A}\mu}\,.
\end{align}
Note that in the strong-washout regime, we can use the zero-temperature approximation because
the distribution functions are Maxwell suppressed, such that $\hat\Sigma^{{\cal A}\mu}(p)=p^\mu/(32\pi)$, what agrees with the corresponding
quantity~(\ref{Sigma:hat}) appearing for decays in the vacuum.

When the RHNs are relativistic, one should also account for the asymmetry
in their different helicity states that is generated from decays and inverse decays
when $f_\ell\not=\bar f_\ell$ or $f_\phi\not=\bar f_\phi$.
This is can be of relevance for leptogenesis from oscillations of RHNs with mass below
the electroweak scale~\cite{Akhmedov:1998qx,Asaka:2005pn,Abada:2015rta,Hernandez:2015wna,Drewes:2016gmt,Antusch:2017pkq,
Ghiglieri:2018wbs,Drewes:2017zyw,Shaposhnikov:2008pf,Eijima:2017anv,Ghiglieri:2017gjz,Eijima:2018qke} but also for heavier RHNs as discussed in Section~\ref{sec:beyond:sw}.
The effect from the helicity
asymmetry is however only material when a sizable fraction of the RHNs are produced
or destroyed associated with the radiation of an extra gauge boson, i.e. in two-by-two
scatterings rather than in one-to-two decay and inverse decay processes. The latter
dominate however in the strong-washout regime with nonrelativistic RHNs, where it is
therefore a good approximation to neglect the helicity asymmetries as we will do throughout
the present discussion.

\subsection{$CP$-violating source}

As for the $CP$-violating source, we note that
the two-loop, vertex-type self energy for the doublet lepton on the CTP,
given by the amputated two-loop diagram in Eq.~(\ref{SDE:lepton:diagrammatic}), is
\begin{align}
\label{Sigmavert:CTP}
{\rm i}\slashed\Sigma_\ell^{{\rm vert}ab}(q)=&-\sum_{cd} cd Y_1^2 {Y_2^*}^2\!\!\!
\int\limits_{q^\prime k k^\prime p}\!\!\delta_{p-q-q^\prime}\delta_{p-k-k^\prime}
P_{\rm R} {\rm i}S_{N2}^{ad}(k-q^\prime)P_{\rm R} {\rm i}{S_\ell^{CP}}^{dc}(k)
P_{\rm L}{\rm i}S_{N1}^{cb}(p)P_{\rm L}
{\rm i}\Delta^{ca}(-k^\prime){\rm i}\Delta^{bd}(q^\prime)\notag\\
+&1\leftrightarrow 2\,.
\end{align}
For simplicity, we now set $M_1$ and $M_2$ to be real what can be achieved by rephasings
of the fields $N_{1,2}$.

The leading order washout terms that are proportional to $q_\ell$ are accounted for by the
contribution from $\slashed\Sigma_\ell^{\rm LO}$ already. We can thus substitute
equilibrium Fermi-Dirac or Bose-Einstein distributions into the expressions for the
fermion and scalar boson propagators. Then, the KMS relation~(\ref{eq:KMS}) implies that
the collision term on the right-hand side of Eq.~(\ref{kin:eq:leptons}) can only depend on the remaining deviation from equilibrium $\delta f_{N1}=f_{N1}-f_{N1}^{\rm eq}$.
Carrying out the sum over the CTP indices and
\begin{itemize}
\item
expanding to linear order in $\delta f_{N1}$,
\item
accounting only for the off-shell contributions from $N_2$ according to the assumption that $M_2\gg T$,
\end{itemize}
we arrive after some laborious rearrangements~\cite{Beneke:2010wd,Garbrecht:2013iga} 
that are shown in Appendix~\ref{app:vertex} at
\begin{align}
{\cal S}^{\rm vert}=&\int\limits_q {\rm tr}
\left[
{\rm i}\slashed\Sigma^{{\rm vert}>}_\ell(q){\rm i}S_\ell^<(q)
-{\rm i}\slashed\Sigma_\ell^{{\rm vert}<}(q){\rm i}S^>_\ell(q)
\right]\notag\\
=&-\left(Y_1^2 {Y_2^*}^2-{Y_1^*}^2 Y_2 \right)\int\limits_{\mathbf q \mathbf q^\prime \mathbf k \mathbf k^\prime \mathbf p}\delta_{p-q-q^\prime}\delta_{p-k-k^\prime}
{\rm tr}\left[
P_{\rm R}\frac{{\rm i}(\slashed k-\slashed q^\prime +M_2)}{(k-q^\prime)^2-M_2^2}
P_{\rm R}\slashed k P_{\rm L} (\slashed p + M_1) P_{\rm L} \slashed q
\right]\notag\\
\times&
\left(1-f_{\ell}(k)+f_{\phi}(k^\prime)\right)
\left(1-f_{\ell}(q)+f_{\phi}(q^\prime)\right)\delta f_{N1}(p)\,.
\label{coll:vert}
\end{align}
In the limit $M_1\gg T$, where the equilibrium distribution functions
for doublet leptons and Higgs bosons may be neglected, this result agrees
with Eq.~(\ref{integral:decayasymmetry}).

Now for the wave-function contribution, we observe that
there is no corresponding explicit two-loop diagram in Eq.~(\ref{SDE:lepton:diagrammatic}).
However, we note that we can approximately solve Eq.~(\ref{SDE:RHN:diagrammatic}) as
\begin{align}
\raisebox{-.1cm}{\includegraphics[scale=.6]{FullRHN.pdf}}
=&\raisebox{-.05cm}{\includegraphics[scale=.6]{TreeRHN.pdf}}+\raisebox{-1.1cm}{\includegraphics[scale=.6]{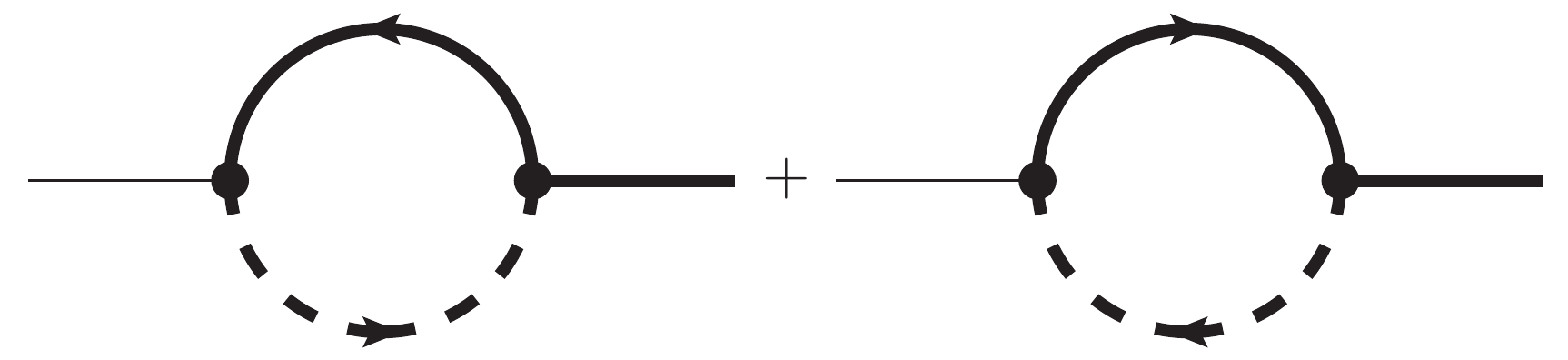}}\notag\\
=&\raisebox{-.05cm}{\includegraphics[scale=.6]{TreeRHN.pdf}}+\raisebox{-1.1cm}{\includegraphics[scale=.6]{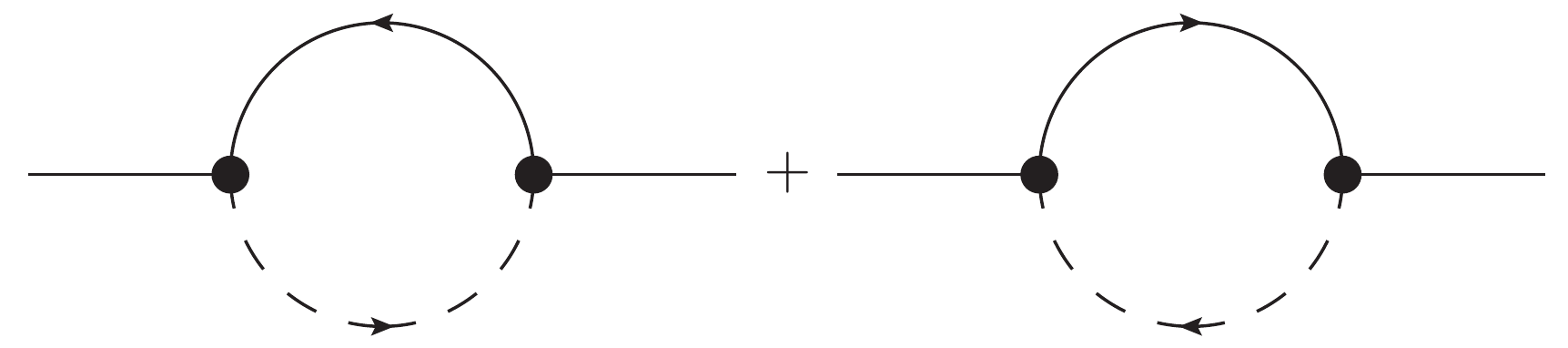}}+\cdots
\label{RHNexpand}
\end{align}
Substituting this approximation to the RHN propagator into the one-loop diagram
in Eq.~(\ref{SDE:lepton:diagrammatic}) then yields a two-loop diagram.
The wave function contribution can then be calculated analogously to the vertex~\cite{Beneke:2010wd}.
In order to compare with the results of Section~\ref{sec:reslg}, we
deviate slightly from that route and
first consider
separately the the correction to the propagator to the RHN, which is
given by
\begin{align}
{\rm i}S^{{\rm wv}ab}_{N ij}=-cd\, {\rm i}S^{ac}_{Ni}\, {\rm i}\slashed\Sigma_{Nij}^{cd}{\rm i}S_{Nj}^{cb}\,,
\end{align}
what is represented diagrammatically by the loop diagrams on the right-hand side of Eq.~(\ref{RHNexpand}).
This correction then enters $\slashed\Sigma_\ell^{<,>}$ that appears in the collision term
of the kinetic equation for the doublet leptons~(\ref{kin:eq:leptons}).
We specifically write down the term
\begin{align}
{\rm i}S^{{\rm wv}>}_{Nij}=
{\rm i}S^>_{Ni}{\rm i}\slashed\Sigma^<_{Nij}{\rm i}S^>_{Nj}
+{\rm i}S^{\bar T}_{Ni}{\rm i}\slashed\Sigma^>_{Nij}{\rm i}S^T_{Nj}
-{\rm i}S^>_{Ni}{\rm i}\slashed\Sigma^T_{Nij}{\rm i}S^T_{Nj}
-{\rm i}S^{\bar T}_{Ni}{\rm i}\slashed\Sigma^{\bar T}_{Nij}{\rm i}S^>_{Nj}\,,
\end{align}
where ${\rm i}S^{{\rm wv}<}_{Nij}$ follows from replacing $<\leftrightarrow>$, $T\leftrightarrow\bar T$. Next, from the KMS relation, we know that
any nonvanishing contributions to the collision term of doublet leptons at
this order must be proportional to deviations from equilibrium.
We therefore expand in $\delta S_{Ni}^{<,>}=\delta S_{Ni}^{<,>}-S_{Ni}^{{\rm eq}<,>}$,
where $\delta S_{Ni}^{<}=\delta S_{Ni}^{>}$, 
\begin{align}
{\rm i}\delta S_{N ij}^{<,>}={\rm i}\delta S_{Ni}\left({\rm i}\slashed\Sigma_{Nij}^>-{\rm i}\slashed\Sigma_{N ij}^T\right){\rm i}S_{N j}^T
-{\rm i}S_{Ni}^{\bar T}\left({\rm i}\slashed\Sigma_{Nij}^>-{\rm i}\slashed\Sigma_{N ij}^{\bar T}\right){\rm i}\delta S_{N j}\,.
\end{align}
In this expression, we have dropped terms that contain products of on-shell
$\delta$-functions pertaining to $N_i$ and $N_j$ with $i\not=j$
that cannot be simultaneously satisfied. The next simplification
is to drop the dispersive part from $\slashed\Sigma_N^T$, such that only
the absorptive, cut part is left. This is in accordance with our approximation
of neglecting corrections to the dispersion relations, i.e. dropping the 
term involving $\slashed\Sigma^{\rm H}$ in Eq.~(\ref{KBE:Wigner:Fermions}).
Furthermore, since
the dispersive parts satisfy $\slashed\Sigma_N^{T,\rm disp}=-\slashed\Sigma_N^{\bar T,\rm disp}$, it can be seen that these lead to contributions to the source term
for the asymmetry that cancel in total. We therefore arrive at
\begin{align}
{\rm i}\delta S_{N ij}^{<,>}\to
{\rm i}\delta S_{Ni}\slashed\Sigma^{\cal A}_{Nij}{\rm i}S_{N j}^T
-{\rm i}S_{Ni}^T\slashed\Sigma^{\cal A}_{Nij}{\rm i}\delta S_{N j}\,.
\end{align}
Note that the equality of the $<,>$ propagators can be verified using the
relation~(\ref{T:Tbar:gr:le}).

Substituting this result into
\begin{align}
\label{Sigma:wv}
{\rm i}\slashed\Sigma^{{\rm wv} ab}_\ell(q)=&\sum\limits_{ij}P_{\rm R} Y_i^* Y_j \int_{p q^\prime} \delta_{p-q-q^\prime}
P_{\rm R}{\rm i}S^{{\rm wv}ab}_{Nij}(p)P_{\rm L}{\rm i}\Delta^{ba}_\phi(q^\prime)\,,
\end{align}
we find for the wave-function contribution to the $CP$-violating source
\begin{align}
{\cal S}^{\rm wv}=&\int\limits_q {\rm tr}
\left[
{\rm i}\slashed\Sigma^{{\rm wv}>}_\ell(q){\rm i}S_\ell^<(q)
-{\rm i}\slashed\Sigma_\ell^{{\rm wv}<}(q){\rm i}S^>_\ell(q)
\right]\notag\\
=&
\int\limits_{p q q^\prime}\delta_{p-q-q^\prime}\sum_{ij}Y_i^* Y_j
{\rm tr}\left[P_{\rm R}{\rm i}\delta S_{Nij}(p) P_{\rm L} \left({\rm i}S_\ell^<(q) {\rm i}\Delta_\phi^<(q^\prime)-{\rm i}S_\ell^>(q) {\rm i}\Delta_\phi^>(q^\prime)\right)\right]\notag\\
=&
-2\int\limits_{p}\sum_{ij}Y_i^* Y_j
{\rm tr}\left[P_{\rm R}{\rm i}\delta S^{<,>}_{Nij}(p) P_{\rm L} \hat{\slashed\Sigma}^{\cal A}\!(p)\right]\notag\\
=&
-8 g_w\int\limits_{\mathbf p} M_1 M_2\, {\rm i}\left[{Y_1^*}^2Y_2^2-Y_1^2{Y_2^*}^2\right]
\frac{\hat\Sigma^{\cal A}_\mu(p)\hat\Sigma^{{\cal A}\mu}(p)}{M_2^2-M_1^2}\delta f_{N1}(\mathbf p) + 1 \leftrightarrow 2\,.
\label{source:wv}
\end{align}
When substituting Eq.~(\ref{Sigma:hat:CTP}) for $\hat\Sigma^{\cal A}$ and noting that the present result
accounts for the nonequilibrium sources from both RHNs as well as that it includes
quantum-statistical factors,
we once more note agreement
with Eq.~(\ref{integral:decayasymmetry}).
%

\subsection{Diagrammatic interpretation}

We note eventually that the expressions derived in the CTP approach justify the
diagrammatic representation in Figure~\ref{fig:integralexpr}. The diagrams obtained
from amputating the lines with a gap correspond to the self energies $\slashed\Sigma$,
and the line with a gap to the propagator that is attached to these in the collision term
of the Kadanoff--Baym equations. This external line is then closed by taking the
spinor trace as well as the momentum integral. The diagrams with closed lines can therefore be
identified with contributions to the 2PI effective action as in Eqs.~(\ref{Gamma2PI}), (\ref{Gamma2})
and~(\ref{Gamma2:diagrammatic}), and the self energies $\slashed\Sigma^{\rm LO}_\ell$, $\slashed\Sigma^{\rm vert}_\ell$ and $\slashed\Sigma_{Nij}$ follow from functional differentiation as in Eq.~(\ref{selferg:fermi}).
In contrast, $\slashed\Sigma^{{\rm wv} ab}_\ell$, Eq.~(\ref{Sigma:wv}), cannot be directly
obtained from the 2PI effective action because the vacuum graph that
would give rise to this contribution by functional differentiation
is not 2PI. Rather, when recalling that the diagrams
in the 2PI effective action are in terms of full propagators, $\slashed\Sigma^{{\rm wv} ab}_\ell$
arises from the leading insertion of $\slashed\Sigma_{Nij}$
into the RHN propagator as indicated in Eq.~(\ref{RHNexpand}).
On the other hand, we may expect that we resolve the issues encountered
in Section~\ref{sec:mixing:variants} concerning the mass-degenerate limit
when appropriately resumming these insertions.
This resummation is readily accomplished by the Schwinger--Dyson equations~(\ref{SDE:RHN:diagrammatic}),
and we will make use of this fact in the following section on resonant leptogenesis.

\section{Resonant effects in out-of-equilibrium decay scenarios}
\label{sec:reslg}

In Section~\ref{sec:kineq:firstprinciples}, we have derived the
$CP$-violating source term in the CTP approach based on first principles of QFT. In view of
the problems in telling apart one-to-two and two-to-two matrix elements
in the Boltzmann approach and thus to comply with the $CPT$ theorem, the CTP approach
appears to be more systematic.
Here, we show how the kinetic equations
derived in the CTP framework in addition lead to a resolution of the questions
pertaining to the resonant limit $M_1\to M_2$ of leptogenesis. In particular,
as discussed in Section~\ref{sec:mixing:variants}, a satisfactory treatment
should address the calculation of the asymmetry in the parametric regime
where the mass splitting is comparable to or smaller than the decay width of the
RHNs.

Within the CTP framework, rather than
inserting the spectral self energy into the propagator of the RHN to first or
any finite order, the solution to the Schwinger--Dyson equations in an appropriate
approximation automatically leads to the all-order resummation of the effects
from the finite lifetime of the RHN.
Compared to the approaches presented Section~\ref{sec:mixing:variants}, that are inconclusive to this end,
the Schwinger--Dyson equation on the CTP has two features that are crucial in order
to lead to correct predictions throughout the parametric range of the seesaw mechanism, including
the resonant regime:
\begin{itemize}
\item
The state of the RHNs and their flavour correlations is solved for consistently. This
is important in the extremely degenerate regime where the correlations turn out
to be of the same magnitude as the deviations of the flavour-diagonal distributions
from equilibrium.
\item
Off-diagonal correlations take a finite time to build up. In the strong washout regime,
under circumstances to be specified in more detail below, this typically is not a concern.
This time-dependence is however captured by the Schwinger--Dyson equations such that
these are applicable also to the weak washout regime or to scenarios of leptogenesis
from the oscillations of relativistic RHNs with masses below the electroweak scale~\cite{Drewes:2016gmt}.
\end{itemize}

Now, in general, the self-consistent solutions to the nonequilbrium portion
of the Wightman functions of the RHNs ${\rm i}\delta S^{<,>}_N$ exhibit a more complicated spinor
structure than the tree-level, equilibrium solutions~(\ref{prop:N:expl}). This is because
the $CP$-violating effects, besides creating charge asymmetries, also lead to asymmetries
in axial charges and densities. The latter are of material importance in models of flavoured
leptogenesis (i.e. when all or an important part of the asymmetry in doublet leptons
is purely flavoured in first place) when the RHNs are relativistic. This combination of circumstances
is characteristic for leptogenesis from oscillations of relativistic RHNs~\cite{Akhmedov:1998qx,Asaka:2005pn,Abada:2015rta,Hernandez:2015wna}.

In order to capture the details of these chirality and helicity effects in the RHN sector,
a decomposition of the spinor fields following the methods developed for electroweak baryogenesis
in Refs.~\cite{Prokopec:2003pj,Prokopec:2004ic} must be applied, which we review in Section~\ref{sec:ewbg} of the present
work. For leptogenesis, this procedure has been carried out and is reported in detail
in Refs.~\cite{Garbrecht:2011aw,Drewes:2016gmt}. Here, for simplicity, we restrict
to the nonrelativistic regime, where the axial and pseudoscalar components of ${\rm i}\delta S^{<,>}_N$
are small compared to the vectorial and scalar ones. We can hence approximate that
${\rm tr}\left[ P_{\rm L,R} {\rm i} \delta S^{<,>}(p) \right]=2 \bar M \delta f_N(\mathbf p) 2\pi \delta(p^2-\bar M^2)$ and
${\rm tr}\left[\slashed V P_{\rm L,R}{\rm i} \delta S^{<,>}(p)\right]=-2 p\cdot V \delta f_N(\mathbf p) 2\pi \delta(p^2-\bar M^2)$ as can be seen from Eqs.~(\ref{prop:N:expl}), where $\bar M=(M_1+M_2)/2$ and we assume $|M_1-M_2|\ll \bar M$.
With these approximations,
integrating~Eq.~(\ref{kineq:fermi:simple}) over $dp^0$ and taking the trace as in Eq.~(\ref{tr:int:n_N}) leads to
\begin{align}
\label{fN:osci}
\frac{d}{dt} f_N(\mathbf p)+\frac{\rm i}{2 p^0}\left[M^2,\delta f_{N}\right]
=-g_w \frac{p\cdot\hat \Sigma^{\cal A}}{p^0} \left\{{\rm Re}[Y^* Y^t],\delta f_N\right\}\,,
\end{align}
where $f_N$ and $\delta f_N$ now take values of matrices in the flavour space of
RHNs, $Y$ is understood as a column vector and the superscript $t$ stands for transposition.
We note that through taking
the trace, we have added together the different polarization states for the right-handed neutrinos.
In the nonrelativistic regime, differences in the decay asymmetry for the two polarization
states in the finite temperature medium remain small such that it is justified not to track the helicity
asymmetry. Corrections to this treatment that become relevant toward the relativistic regime are
covered by the derivations of Refs.~\cite{Garbrecht:2011aw,Drewes:2016gmt} (see also Refs.~\cite{Akhmedov:1998qx,Asaka:2005pn,Abada:2015rta,Hernandez:2015wna,Drewes:2016gmt,Ghiglieri:2018wbs,Shaposhnikov:2008pf,Eijima:2017anv,Ghiglieri:2017gjz,Eijima:2018qke} that address this matter using canonical and finite-temperature field-theory approaches in contrast to the CTP method). We can view
Eq.~(\ref{fN:osci}) as a reduced version (through taking a trace and applying the nonrelativistic approximation) of the Schwinger--Dyson equation~(\ref{SDE:RHN:diagrammatic})
that carries out the necessary resummations (i.e. the one-loop insertions
into the RHN propagator accounting for the finite width) for the mass-degenerate regime.

To account for the expansion of the Universe, we apply the rule~(\ref{replace:rule:expansion})
to Eq.~(\ref{fN:osci}), which yields
\begin{align}
\label{sterile:oscillations}
\bar M \frac{d}{dz}\delta  f_N+\frac{a_{\rm R} z}{2p^0 \bar M}{\rm i}[M^2,\delta f_N]+\bar M\frac{d}{dz}f_N^{{\rm eq}}
=&
- \frac{a_{\rm R}}{\bar M} z g_w \bigg\{{\rm Re}[Y^* Y^t]\frac{p\cdot\hat\Sigma^{\cal A}_{N}}{p^0},\delta f_N\bigg\}
\,,
\end{align}
where $f_N^{\rm eq}$ is the equilibrium Fermi-Dirac distribution of the RHNs. We have explicitly
decomposed the temporal derivative acting on $f_N=f_N^{\rm eq}+\delta f_N$, such that
the term $d/dz\,f_N^{\rm eq}$ mediates the deviation from equilibrium that drives
leptogenesis. Note that Eq.~(\ref{sterile:oscillations}) is structurally similar to
the equation that has been studied in Ref.~\cite{Cirigliano:2009yt} for a system
of mixing scalar particles.

Now, since $\delta f_N$ takes the values of Hermitian matrices,
it has four degrees of freedom, for which we
can write down coupled first-order differential equations~\cite{Iso:2013lba,Iso:2014afa,Garbrecht:2014aga}.
The matrix associated with the homogeneous
part of these differential equations has
four real eigenvalues, which depend on $M_1^2-M_2^2$
as well as ${\rm Re}[Y^*Y^t]p\cdot\hat\Sigma_N^{\cal A}(p)$.
Provided these eigenvalues are large compared to $(\bar M d/dz f_N^{\rm eq})/f_N^{\rm eq}$,
we can neglect the first term in Eq.~(\ref{sterile:oscillations}).
We note that for two RHNs, these eigenvalues
can be worked out analytically~\cite{Garbrecht:2014aga}, but simple sufficient criteria
are that $R^{1/4}\gg H$ or $|M_1^2-M_2^2|/\bar M\gg H$, implying
that the characteristic scales of relaxation toward equilibrium or flavour oscillations
are faster than the Hubble rate that sets the scale for the temporal derivatives, where $R$ is the regulator defined in Eq.~\eqref{epsilon:recast} below.
As a consequence, the
system can then be solved algebraically. Before we carry out this
task, we note that one may choose initial conditions where $d/dt\,f_{Nij}\sim (M_i^2-M_j^2)/\bar M f_{Nij}$, i.e. where the derivative is dominated by flavour oscillations. However, such initial
conditions are not realized in the strong-washout scenario because when  $|M_i^2-M_j^2|/(M_i+M_j)\gg H$,  the RHNs
continuously drop out-of-equilibrium over many oscillation times such that no coherent oscillations will occur~\cite{Garbrecht:2014aga}. Even if one chooses initial conditions leading to coherent
oscillations, these will average about the solution to the algebraic system such that they can be neglected in calculations of the resulting lepton asymmetry~\cite{Garbrecht:2011aw}. Nonetheless, we note that outside the strong-washout regime,
the time-dependence of the flavour correlations of the  RHNs have to be accounted for by keeping the
time-derivative in Eq.~(\ref{sterile:oscillations}).

Dropping the derivative in Eq.~(\ref{sterile:oscillations}) as per the discussion above, we obtain
the algebraic solution~\cite{Garbrecht:2014aga}
\begin{align}
\label{deltan:latetime}
\delta f_{N ij}=&\frac{\bar M}{2 D}([YY^\dagger]_{ij}+[Y^*Y^t]_{ij})
([YY^\dagger]_{ii} +[YY^\dagger]_{jj})
\\\notag
\times&
[\bar M^2 \bar\Gamma([YY^\dagger]_{ii} +[YY^\dagger]_{jj})-{\rm i}(M_i^2-M_j^2)]
\times\frac{\bar M^2}{a_{\rm R} z}\frac{d}{dz} f_N^{\rm eq}\,,
\end{align}
where
\begin{align}
D=&[YY^\dagger]_{11} [YY^\dagger]_{22} (M_1^2-M_2^2)^2
\\\notag
+&\bar M^4\bar\Gamma^2([YY^\dagger]_{11} +[YY^\dagger]_{22})^2([YY^\dagger]_{11}[YY^\dagger]_{22}-{\rm Re}\{[YY^\dagger]_{12}\}^2)\,,
\end{align}
and $\bar\Gamma=1/(8\pi)$.
Substituting this into the source term, we obtain 
\begin{align}
{\cal S}^{\rm wv}=&
-2\int\limits_{p}\sum_{ij}Y_i^* Y_j
{\rm tr}\left[P_{\rm R}{\rm i}\delta S^{<,>}_{Nij}(p) P_{\rm L} \hat{\slashed\Sigma}^{\cal A}\!(p)\right]
\label{source:wv:reslg}\,,
\end{align}
which generalizes the results~(\ref{integral:decayasymmetry}) and (\ref{source:wv}) and now
also applies in the extremely degenerate regime.

We also note that in Refs.~\cite{Dev:2014laa,Dev:2014wsa} it is argued that besides the $CP$-violating source
calculated in this section, the source according to Eq.~(\ref{squamp:LSZ}) is a separate and distinct contribution.
Obtaining the source term from the 2PI effective action, we find however that the source terms~(\ref{source:wv})
and~(\ref{source:wv:reslg}) are the same contribution and agree within the range of applicability of Eq.~(\ref{source:wv}),
i.e. when $|M_1-M_2|\gg |\Gamma_{ij}|$. Certainly, further investigations into this matter, as have been initiated in Ref.~\cite{Kartavtsev:2015vto}, would therefore be of interest.

In order to verify that the above results are indeed recovered from this more general expression,
we note that outside of the extremely degenerate regime the diagonal components of
the matrix of the nonequilibrium RHN neutrino distributions dominate the off-diagonal
ones. These are then related to the rate at which the RHNs drop out of equilibrium as
\begin{align}
\frac{1}{|Y_i|^2}\frac{\bar M^2}{a_{\rm R} z}\frac{d}{dz} f_N^{\rm eq}=2 g_w \frac{p\cdot\hat\Sigma^{\cal A}}{p^0}\delta f_{Ni}\,.
\end{align}
Substituting this into Eq.~(\ref{deltan:latetime}), we obtain
\begin{align}
\delta f_{Nij}=-\frac{\rm i}{2}\frac{Y_i Y_j^*+Y_i^* Y_j}{M_i^2-M_j^2}2 g_w \frac{p\cdot\hat\Sigma^{\cal A}}{p^0}\bar M \left(\delta f_{Ni}(\mathbf p)-\delta f_{Nj}(\mathbf p)\right)\,,
\end{align}
which, when used in Eq.~(\ref{source:wv:reslg}) leads to
\begin{align}
{\cal S}^{\rm wv}={\rm i}\left[{Y_1^*}^2Y_2^2-Y_1^2 {Y_2^*}^2\right]
\int\limits_{\mathbf p}\frac{\bar M^2}{M_1^2-M_2^2}8 \hat\Sigma^{{\cal A}0}\hat\Sigma^{{\cal A}0}
\left(\delta f_{N1}(\mathbf p)-\delta f_{N2}(\mathbf p)\right)\,,
\end{align}
in agreement with the results~(\ref{integral:decayasymmetry}) and (\ref{source:wv}).

Noting that in the strong-washout regime, the evolution of the lepton asymmetry can be
expressed as in Eq.~(\ref{Delta:ell:swo}) (There is a different sign in that equation compared to Ref.~\cite{Garbrecht:2014aga}.) below and comparing with Eq.~(\ref{deltan:latetime}) and~(\ref{source:wv:reslg}), we find for the decay asymmetry~\cite{Garbrecht:2014aga} (As there is no mass hierarchy, this
is the average decay asymmetry for both RHNs.)
\begin{align}
\label{epsilon:flavoured}
\varepsilon=&{\rm i}\frac{{Y_1^*}^2Y_2^2-Y_1^2{Y_2^*}^2}{16 \pi}\frac{M_1^2-M_2^2}{D}\bar M^2  \left(|Y_1|^2+|Y_2|^2\right)\,.
\end{align}
For the sake of comparison between different approaches to resonant leptogenesis
that can be found in the literature, we rearrange this expression for the asymmetry into
\begin{align}
\label{epsilon:recast}
\varepsilon=&
{\rm i}\frac{{Y_1^*}^2Y_2^2-{Y_2^*}^2Y_1^2 }{16\pi}\left(\frac{1}{|Y_1|^2}+\frac{1}{|Y_2|^2}\right)\frac{\bar M^2 (M_1^2-M_2^2)}{(M_1^2-M_2^2)^2+R}\,,
\end{align}
where
\begin{align}
\label{R:strongwashout}
R=\frac{\bar M^4}{64\pi^2}\frac{\left(|Y_1|^2+|Y_2|^2\right)^2}{|Y_1|^2|Y_2|^2}\left({\rm Im}[Y_1 Y_2^*]\right)^2\,.
\end{align}

First, we compare with the result from Refs.~\cite{Pilaftsis:1997jf,Pilaftsis:2003gt}, where the regulator is obtained in the standard
$S$-matrix formalism by using a resummed form for the RHN propagator,
as it is discussed here in the derivation of Eq.~\eqref{squamp:LSZ}.
For two RHNs, the sum of the decay asymmetries is found to be
\begin{align}
\varepsilon=&
{\rm i}\frac{{Y_1^*}^2Y_2^2-Y_1^2{Y_2^*}^2}{16 \pi}
\left(\frac{\bar M^2 (M_1^2-M_2^2)}{(M_1^2-M_2^2)^2+\frac{|Y_2|^4}{64\pi^2}\bar M^4}\frac{1}{|Y_1|^2}
+\frac{\bar M^2 (M_1^2-M_2^2)}{(M_1^2-M_2^2)^2+\frac{|Y_1|^4}{64\pi^2}\bar M^4}\frac{1}{|Y_2|^2}\right)\,.
\end{align}
While this expression cannot be cast into the form of Eq.~\eqref{epsilon:recast}, it nonetheless
may easily be compared with that equation in conjunction with the regulator~\eqref{R:strongwashout}, such that one notes
disagreement when $(M_1^2-M_2^2)^2$ is of order of
$R$, $|Y_{1,2}|^4\bar M^4/(64\pi^2)$ or smaller.

\begin{figure}
\begin{center}
\parbox{8cm}{\parbox{.7cm}{\vskip-6cm$\frac{|Y_2|}{|Y_1|}$}\includegraphics[width=6cm]{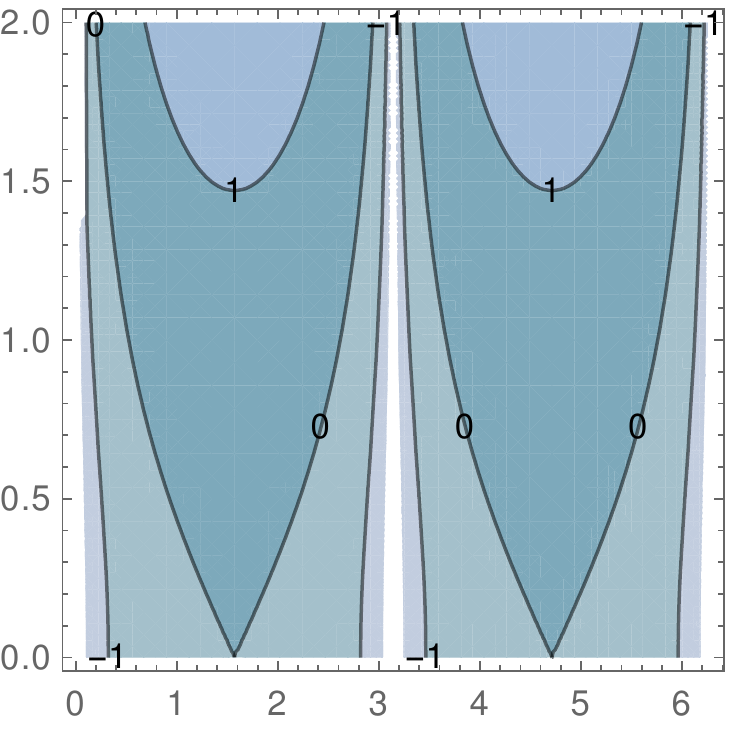}\\[-3mm]$\phantom{X}$\hspace{3.5cm}$\varphi$}
\parbox{8cm}{\parbox{.7cm}{\vskip-6cm$\frac{|Y_2|}{|Y_1|}$}\includegraphics[width=6cm]{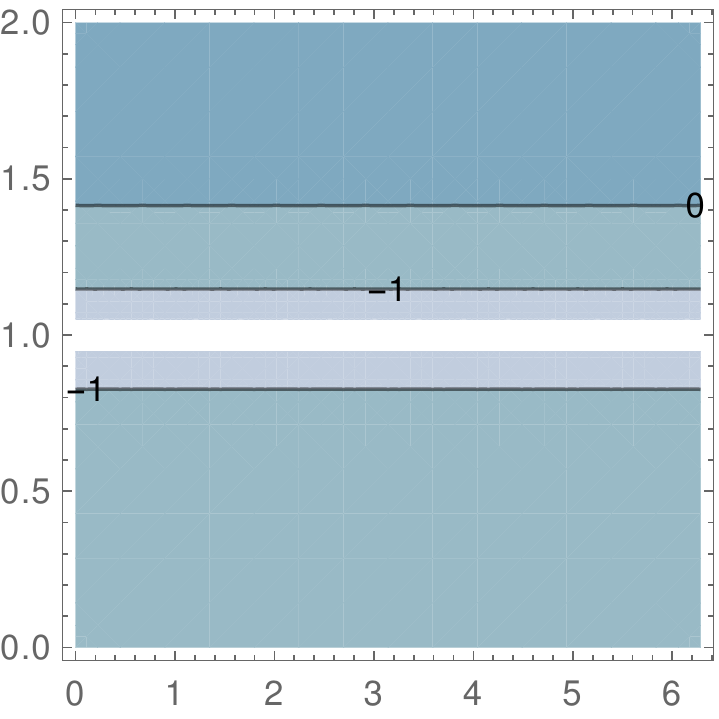}\\[-3mm]$\phantom{X}$\hspace{3.5cm}$\varphi$}
\end{center}
\caption{\label{fig:R}Regulators $(64 \pi^2/\bar M^4) R$ for the strong-washout scenario from Eq.~\eqref{R:strongwashout} and for RHN oscillations in vacuum from Eq.~\eqref{regulator:Broncano} (right panel). We have set $Y_1=|Y_1|$ and $Y_2={\rm e}^{{\rm i}\varphi}|Y_2|$.}
\end{figure}

In Ref.~\cite{Anisimov:2005hr}, it is argued that the resummation needs to
take account of the mixing of the RHNs in a different way, leading to the
regulator
\begin{align}
\label{regulator:Broncano}
R=\frac{\bar M^4}{64\pi^2}\left(|Y_1|^2-|Y_2|^2\right)^2
\,.
\end{align}
This result turns out to be in agreement with what is
found using the Hamiltonian description of RHN oscillations in
vacuum from Refs.~\cite{Flanz:1994yx,Flanz:1996fb} what
has lead us to Eq.~\eqref{reslg:Hamiltonian}. In Figure~\ref{fig:R},
we compare the Regulators from Eqs.~\eqref{R:strongwashout} and~\eqref{regulator:Broncano}. As stated above, while the regulator~\eqref{R:strongwashout}
is the one that holds for strong washout, outside of that regime
the oscillation equations from Section~\ref{sec:reslg} should be solved in order to
determine the asymmetry correctly~\cite{Garbrecht:2011aw}.

\section{Analytic approximation to the strong washout solution}
\label{sec:pedestrian}

\subsection{Minimal out-of-equilibrium decay and strong washout}

Baryogenesis calculations often rely on numerical solutions to the
fluid equations for the evolution of the asymmetry. Notably,
the solutions to 
Eqs.~(\ref{fluid:eq:leptogenesis}) with the rates~(\ref{rates:exp:bkg}) for leptogenesis have relatively simple
analytic approximations, and the derivation of these offers some insights
into the mechanism~\cite{Kolb:1983ni}.

Provided $(Y_{N_1}-Y_{N1}^{\rm eq})\ll Y_{N1}^{\rm eq}$ throughout the times
we are interested in, we can readily write down the approximate
solution to Eq.~(\ref{fluid:eq:leptogenesis:N1}) as
\begin{align}
\label{Delta:YN}
(Y_{N_1}-Y_{N1}^{\rm eq})=-\frac{1}{\bar\gamma}\frac{d}{dz} Y_{N1}^{\rm eq}\,.
\end{align}
Our parametrization is chosen such that a derivative with respect to
$z$ can be counted as order one in the radiation-dominated Universe, while
the parameter $\bar\gamma$ is given in Eq.~(\ref{gammabar}). Above inequality
is therefore satisfied if $\bar\gamma(z=1)\gg 1$, (what as a more physical relation implies
$\gamma\big|_{T=M_1}\gg H$, i.e. the reaction rate is much faster than the Hubble rate
when $z\sim 1$).
The time where $z\sim 1$ is  of relevance because this is where a Fermi-Dirac distribution
for fermions of mass $M_1$
in comoving momentum that is subject to redshift deviates from the equilibrium form and
because the lepton-number violating decay and inverse decay rates acquire a factor of
exponential Maxwell suppression $\sim\exp(-M_1/T)$, such that these processes freeze out soon
after.

Substituting the approximation~(\ref{Delta:YN}) into Eq.~(\ref{fluid:eq:leptogenesis:ell}), we next obtain
\begin{align}
\label{Delta:ell:swo}
\frac{d Y_{\Delta \ell}}{dz}=&
-\varepsilon\frac{1}{g_w}\frac{d}{dz} Y_{N1}^{\rm eq}
-\bar W_{\left(\frac32\right)}Y_{\Delta \ell}\,,
\end{align}
where $\bar W_{\left(\frac32\right)}=\frac32\bar W$,
which has the formal solution
\begin{align}
\label{YDeltaL:formal}
Y_{\Delta \ell}(z)=-\int\limits_0^z dz^\prime \varepsilon\frac{\bar\gamma}{g_w}\frac{d}{dz^\prime} Y_{N1}^{\rm eq}(z^\prime)
\exp\left\{-\int\limits_{z^\prime}^z d z^{\prime\prime} \bar W_{\left(\frac32\right)}(z^{\prime\prime})\right\}\,.
\end{align}
This integral can be approximately evaluated using Laplace's method, where
besides the explicit exponential in the above integrand, there is also an exponential
factor contained within $Y_{N1}^{\rm eq}$, cf. Eq.~(\ref{YN1:eq}). The full exponent
is extremal for
\begin{align}
\label{exp:extremal}
\bar W_{\left(\frac32\right)}(z_{\rm f})=1\,,
\end{align}
where we attach a subscript $\rm f$ to the solution for $z$ because around this
point in time, lepton-number violating interactions freeze out. It is as useful as
customary to define the washout strength
\begin{align}
K=\frac{\Gamma_{N_1 \to \ell \phi^*,\ell^{CP} \phi}}{H}\Big|_{T=M_1}
=\frac{|Y_1|^2 M_1}{8\pi H|_{T=1}}=\frac{|Y_1|^2}{8\pi}\frac{a_{\rm R}}{M_1}\,,
\end{align}
implying that for $K\gg 1$, the lepton-number violating interactions are close to
equilibrium before they freeze out, what corresponds to strong washout. We can then express the Laplace approximation
to Eq.~(\ref{YDeltaL:formal}) as
\begin{align}
\label{YDelta:zf}
Y_{\Delta \ell}=\varepsilon \frac{15\sqrt 3}{g_w g_\star\sqrt{K}}\pi^{-\frac94}2^{-\frac34}z_{\rm f}^\frac14 \exp\left\{-\frac{\rm z_{\rm f}}{2}-\int\limits_{z_{\rm f}}^\infty dz^\prime \bar W_{\left(\frac32\right)}(z^\prime)\right\}\,.
\end{align}

When substituting back the solution to Eq.~(\ref{exp:extremal}) [The integral in the
exponent of Eq.~(\ref{YDelta:zf}) evaluates to \emph{one} for $z_{\rm f}\gg 1$.], we obtain
\begin{align}
Y_{\Delta\ell}=\varepsilon\frac{5 \sqrt 2}{{\rm e} g_\star g_w K \pi^\frac32 z_{\rm f}}\,.
\end{align}
While larger $K$ increase $z_{\rm f}$, this only occurs logarithmically slow. Apart from this
small dependence, most significant about this result is its behaviour $\sim 1/K$, implying that
the outcome of leptogenesis in the strong-washout regime has a simple dependence on two parameters: the decay asymmetry $\varepsilon$ and the washout strength $K$. A more accurate analytic approximation is derived in Ref.~\cite{Buchmuller:2004nz}.

We note that eventually, sphaleron processes convert baryon-minus-lepton number $B-L$ to baryon number $B$
as~\cite{Khlebnikov:1996vj,Laine:1999wv}
\begin{align}
Y_B=4\frac{77 T^2+27 \langle\sqrt2|\phi|\rangle^2}{869 T^2+333 \langle\sqrt2|\phi|\rangle^2}Y_{B-L}\,,
\end{align}
where $\langle |\phi| \rangle$ is the expectation value of the Higgs field through the
electroweak crossover, and 
for simplicity, we assume here that the asymmetry is unflavoured, i.e. the lepton number is the same in all generations of SM leptons. The generalized expression for the more realistic case of flavoured asymmetries
can also be found in Ref.~\cite{Laine:1999wv}. Taking from Ref.~\cite{DOnofrio:2014rug} the temperature
of $131\,{\rm GeV}$ for sphaleron freeze out and $\langle\sqrt2| \phi|\rangle\sim 170\,{\rm GeV}$
at that temperature, one finds $Y_B\approx0.343\, Y_{B-L}$, to be compared with the
relation $Y_B\approx28/79\, Y_{B-L}\approx 0.354\, Y_{B-L}$~\cite{Harvey:1990qw} in the symmetric
phase where $\langle\phi\rangle=0$. Note also that because of our counting of the
weak isodoublet charges, $Y_{B-L}=2\,Y_{\Delta\ell}$. The above result for $Y_B$ can be
compared with the observed value~(\ref{BAU:BBN}) when switching from entropy to photon
normalization.

\subsection{Beyond minimal strong washout: weak washout and spectator effects}
\label{sec:beyond:sw}

Up to this point, the present work has been concerned  with the application of the CTP approach
to the minimal realization of leptogenesis in the strong-washout regime, which may be considered as the archetypical example of baryogenesis from
out-of equilibrium decays. One exception, where we go into more particular detail, is
the discussion on resonant leptogenesis, where the CTP methods establish the connection between
decay asymmetries from $S$-matrix elements and those computed from the
oscillations of the RHNs, thus leading to an accurate calculation of the asymmetry produced in the mass-degenerate regime. Nonetheless, also that
discussion may be generalized to other scenarios relying on the mixing of 
almost mass-degenerate states~\cite{Garbrecht:2012qv,Garbrecht:2012pq,Garbrecht:2014iia}.
Moreover, in Section~\ref{sec:ewbg}, we draw some parallels with electroweak baryogenesis, such that resonant leptogenesis seems to be a fitting topic
within the present review. Here, in addition, we give a qualitative overview
of the use of the CTP techniques in leptogenesis from RHNs with masses
way above the electroweak scale beyond the minimal scenario of strong washout.

In principle, both Hamiltonian as well as CTP formulations yield the time evolution of the
nonequilibrium quantum state. Thus, Boltzmann kinetic equations can then be derived from first principles.
In practice, certain reaction rates can either be derived using these techniques of real-time evolution
but alternatively, also equilibrium field theory may be a useful method. Of interest is here the
inclusion of corrections due to the radiation of gauge bosons or top quarks. In the norelativistic strong-washout regime, these are perturbatively suppressed, and the washout and decay rates for
RHNs have been calculated~\cite{Salvio:2011sf,Laine:2011pq}. Notably, using an effective theory where the RHNs are integrated out,
also the perturbative corrections to the decay asymmetry have been found in that limit~\cite{Biondini:2013xua,Biondini:2015gyw,Biondini:2016arl}.

On the other hand, the production of RHNs in the relativistic regime has received interest in
the context of leptogenesis from light RHNs with ${\rm GeV}$-scale masses~\cite{Akhmedov:1998qx,Asaka:2005pn,Abada:2015rta,Hernandez:2015wna,Drewes:2016gmt,Antusch:2017pkq,Ghiglieri:2018wbs,Drewes:2017zyw}. Calculations
of the reaction rate using thermal field theory include Refs.~\cite{Anisimov:2010gy,Laine:2011pq,Besak:2012qm,Laine:2013lka,Ghisoiu:2014ena}, whereas the CTP approach
has been applied to this matter in Refs.~\cite{Garbrecht:2013gd,Garbrecht:2013urw}.

In the context of leptogenesis from light RHNs, it has also been found necessary to track the helicity asymmetries within the RHNs, that
correspond to generalized lepton asymmetries in the relativistic limit.
Consequently, one can then distinguish
between lepton-number conserving and lepton-number violating reactions~\cite{Antusch:2017pkq,
Ghiglieri:2018wbs,Drewes:2017zyw,Shaposhnikov:2008pf,Eijima:2017anv,Ghiglieri:2017gjz,Eijima:2018qke}, where the latter are mediated by the Majorana mass that is small compared
to the temperature. Moreover,
both types of rates play a role in the source term for the asymmetry~\cite{Garbrecht:2019zaa,Hambye:2016sby}. Within the CTP approach,
accounting for helicity asymmetries can be dealt with by accounting for
the axial and pseudoscalar density (in contrast to the approximations made in Section~\ref{sec:reslg}), as it is worked out in Refs.~\cite{Drewes:2016gmt,Antusch:2017pkq} (cf. also Section~\ref{sec:fermiontransport}).

For leptogenesis with heavy RHNs with masses way above the electroweak scale, these findings
have important implications~\cite{Garbrecht:2019zaa}. Provided asymmetries that are produced at early times, when the RHNs are relativistic, persist until lepton-number violating processes freeze out, tracking the helicity asymmetries within the RHNs and distinguishing lepton-number conserving from lepton-number violating rates is
important for the dynamics of washout as well as for the source term. In particular, the lepton-number
violating source is a product of lepton-number conserving and lepton-number violating factors,
where the former are enhanced over the latter in the relativistic regime. This also implies
that the decay asymmetry in vacuum is to be replaced by a temperature-dependent asymmetry. Early asymmetries can persist either
in the case of weak washout or in the presence of partially equilibrated spectator fields.
Spectators are SM degrees of freedom
that can store baryon or lepton number but not suffer direct washout from the RHNs.
Nonetheless, when there is chemical equilibrium between the lepton asymmetry and the spectators,
the washout in the leptons is instantaneously communicated to the spectators. In the case of partial
equilibration however, large asymmetries that may be present at early times can partly
persist in the spectator fields, even when the lepton asymmetries suffer strong washout.
For example, at temperatures
around $10^{13} {\rm GeV}$, lepton asymmetries can be transferred via the Higgs fields and bottom-Yukawa couplings to $b$-quarks. Furthermore, the weak sphaleron communicates lepton asymmetries to the quarks, cf. Figure~\ref{fig:sphalerons}. Both of these processes
are only partly in equilibrium though. (Top-quark Yukawa and strong sphaleron interactions are fully
equilibrated at these temperatures.)

\begin{figure}
\begin{center}
\includegraphics[width=9cm]{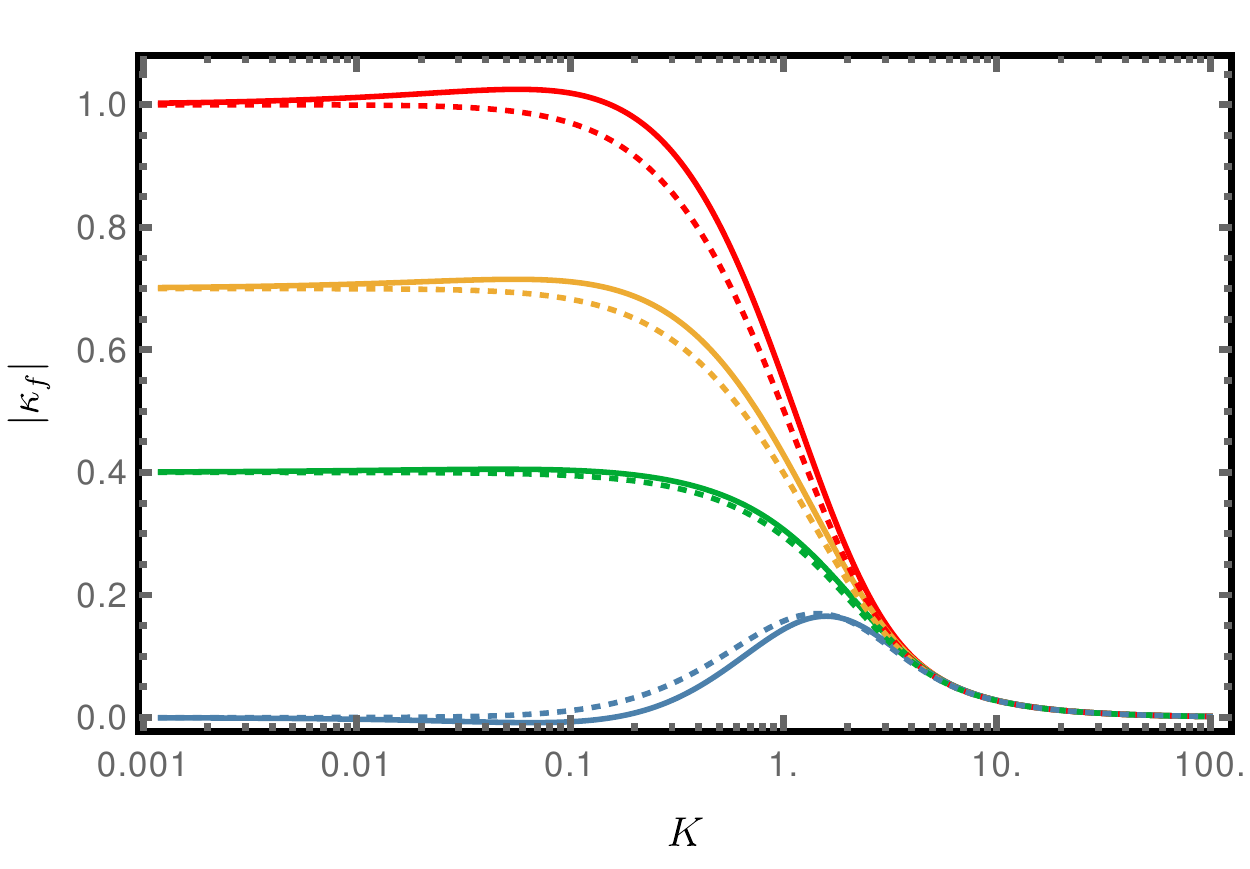}
\includegraphics[width=9cm]{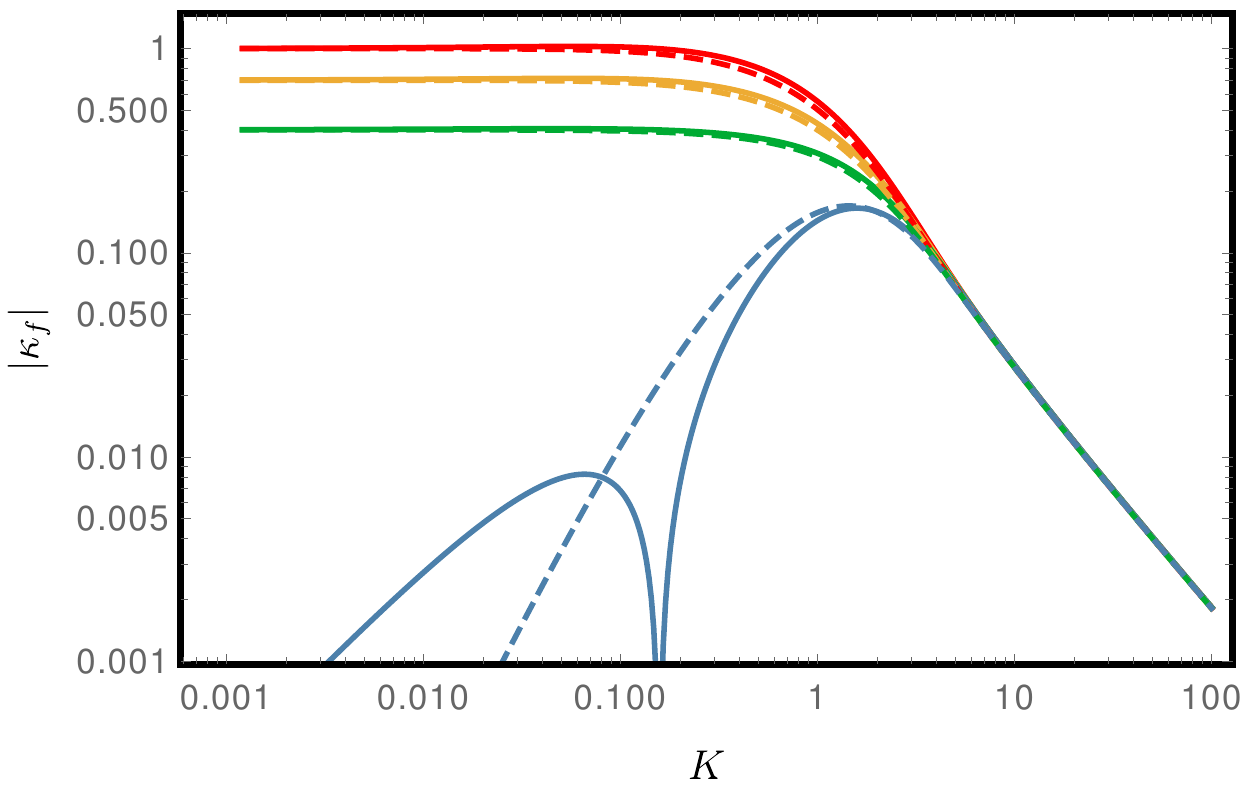}
\end{center}
\caption{\label{fig:efficiency}The efficiency factor $\kappa_f$ as a function of the washout strength $K$.
Solid: fully relativistic result, dashed: nonrelativistic approximation. The initial conditions for $z\to 0$ are $Y_{N1}/Y_{N1}^{\rm eq}=1$ (red),  $Y_{N1}/Y_{N1}^{\rm eq}=0.7$ (orange), $Y_{N1}/Y_{N1}^{\rm eq}=0.4$ (green),  $Y_{N1}/Y_{N1}^{\rm eq}=0$ (blue).}
\end{figure}

In Figure~\ref{fig:efficiency}, we show the impact of asymmetries that have been generated at early times in a model without
spectators for different strengths of washout and for different initial
densities of RHNs~\cite{Garbrecht:2019zaa}. We quantify this in terms of the efficiency factor $\kappa_f$, as defined through the relation
\begin{align}
Y_{B-L}(z\to \infty)=-\varepsilon Y_{N1}^{\rm eq}(z=0)\,\kappa_f\,.
\end{align}
Here,  $\varepsilon$ is the decay asymmetry~\eqref{epsilon:simplemodel} in vacuum.
The results are compared with those obtained from the solutions to
Eqs.~(\ref{fluid:eq:leptogenesis}) given the rates~(\ref{rates:exp:bkg}), i.e. the nonrelativistic approximation.
As explained above, $\kappa_f$ now depends on the dynamics of the washout (because lepton-number violation is relatively suppressed compared to the other reaction rates for
relativistic RHNs) as well as on the temperature dependence of the decay asymmetry.
In the strong-washout regime, where $K\gtrsim1$, the inclusion of the early-time asymmetries
is of negligible relevance (just as the dependence on different initial densities for
the RHNs is). The picture changes however in the weak-washout regime, $K\lsim 1$.
For nonvanishing initial abundance of RHNs, these particles become overabundant $(Y_{N1}(z)>Y_{N1}^{\rm eq}(z)$)
as the temperature drops below their mass. Due to the enhanced decay asymmetry
at high temperature, the early time asymmetries add to those produced at late times,
thus enhancing the overall efficiency factor. In contrast, for vanishing initial abundance
of RHNs, these particles are underabundant when the early time-asymmetries are produced,
thus leading in part to a cancellation with the late time asymmetries that are produced when the RHNs are overabundant.

\begin{figure}
\begin{center}
\includegraphics[width=7.8cm]{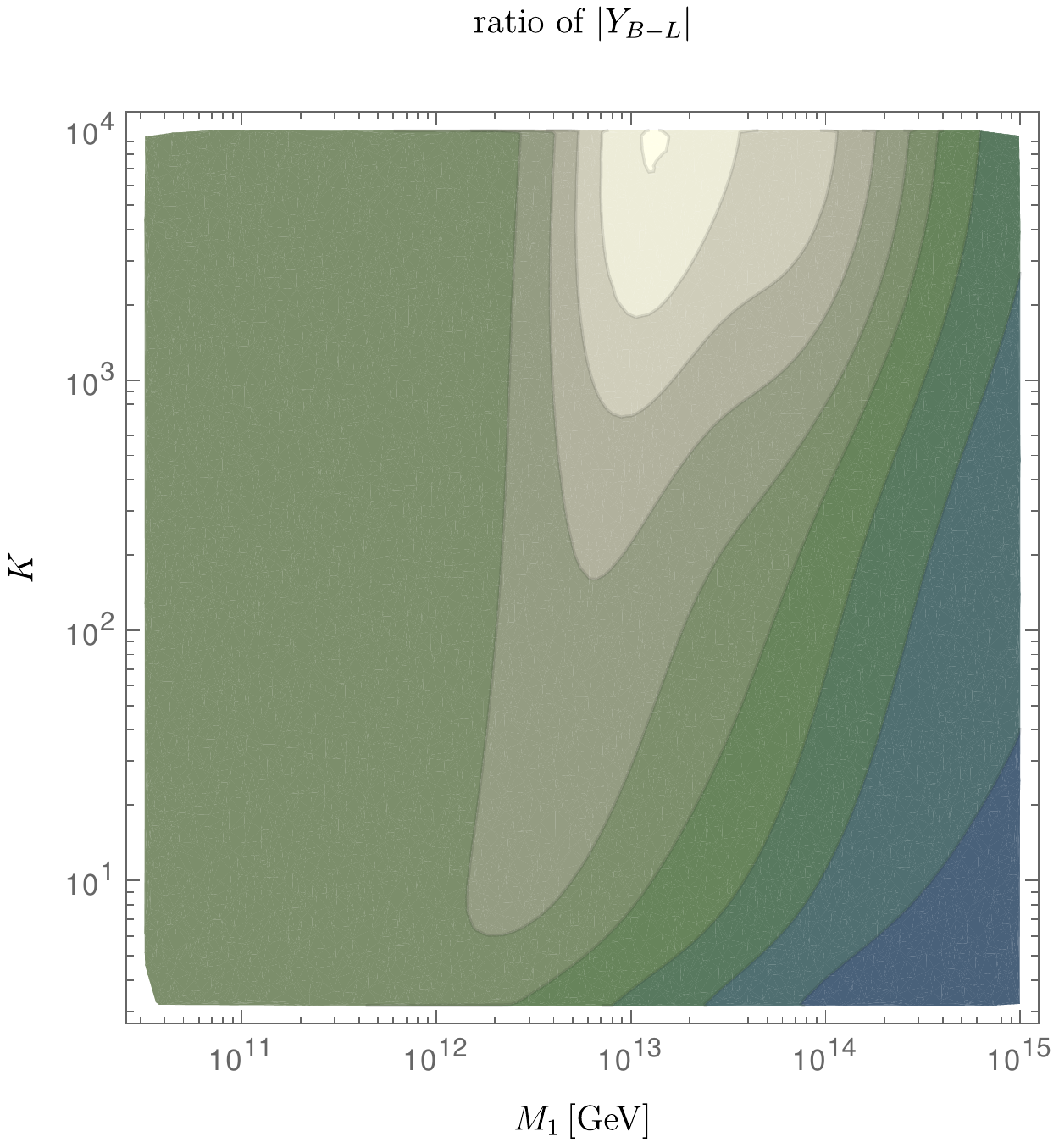}
\includegraphics[width=1cm]{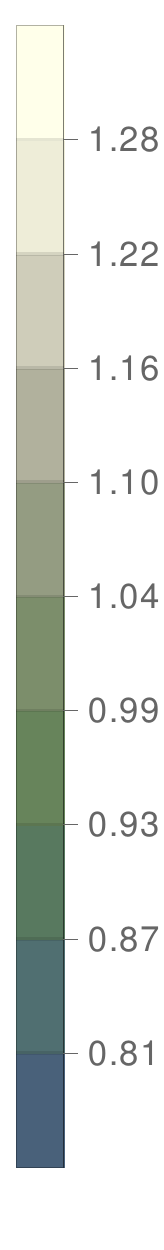}
\includegraphics[width=7.8cm]{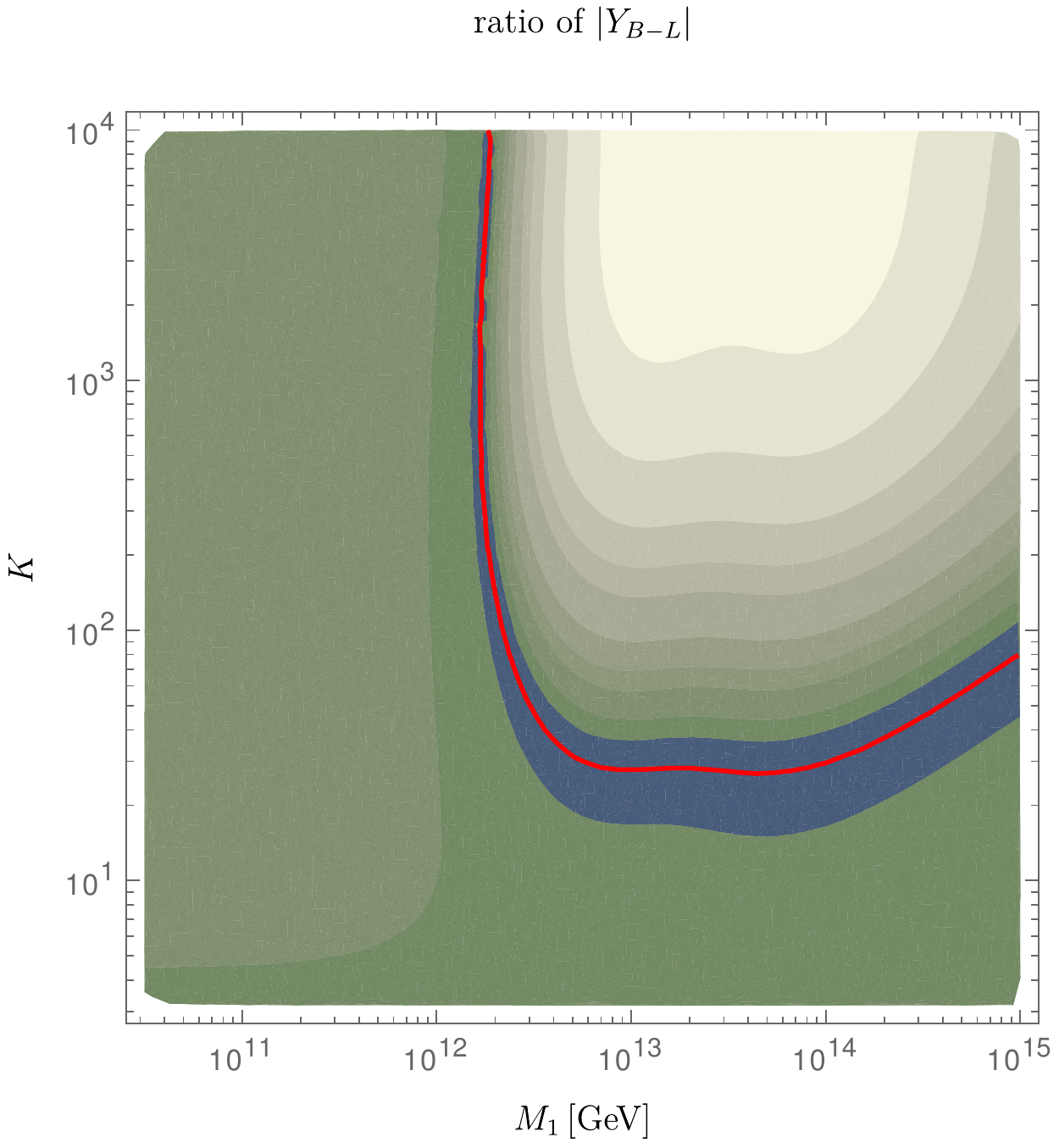}
\includegraphics[width=1.13cm]{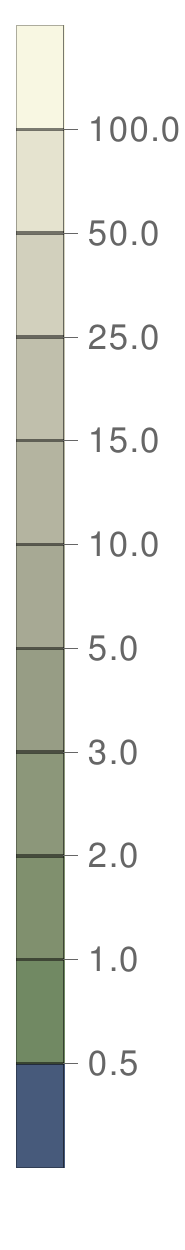}
\end{center}
\caption{\label{fig:spectators}
Ratio of the final asymmetries for partially equilibrated spectators (through bottom-Yukawa and weak sphaleron interactions) and for fully equilibrated spectators. Left panel: thermal initial abundance of RHNs, right panel: vanishing initial abundance. The red line
indicates the sign change in the asymmetry.
}
\end{figure}

The effect of spectators for leptogenesis at temperatures around $10^{13} {\rm GeV}$ is shown in Figure~\ref{fig:spectators}, cf. Refs.~\cite{Garbrecht:2014kda,Garbrecht:2019zaa} for further details.
Using the fully relativistic rates, we compare the final asymmetries
accounting for the partial equilibration of the bottom-Yukawa and weak sphaleron
interactions and compare these with the results obtained when these reactions are
imposed to be fully equilibrated. For small values of $M_1$, the assumption of full equilibration works well, while sizable deviations appear for larger  $M_1$ and in
particular for strong washout. For vanishing initial abundances of RHNs, the initial deviation from equilibrium is large, such that substantial early-time asymmetries
can be produced that are partly protected from washout within the spectator fields.
Since these early-time asymmetries originate from underabundant RHNs, they are opposite in
sign to the late-time asymmetries, such that the overall sign of the asymmetry
can change throughout the parameter space.
It may be of interest to investigate in the future whether
the upper bound on the light neutrino masses~\cite{Buchmuller:2002rq,Buchmuller:2002jk,Buchmuller:2003gz} mentioned in Section~\ref{sec:RIS} is affected by such effects.

\section{Baryogenesis at phase boundaries}
\label{sec:ewbg}

While it remains plausible that the deviation from equilibrium necessary for baryogenesis
is due to the initial conditions that may have been physically established e.g.
at the end of inflation, it may be appealing to consider that all relevant degrees of freedom have effectively been in thermal equilibrium at some point before baryogenesis.
Then, the asymmetry would depend only on the particle physics model,
as expressed through its Lagrangian, embedded in a hot big bang scenario and not additionally on the dynamics of the very early Universe. (Nonetheless, certain nonequilibrium initial conditions such as vanishing initial densities in leptogenesis from oscillations of light RHNs~\cite{Akhmedov:1998qx,Asaka:2005pn} may appear generic enough to directly connect the resulting asymmetry with the parameters of the particle physics model.)
In the strong-washout regime of baryogenesis from decays or inverse decays, the system is kept close to equilibrium at early times due to fast reaction rates and the expansion of the Universe eventually drives the decaying species out of equilibrium.
Another possibility for cosmology to create nonequilibrium environments for initial states that have been very close to equilibrium initially are phase boundaries.
These may be present e.g. in the form of domain walls or bubble walls in first-order
phase transitions (i.e. where bubbles containing the true ground state nucleate and expand
into the phase of the false ground state). Across such boundaries, particle masses may
change, leading to a deflection of their trajectories already at the classical level.
In combination with quantum interference, this may lead to $CP$-violating currents. In this section,
we discuss the computation of such currents using CTP methods. A detailed review on these
matters is provided by Ref.~\cite{Konstandin:2013caa}.

Before discussing these technicalities, we very briefly explain the basic picture
of electroweak baryogenesis that is covered in more detail in the review articles~\cite{Trodden:1998ym,Morrissey:2012db,Konstandin:2013caa,White:2016nbo}. The possibility of baryon number generation during the electroweak
phase transition was first proposed in Ref.~\cite{Kuzmin:1985mm}. Early work
on the presently
favoured scenario where axial asymmetries generated in the wall of the first order
phase transition diffuse ahead of it, where they get converted into baryons by sphaleron
processes is reported in Ref.~\cite{Cohen:1994ss}, while the remaining papers on electroweak baryogenesis referred to in this section further develop this mechanism
in numerous details. According to these works, electroweak baryogenesis proceeds
as follows (cf. Figure~\ref{fig:ewbg}):
\begin{figure}[t]
\begin{center}
\includegraphics[width=9cm]{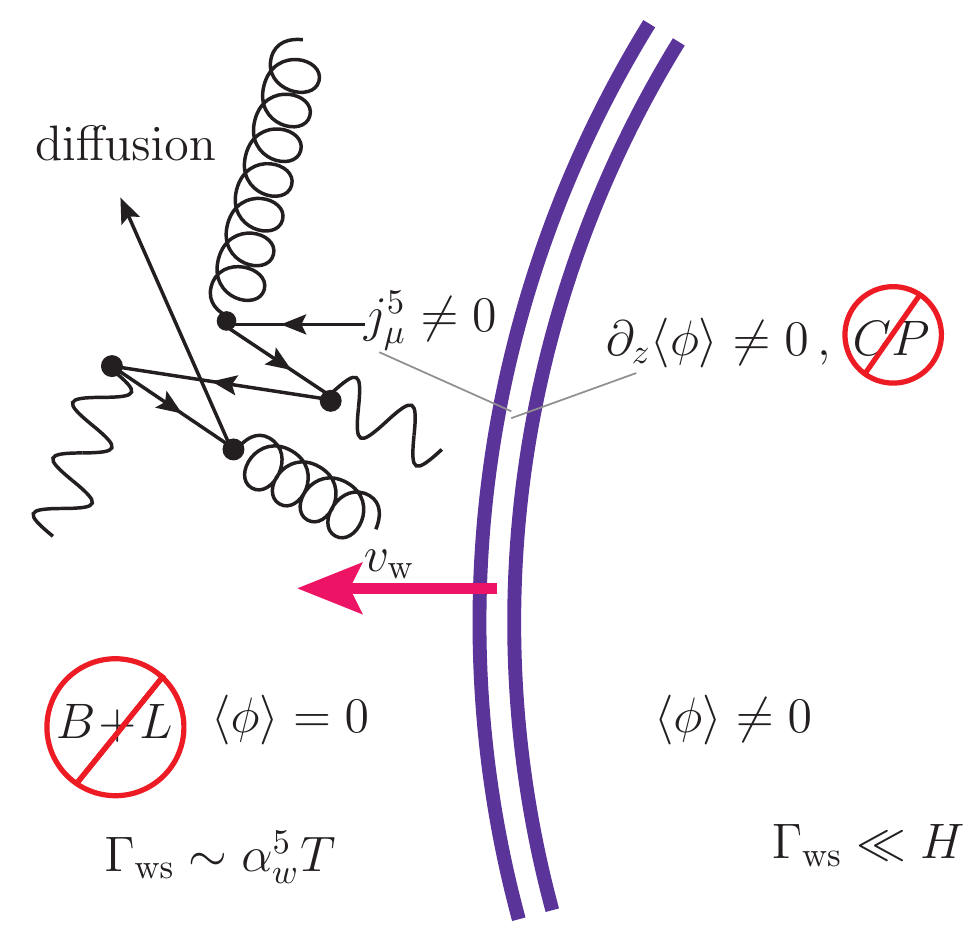}
\end{center}
\caption{\label{fig:ewbg}Schematic picture of electroweak baryogenesis.
During a first order phase transition, bubbles with Higgs-field expectation values
$\langle \phi \rangle\not=0$ (broken phase) expand into the symmetric phase, where
$\langle \phi \rangle=0$, with a wall-velocity $v_{\rm w}$. In conjunction with $CP$-violating interactions, the gradients of the Higgs expectation value $\partial_z\langle \phi \rangle$ inside the bubble wall induce a force term which, for
fermion fields, leads to a chiral current $j_\mu^5$. This leads to chiral charge densities that are distributed primarily via scatterings with gauge bosons, which can be effectively described by diffusion, ahead of the
wall. There, weak
sphalerons turn the chiral charges into baryon-plus-lepton number $B+L$ at
the rate $\Gamma_{\rm ws}$. For a so-called strong first order phase transition, inside the
wall, $\Gamma_{\rm ws}\ll H$, such that the baryon number is frozen in.}
\end{figure} 
\begin{itemize}
\item
Electroweak symmetry breaking occurs through a first order phase transition.
This does not happen in the SM, where it would require the mass of the
Higgs boson to be below $70\,{\rm GeV}$~\cite{Kajantie:1996mn,Rummukainen:1998as}. One therefore has to resort to
extensions of the SM, that can be probed by collider experiments at the electroweak scale.
\item
In the bubble wall, the gradients of the scalar field expectation values
in the Higgs sector generate $CP$-violating currents as we discuss in Section~\ref{sec:fermiontransport}.
\item
The particle flows thus generated undergo rescatterings such that their motion effectively is diffusive. Of particular importance is the diffusion
ahead of the bubble wall into the symmetric phase,
where baryon-number violating processes are
occurring. Moreover, the charges get transferred into
other particles, most importantly the left-handed fermions of the SM, by
scatterings mediated e.g. by Yukawa couplings and strong sphaleron (thermal
QCD instantons) processes. The left-handed charge is then transferred
by anomalous processes, so-called weak sphalerons, into baryon-plus-lepton number.
We review these matters briefly in Section~\ref{sec:CPconservingprocesses}.
\item
Eventually, the bubble containing the broken electroweak phase
captures the baryon-plus-lepton charge that is present in its wake. In the
broken phase, sphaleron processes freeze out (provided the phase transition
is strong enough) such that the captured baryon charge can possibly explain
the matter--antimatter asymmetry.
\end{itemize}

\subsection{Classical forces}

The action of a classical point particle with spacetime-dependent mass $m(x)$ is given by
\begin{align}
S=\int\limits_{\tau_A}^{\tau_B}d\tau \left[-m(x) c^2\right]\,,
\end{align}
where $\tau\in[\tau_A,\tau_B]$ is proper time and where we
have reintroduced  the speed of light $c$ for the time being. Imposing stationarity
under variations with respect to $\delta x^\mu(\tau)$ (and minding that
$d \tau^2=dx_\mu dx^\mu/c^2$) leads to the equation
\begin{align}
\label{force:class}
\frac{d}{d\tau}p^\mu=c^2 \frac{dm}{dx_\mu}\,,
\end{align}
where $p^\mu=m\,dx^\mu/d\tau$. We can therefore identify
$c^2 dm/dx_\mu$ with a classical four-force. Alternatively,
we may also derive this results by
\begin{align}
p^\mu \frac{d}{d\tau} p_\mu=
\frac12\frac{d}{d\tau} p^2=\frac12\frac{d}{d\tau} m^2 c^2=m c^2\frac{d m}{dx^\mu}\frac{d x^\mu}{d\tau}=c^2 p^\mu \frac{d m}{dx^\mu}\,.
\end{align}

At the level of kinetic theory, we recover the classical force when considering a distribution
function $g(x,p)$ and applying the Liouville theorem,
\begin{align}
\label{Liouville:relativistic}
\frac{d}{d\tau}g(x,p)=\frac{1}{m}p^\mu\frac{\partial g(x,p)}{\partial x^\mu}+\frac{d p^\mu}{d\tau}\frac{\partial g(x,p)}{\partial p^\mu}=0\,,
\end{align}
where $p^\mu=m dx^\mu/d\tau$.
Substituting Eq.~(\ref{force:class}), we obtain the driving force that the mass gradients apply to
the distributions. The latter are often of the quasi-particle form, i.e. they approximately
fulfill the on-shell relation
\begin{align}
g(x,p)=2\pi\delta(p^2-m^2)f(x,\mathbf p)\,.
\end{align}
Kinetic equations in the standard form are then obtained when taking the zeroth moment of Eq.~(\ref{Liouville:relativistic})
by integration over $2 \,dp^0/(2\pi)$, such that
\begin{align}
\label{Liouville:relativistic:2}
\frac{1}{p^0}\left(u^\mu \frac{\partial f(x,\mathbf p)}{\partial x^\mu}+\frac{d p^\mu}{d \tau}\frac{\partial f(x,\mathbf p)}{\partial p^\mu}\right) =0\,,
\end{align}
where $u^\mu=dx^\mu/d\tau$. The term in brackets is the same as the left-hand side of the Boltzmann
equation~(\ref{Eq:Boltzmann}).

\subsection{Transport equations on the CTP}
\label{sec:transport:scalar:CTP}

We next aim to work out how the force term emerges in a calculation
based on the CTP approach. For simplicity, we first consider scalar particles.
Proceeding in parallel to the fermionic case discussed in Section~\ref{sec:CTP},
we take the $>\equiv-+$ and $<\equiv+-$ components of the Schwinger--Dyson equation~(\ref{SDE:scalar})
for scalar fields and obtain~\cite{Prokopec:2003pj,Prokopec:2004ic} (setting $c=1$ again)
\begin{align}
\left[p^2-\frac14 \partial^2+{\rm i}p\cdot \partial -m^2 {\rm e}^{-\frac{\rm i}{2}\diamond}\right]\Delta^{<,>}-\Pi^{\rm H} {\rm e}^{-\frac{\rm i}{2}\diamond}\Delta^{<,>}
-\Pi^{<,>} {\rm e}^{-\frac{\rm i}{2}\diamond}\Delta^{\rm H}=\frac12\left(\Pi^>{\rm e}^{-\frac{\rm i}{2}\diamond} \Delta^<-\Pi^<{\rm e}^{-\frac{\rm i}{2}\diamond} \Delta^>\right)\,.
\end{align}
The Hermitian part of this yields the kinetic equation
\begin{align}
p^\mu\partial_\mu{\rm i}\Delta^{<,>} +(m^2+\Pi^{\rm H})\sin\diamond\,\Delta^{<,>}
+\Pi^{<,>}\sin\diamond\,\Delta^{\rm H}=\frac{1}{2}\left({\rm i}\Pi^>\cos\diamond\,{\rm i}\Delta^<
-{\rm i}\Pi^<\cos\diamond\,{\rm i}\Delta^>\right)
\end{align}
and the Antihermitian part the constraint equation
\begin{align}
\label{constraints:scalar}
\left[p^2-\frac14 \partial^2 -(m^2+\Pi^{\rm H})\cos\diamond\right]\Delta^{<,>}-\Pi^{<,>}\cos\diamond\,\Delta^{\rm H}=-\frac{\rm i}{2}\left({\rm i}\Pi^>\sin\diamond\,{\rm i}\Delta^<
-{\rm i}\Pi^<\sin\diamond\,{\rm i}\Delta^>\right)\,.
\end{align}

Next, for the similar reasons as given for the fermions in Section~\ref{sec:kineq:firstprinciples}, we
neglect the terms involving $\Pi^{\rm H}$ and $\Delta^{\rm H}$ in the kinetic equation and expand
it to first order in gradients, which yields
\begin{align}
p^\mu\partial_\mu{\rm i}\Delta^{<,>} +\frac12\frac{\partial m^2}{\partial x^\mu}\frac{\partial {\rm i}\Delta^{<,>}}{\partial p_\mu}=\frac{1}{2}\left({\rm i}\Pi^>{\rm i}\Delta^<
-{\rm i}\Pi^<{\rm i}\Delta^>\right)\,.
\end{align}
Comparing the left-hand side of this CTP result with Eq.~(\ref{Liouville:relativistic})
from classical kinetic theory, we can identify
$\partial m/\partial x_\mu$ with the four-force $d p^\mu/d\tau$.

The constraint equation~(\ref{constraints:scalar}), when truncated at zeroth order in gradients, is consistent with the
tree-level solutions~(\ref{prop:phi:expl}). Substituting these into the kinetic equations and taking the zeroth moment therefore leads to
\begin{align}
\label{transport:scalar}
\frac{1}{2 p^0}\left(p^\mu\partial_\mu f(x,\mathbf p)+m\frac{\partial m}{\partial x^\mu}\frac{\partial f(x,\mathbf p)}{\partial p_\mu}\right)
=-\frac{1}{2}\int\limits_0^\infty\frac{dp^0}{2\pi}\left({\rm i}\Pi^>{\rm i}\Delta^<
-{\rm i}\Pi^<{\rm i}\Delta^>\right)\,,
\end{align}
where the left-hand side agrees up to a prefactor with the result~(\ref{Liouville:relativistic:2}) from classical
kinetic theory.

\subsection{Quantum transport of fermions}
\label{sec:fermiontransport}

For the time-dependent case of leptogenesis, we have simplified in Section~\ref{sec:reslg}
the spinor structure
making use of nonrelativistic approximations for the RHNs. Here, we drop
the nonrelativistic approximation and instead make use of the
symmetry of the problem parallel to the boundary, that is approximated as planar. This reduces the
spinorial problem to eight real degrees of freedom (from in general sixteen complex ones
for a four-by-four matrix in spinor space).

Further, the calculation in above Section~\ref{sec:transport:scalar:CTP} for scalar fields
is truncated at first order in gradients and does not account for flavour mixing in the presence
of several species. Mixing particles often observe a global ${\rm U}(1)$ symmetry, such that
$CP$ violation may only be present
in terms of flavoured asymmetries with cancelling contributions to the total
${\rm U}(1)$ charge. For fermions, the Dirac mass terms violate chiral symmetry such that $CP$ violation can manifest
itself even in single-flavour systems in terms of an axial asymmetry. We therefore consider
in the following systems of mixing fermion flavours that exhibit both types of asymmetry.

Methodically, we follow here the developments on the spinor decomposition
and gradient expansion in
Refs.~\cite{Joyce:1999uf,Kainulainen:2001cn,Prokopec:2003pj,Prokopec:2004ic,Konstandin:2004gy,Konstandin:2005cd} on electroweak baryogenesis. These have been adapted to leptogenesis in Ref.~\cite{Garbrecht:2011aw}, thus leading to a relativistic generalization of the
results in Section~\ref{sec:reslg}.
Starting point are the Kadanoff--Baym equations for fermions in Wigner space~(\ref{KBE:Wigner:Fermions})
that we now aim to decompose within the background of the phase boundary into
kinetic and constraint equations. It should be noted that many phenomenological papers
compute the source term using the methods of Refs.~\cite{Riotto:1998zb,Lee:2004we},
that however do not rely on a detailed spinor decomposition. Also, the gradients are
not expanded from the full spacetime dependent mass terms
but are inserted perturbatively into the collision term.
As a consequence, an apparent difference is that the sources computed in Refs.~\cite{Joyce:1999uf,Kainulainen:2001cn,Prokopec:2003pj,Prokopec:2004ic,Konstandin:2004gy,Konstandin:2005cd} do not rely on a collision term, while the results from Refs.~\cite{Riotto:1998zb,Lee:2004we}
crucially depend on the finite width of the mixing particles.

In order to facilitate the decomposition the fermionic Kadanoff--Baym equations, we first
make use of the symmetry of the problem. Without loss of generality, the bubble wall
is assumed to propagate in $z$-direction, that we identify with the three-component
of the position vector. (That is to say we write $z=x^3$ and $k^z=k^3$ but maintain
writing $\gamma^3$ for notational appeal.) Angular momentum pointing in that direction is therefore conserved,
such that for fermions, the $z$-component of the spin operator
\begin{align}
S^z=\frac{1}{\tilde k^0}\left(\gamma^0 k^0 -\gamma^1 k^1 -\gamma^2 k^2\right)
\end{align}
where
\begin{align}
\tilde k^0={\rm sign} k^0\sqrt{(k^0)^2 -(k^1)^2-(k^2)^2}
\end{align}
is a good quantum number. The Wightman function can hence be decomposed as
\begin{align}
{\rm i} S^{<,>}=\sum\limits_{s=\pm 1}{\rm i}S_s^{<,>}\,,
\end{align}
where
\begin{align}
\label{spin:decomposition}
{\rm i} S_s^{<,>}=-P_s\left[s\gamma^3\gamma^5 g_0^{s<,>}-s\gamma^3 g_3^{s<,>}+\mathbbm 1 g_1^{s<,>}-{\rm i} \gamma^5 g_2^{s<,>}\right]\,,
\end{align}
with the spin projector
\begin{align}
P_s=\frac 12(\mathbbm 1 + s S^z)\,.
\end{align}
The spinor structures in Eq.~(\ref{spin:decomposition}) are chosen such as to commute with the spin operator. 

As commented in Section~\ref{sec:CTP}, it is of interest to extract currents from the
fermionic Wightman functions. These currents can readily be expressed in terms of the
decomposition~(\ref{spin:decomposition}):
\begin{align}
\label{currents:densities:speven}
g_0^++g_0^-=&-\frac12\sum_{s=\pm}{\rm tr}\left[s \gamma^3\gamma^5{\rm i} S_s^{<,>}\right]\,,\quad\textnormal{charge density,
}\notag\\
g_3^++g_3^-=&-\frac12\sum_{s=\pm}{\rm tr}\left[s \gamma^3{\rm i} S_s^{<,>}\right]\,,\quad\textnormal{axial charge density,
}\notag\\
g_1^++g_1^-=&-\frac12\sum_{s=\pm}{\rm tr}\left[ {\rm i} S_s^{<,>}\right]\,,\quad\textnormal{scalar density,
}\notag\\
g_2^++g_2^-=&-\frac12\sum_{s=\pm}{\rm tr}\left[{\rm i} \gamma^5 {\rm i} S_s^{<,>}\right]\,,\quad\textnormal{pseudoscalar density,
}
\end{align}
and
\begin{align}
\label{currents:densities:spodd}
g_0^+-g_0^-=&-\frac12\sum_{s=\pm}{\rm tr}\left[\gamma^3\gamma^5{\rm i} S_s^{<,>}\right]\,,\quad\textnormal{axial current density,
}\notag\\
g_3^+-g_3^-=&-\frac12\sum_{s=\pm}{\rm tr}\left[\gamma^3{\rm i} S_s^{<,>}\right]\,,\quad\textnormal{current density,
}\notag\\
g_1^+-g_1^-=&-\frac12\sum_{s=\pm}{\rm tr}\left[s {\rm i} S_s^{<,>}\right]\,,\quad\textnormal{spin density, 
}\notag\\
g_2^+-g_2^-=&-\frac12\sum_{s=\pm}{\rm tr}\left[s {\rm i} \gamma^5 {\rm i} S_s^{<,>}\right]\,,\quad\textnormal{axial spin density.
}
\end{align}

It is further useful to define
\begin{align}
\hat k^0=\tilde k^0 -\frac{\rm i}{2}\frac{k^0\partial_t+\mathbf k_\parallel\cdot\nabla_\parallel}{\tilde k^0}\,,\qquad \hat k^z=k^z-\frac{\rm i}{2}\frac{\partial}{\partial z}\,,\qquad
\mathbf{k}_\parallel\cdot\nabla_\parallel=k^1\partial_1+k^2\partial_2
\,.
\end{align}
Multiplication of the Kadanoff--Baym equation~(\ref{KBE:Wigner:Fermions}) by
\begin{align}
\frac12\{\mathbbm 1,s\gamma^3\gamma^5,-{\rm i}s\gamma^3,-\gamma^5\}
\end{align}
and taking the trace, one obtains
\begin{subequations}
\label{KBE:fermi:decmop}
\begin{align}
2{\rm i}\hat k^0 g_0^s-2{\rm i}s\hat k^z g_3^s- 2{\rm i} M^{\rm H} {\rm e}^{-\frac{\rm i}{2} \overleftarrow\partial\cdot\overrightarrow\partial_{\!k}} g_1^s-2{\rm i}M^{\rm A}{\rm e}^{-\frac{\rm i}{2} \overleftarrow\partial\cdot\overrightarrow\partial_{\!k}} g_2^s=&0\,,\\
2{\rm i}\hat k^0 g_1^s-2 s\hat k^z g_2^s- 2{\rm i} M^{\rm H} {\rm e}^{-\frac{\rm i}{2} \overleftarrow\partial\cdot\overrightarrow\partial_{\!k}} g_0^s+2 M^{\rm A}{\rm e}^{-\frac{\rm i}{2} \overleftarrow\partial\cdot\overrightarrow\partial_{\!k}} g_3^s=&0\,,\\
2{\rm i}\hat k^0 g_2^s+2s\hat k^z g_1^s- 2 M^H {\rm e}^{-\frac{\rm i}{2} \overleftarrow\partial\cdot\overrightarrow\partial_{\!k}} g_3^s-2{\rm i}M^{\rm A}{\rm e}^{-\frac{\rm i}{2} \overleftarrow\partial\cdot\overrightarrow\partial_{\!k}} g_0^s=&0\,,\\
2{\rm i}\hat k^0 g_3 ^s-2{\rm i}s\hat k^z g_0^s+ 2 M^{\rm H} {\rm e}^{-\frac{\rm i}{2} \overleftarrow\partial\cdot\overrightarrow\partial_{\!k}} g_2^s-2 M^{\rm A}{\rm e}^{-\frac{\rm i}{2} \overleftarrow\partial\cdot\overrightarrow\partial_{\!k}} g_1^s=&0\,.
\end{align}
\end{subequations}
Here, we have set the collision term in Eq.~(\ref{KBE:Wigner:Fermions}) to zero. We will comment
on this approximation in relation to other approaches pursued in phenomenological calculations
toward the end of this section.

Before proceeding to decompose into kinetic and constraint equations, we change to working with
quantities that have definite properties under transformations of the mass matrix $M$.
Let $M_d=U M V^\dagger$ be diagonal (such that both $U M M^\dagger U^\dagger$ and $V M^\dagger M V^\dagger$ are diagonal as well) and define
\begin{subequations}
\begin{align}
X=&P_{\rm L}\otimes V + P_{\rm R}\otimes U=\frac12\left[\mathbbm 1\otimes(V+U)-\gamma^5\otimes(V-U)\right]\,,\\
Y=&P_{\rm L}\otimes U + P_{\rm R}\otimes V=\frac12\left[\mathbbm 1\otimes(V+U)+\gamma^5\otimes(V-U)\right]\,.
\end{align}
\end{subequations}
Then, $\widehat M_d=X \widehat M Y^\dagger$ is diagonal, where we recall the definitions~(\ref{M:decomp})  and~(\ref{M:spinor}). When transforming the Dirac operator accordingly, such
as to obtain a diagonal mass, the Wigner function transforms as
\begin{align}
S_d=Y S X^\dagger\,.
\end{align}
As a consequence, the functions $g^s_i$ transform in a definite manner when arranged in
chirality-blocks
\begin{align}
\label{chirality:blocks}
g^s_{\rm L,R}=g^s_0\mp g^s_3\,,\; g_N=g^s_1+{\rm i}g^s_2\,,\;g_N^\dagger=g^s_1-{\rm i}g^s_2\,,  
\end{align}
such that these transform to the mass-diagonal basis as
\begin{align}
g_{\rm R}^d=V g_{\rm R} V^\dagger\,,\;g_{\rm L}^d=U g_{\rm L} U^\dagger,\;
g_N^d=U g_N V^\dagger\,,\; g_N^{d\dagger}= V g_N^\dagger U^\dagger\,.
\end{align}
Taking the according linear combinations of Eqs.~(\ref{KBE:fermi:decmop}), the
Kadanoff--Baym equations then read
\begin{subequations}
\begin{align}
{\rm i}\left(\hat k^0 g^s_{\rm R}-s \hat k^z g^s_{\rm R} -M^\dagger {\rm e}^{-\frac{\rm i}{2} \overleftarrow\partial\cdot\overrightarrow\partial_{\!k}} g^s_N\right)&=0\,,\\
{\rm i}\left(\hat k^0 g^s_{\rm L}+s \hat k^z g^s_{\rm L} -M {\rm e}^{-\frac{\rm i}{2} \overleftarrow\partial\cdot\overrightarrow\partial_{\!k}} {g^s_N}^\dagger\right)&=0\,,\\
{\rm i}\left(\hat k^0 g^s_N+s \hat k^z g^s_N -M {\rm e}^{-\frac{\rm i}{2} \overleftarrow\partial\cdot\overrightarrow\partial_{\!k}} g^s_{\rm R}\right)&=0\,,\\
{\rm i}\left(\hat k^0 {g^s_N}^\dagger-s \hat k^z {g^s_N}^\dagger -M^\dagger {\rm e}^{-\frac{\rm i}{2} \overleftarrow\partial\cdot\overrightarrow\partial_{\!k}} g^s_{\rm L}\right)&=0\,.
\end{align}
\end{subequations}
Taking the Hermitian part leads to the kinetic equations
\begin{subequations}
\begin{align}
\label{eq:gR}
\frac{k^0\partial_t+\mathbf k_\parallel\cdot\nabla_\parallel}{\tilde k^0} g^s_{\rm R}
-s \frac{\partial}{\partial z} g^s_{\rm R} - {\rm i} M^\dagger {\rm e}^{-\frac{\rm i}{2} \overleftarrow\partial\cdot\overrightarrow\partial_{\!k}} g^s_N+{\rm i} {g^s_N}^\dagger{\rm e}^{\frac{\rm i}{2} \overleftarrow\partial_{\!k}\cdot\overrightarrow\partial} M=&0\,,\\
\label{eq:gL}
\frac{k^0\partial_t+\mathbf k_\parallel\cdot\nabla_\parallel}{\tilde k^0} g^s_{\rm L}
-s \frac{\partial}{\partial z} g^s_{\rm L} - {\rm i} M {\rm e}^{-\frac{\rm i}{2} \overleftarrow\partial\cdot\overrightarrow\partial_{\!k}} {g^s_N}^\dagger+{\rm i} {g^s_N} {\rm e}^{\frac{\rm i}{2} \overleftarrow\partial_{\!k}\cdot\overrightarrow\partial} M^\dagger=&0\,,\\
\label{eq:gN}
\frac{k^0\partial_t+\mathbf k_\parallel\cdot\nabla_\parallel}{\tilde k^0} g^s_N
+2{\rm i} s k^z g^s_N - {\rm i} M {\rm e}^{-\frac{\rm i}{2} \overleftarrow\partial\cdot\overrightarrow\partial_{\!k}} g^s_{\rm R}+{\rm i} g^s_{\rm L}{\rm e}^{\frac{\rm i}{2} \overleftarrow\partial_{\!k}\cdot\overrightarrow\partial} M=&0\,,\\
\label{eq:gNdagger}
\frac{k^0\partial_t+\mathbf k_\parallel\cdot\nabla_\parallel}{\tilde k^0} {g^s_N}^\dagger
-2{\rm i} s k^z {g^s_N}^\dagger - {\rm i} M^\dagger {\rm e}^{-\frac{\rm i}{2} \overleftarrow\partial\cdot\overrightarrow\partial_{\!k}} g^s_{\rm L}+{\rm i} g^s_{\rm R} {\rm e}^{\frac{\rm i}{2} \overleftarrow\partial_{\!k}\cdot\overrightarrow\partial} M^\dagger=&0\,.
\end{align}
\end{subequations}

Next, we move to a frame that is comoving with the phase boundary such that we
can replace $\partial_t\to 0$ and make use of the isotropy parallel to the
wall implying $\mathbf k_\parallel\cdot \nabla\to 0$. From Eqs.~(\ref{eq:gN},\ref{eq:gNdagger}),
it then follows that
\begin{align}
g_N^s=&\frac{1}{2 s k^z}\left(M {\rm e}^{-\frac{\rm i}{2} \overleftarrow\partial\cdot\overrightarrow\partial_{\!k}} g^s_{\rm R} -g^s_{\rm L}{\rm e}^{\frac{\rm i}{2} \overleftarrow\partial_{\!k}\cdot\overrightarrow\partial} M\right)\,,
\end{align}
which, substituted in Eqs.~(\ref{eq:gR}) and~(\ref{eq:gL}) leads to
\begin{subequations}
\label{kineq:LR}
\begin{align}
\frac{\partial}{\partial z} g_{\rm L}^s -\frac{\rm i}{2} M {\rm e}^{-\frac{\rm i}{2} \overleftarrow\partial\cdot\overrightarrow\partial_{\!k}} \frac{1}{k^z}\left(g^s_{\rm R}  {\rm e}^{\frac{\rm i}{2} \overleftarrow\partial_{\!k}\cdot\overrightarrow\partial} M^\dagger -M^\dagger {\rm e}^{-\frac{\rm i}{2} \overleftarrow\partial\cdot\overrightarrow\partial_{\!k}}g^s_{\rm L}\right)&\notag\\
+\frac{\rm i}{2}\frac{1}{k^z}\left(M {\rm e}^{-\frac{\rm i}{2} \overleftarrow\partial\cdot\overrightarrow\partial_{\!k}} g^s_{\rm R}-g_{\rm L}^s{\rm e}^{\frac{\rm i}{2} \overleftarrow\partial_{\!k}\cdot\overrightarrow\partial} M\right){\rm e}^{\frac{\rm i}{2} \overleftarrow\partial_{\!k}\cdot\overrightarrow\partial}M^\dagger&=0\,,\\
\frac{\partial}{\partial z} g_{\rm R}^s +\frac{\rm i}{2} M^\dagger {\rm e}^{-\frac{\rm i}{2} \overleftarrow\partial\cdot\overrightarrow\partial_{\!k}} \frac{1}{k^z} \left(M {\rm e}^{-\frac{\rm i}{2} \overleftarrow\partial\cdot\overrightarrow\partial_{\!k}} g^s_{\rm R}-g_{\rm L}^s{\rm e}^{\frac{\rm i}{2} \overleftarrow\partial_{\!k}\cdot\overrightarrow\partial} M\right)&\notag\\
-\frac{\rm i}{2}\frac{1}{k^z} \left(g^s_{\rm R}  {\rm e}^{\frac{\rm i}{2} \overleftarrow\partial_{\!k}\cdot\overrightarrow\partial} M^\dagger -M^\dagger {\rm e}^{-\frac{\rm i}{2} \overleftarrow\partial\cdot\overrightarrow\partial_{\!k}}g^s_{\rm L}\right){\rm e}^{\frac{\rm i}{2} \overleftarrow\partial_{\!k}\cdot\overrightarrow\partial}M&=0\,.
\end{align}
\end{subequations}
These equations are symmetric under ${\rm R}\leftrightarrow {\rm L}$ and $M\leftrightarrow M^\dagger$. Expanding
to second order in gradients (i.e. expanding the exponentiated operator up to second order), we therefore only quote the equation for $g_{\rm L}$:
\begin{align}
&k^z\frac{\partial}{\partial z}g_{\rm L}^s+\underbrace{\frac{\rm i}{2}\left[MM^\dagger,g^s_{\rm L}\right]}_{\textnormal{mixing term}}\notag\\
&\hskip-.4cm\underbrace{-\frac14 \left\{\left(M M^\dagger\right)^\prime,\partial_{k^z} g^s_{\rm L}\right\}}_{\textnormal{classical force}}
\underbrace{-\frac{1}{4 k^z}\left(M^\prime g^s_{\rm R} M^\dagger+ M g^s_{\rm R} M^{\prime\dagger}\right)
+\frac{1}{4 k^z}\left(M^\prime M^{\dagger}g^s_{\rm L}+g^s_{\rm L} M M^{\prime\dagger}\right)}_{\textnormal{gradient-mixing terms}}\notag\\
+&\underbrace{\frac{\rm i}{8}\left(M^{\prime\prime} M^\dagger \partial_{k^3}\frac{g^s_{\rm L}}{k^z}- \partial_{k^z}\frac{g^s_{\rm L}}{k^z} M M^{\prime\prime\dagger}\right)
-\frac{\rm i}{8}\left(M^{\prime\prime} \partial_{k^z} \frac{g^s_{\rm R}}{k^z} M^\dagger - M \partial_{k^3} \frac{g^s_{\rm R}}{k^z} M^{\prime\prime\dagger}\right)}_{\textnormal{semiclassical force}}\notag\\
-&\frac{\rm i}{16}\left[\left(M M^\dagger\right)^{\prime\prime},\partial^2_{k^z} g^s_{\rm L}\right]
+\frac{\rm i}{8 k^z}\left[M^\prime M^{\prime\dagger},\partial_{k^z}g^s_{\rm L}\right]=0\,.
\label{eq:transport:gradients}
\end{align}
For reference in the further discussion, we label here some of the particular terms with
the tags under the braces.

\paragraph{Semiclassical force for a single flavour}

Considering a single flavour, it turns out that there is no $CP$-violating source at
first order in derivatives. At second order, the terms involving commutators in Eq.~(\ref{eq:transport:gradients}) vanish, and we are left with the semiclassical force~\cite{Joyce:1994fu,Joyce:1994zn,Joyce:1999uf,Kainulainen:2001cn,Kainulainen:2002th,Konstandin:2004gy}.
The relevant linear combination of the left and right chiral as well as spin-dependent
quantities is the one that leads to the axial current
\begin{align}
{\rm tr}\left[\gamma^3 \gamma^5 {\rm i}S^{<,>}\right]
=2\left(g_0^+-g_0^-\right)={j^5}^z\,.
\end{align}
Taking the corresponding combination from Eq.~(\ref{eq:transport:gradients})
and dropping the terms that do not contribute leads to
\begin{align}
\label{transport:singleflavour}
k^z\frac{\partial}{\partial z}2(g_0^+-g_0^-)-(|m|^2)^\prime\partial_{k^z}(g_0^+-g_0^-)
-\frac12\left(|m|^2\vartheta^\prime\right)^\prime
\partial_{k^z}\frac{g_3^+-g_3^-}{k^z}=0\,,
\end{align}
where we have parametrized $m=|m|\exp({\rm i}\vartheta)$, such that
\begin{align}
\left(m^{\prime\prime} m^* - m {m^*}^{\prime\prime}\right)=2{\rm i}\left(|m|^2\vartheta^\prime\right)^\prime\,.
\end{align}

To further evaluate Eq.~(\ref{transport:singleflavour}), we integrate over $dp^0$ in order arrive at a compact form of the fluid equations. We note that
\begin{subequations}
\begin{align}
\label{g0:moment}
\int\limits_{\-\infty}^{\infty}\frac{dp^0}{2\pi} g_0^s=&\frac12 \left|\frac{\tilde p^0}{p^0}\right|
\left(f^s(\mathbf p)-\bar f^s(\mathbf p)\right)\,,\\
\label{g3:moment}
\int\limits_{\-\infty}^{\infty}\frac{dp^0}{2\pi} g_3^s=&
\frac{s p^3}{2|p^0|}\left(f^s(\mathbf p)+\bar f^s(\mathbf p)\right)\,.
\end{align}
\end{subequations}
We further define
\begin{align}
h^{s}(\mathbf p)=f^s(\mathbf p)-\bar f^s(\mathbf p)
\end{align}
appearing in Eq.~(\ref{g0:moment}) as the charge distribution function for particles with a certain spin orientation perpendicular to the boundary.
Taking the zeroth moment of~Eq.~(\ref{transport:singleflavour}) eventually leads to
\begin{align}
\label{semiclassical:force}
\frac{\tilde k^0}{k^0}\left(k^z\frac{\partial}{\partial z}h^s-\frac12\left(|m|^2\right)^\prime\partial_{k^z}h^s\right)
-\frac{s}{2}\left(|m|^2\vartheta^\prime\right)^\prime\frac{2}{|k^0|}\partial_{k^z}f^s=0\,,
\end{align}
where the second term acts as a semiclassical force. This interpretation follows by comparison
with e.g. Eq.~(\ref{transport:scalar}).
Therefore, in the wall frame, there are spin-dependent charge distributions present, and these are opposite
for the two different spin states. In the plasma frame, this leads to an axial current density, i.e. $k^z\frac{\partial}{\partial z}(h^+ + h^-)\to\partial_\mu j^{5\mu}$ when replacing this term by a Lorentz-covariant form.

\paragraph{Two flavours: resonant enhancement}

For two fermion flavours mixing in the background of the bubble wall,
an axial current can also be generated when neglecting the
semiclassical force term, which can be justified due to the
possible resonant enhancement even in the presence of a moderate
mass degeneracy. This axial current arises from the interplay of the
mixing terms and the classical force. Computing it requires a
numerical solution of Eq.~(\ref{eq:transport:gradients}) or an analytic
approximation.

In order to obtain the latter, we recall that in Section~\ref{sec:reslg} we
found suitable solutions for the system of mixing RHNs
in leptogenesis by neglecting the derivative terms in the kinetic
equations, provided the oscillation frequency or the damping rate exceed
the time-dependence of the system. In the present case of mixing in
the background of a phase boundary, we may use a corresponding approximation
provided that the oscillation length $\sim 1/|m_1-m_2|$ is small compared to
the size of the boundary wall $\ell_{\rm w}$. Besides  the wall width,
also collision terms that are neglected in Eqs.~(\ref{KBE:fermi:decmop}) should
be expected to cap the resonant enhancement, cf. Section~\ref{sec:reslg} on leptogenesis.

Even when applying these simplifications, the result for the axial
current generated from a general space-dependent
mass matrix remains involved. In order to compare with the relevant results
derived for resonant mixing scenarios that have drawn some attention in
the historic context of the Minimally Supersymmetric Standard Model~\cite{Riotto:1998zb,Cline:2000nw,Carena:2002ss,Lee:2004we,Konstandin:2005cd,Cirigliano:2006wh},
we choose a mass matrix similar to what one encounters in that setup,
\begin{align}
M=\left(\begin{array}{cc}m_1 & {\rm e}^{{\rm i}\varphi} v_b \\ v_a & m_2\end{array}\right)
\to \widehat M=\left(\begin{array}{cc}m_1 & v_a P_{\rm L}+{\rm e}^{{\rm i}\varphi}v_b P_{\rm R}\\v_a P_{\rm R} + v_b {\rm e}^{-{\rm i}\varphi} P_{\rm L}& m_2\end{array}\right)
=\left(\begin{array}{cc}m_1 & 0 \\ 0 & m_2 \end{array}\right)+\delta \widehat M
\,.
\end{align}
The spatial dependence resides in $v_{a,b}$ that smoothly change from zero
in the phase of electroweak symmetry across the bubble wall to nonvanishing
values in the broken phase. The masses $m_{1,2}$ are assumed constant.

We then take Eq.~(\ref{eq:transport:gradients}) as well as the corresponding
one obtained via $g^s_{\rm L}\leftrightarrow g^s_{\rm R}$ and $M\leftrightarrow M^\dagger$ and
neglect the derivatives with respect to $z$ acting on $g^s_{\rm L,R}$. As a zeroth order input, we
set ${g_{0,3}^{(0)s}}_{ii}\not=0$ and ${g_{0,3}^{(0)s}}_{ij}=0$ for $i\not=j$, where the relation
with $g^s_{\rm L,R}$ is given by Eq.~(\ref{chirality:blocks}). Then, we compute
the off-diagonal elements ${g_{0,3}^{(1)s}}_{ij}$ for $i\not=j$ as a first perturbation
from substituting these into the mixing term and substituting
${g_{0,3}^{(0)s}}$ into the gradient-mixing and the classical-force terms while neglecting
all other terms appearing in the kinetic equation. 
We eventually substitute $g_{0,3\,ij}^{(1)s}$
into the gradient-mixing and the classical-force terms to obtain from the
diagonal elements the divergence of the
current
\begin{align}
k^z\frac{\partial}{\partial z}g_{0\,11}=&
k^z\frac{\partial}{\partial z}\left(g_{\rm L}+g_{\rm R}\right)_{11}
=-k^z\frac{\partial}{\partial z}\left(g_{\rm L}+g_{\rm R}\right)_{22}\notag\\
=&\frac12\frac{m_1 m_2  v_a^\prime v_b^\prime \sin\varphi}{m_1^2-m_2^2}
\left(
\frac{1}{k^z}\frac{\partial}{\partial k^z} g_{3\,11}-\frac{1}{2 {k^z}^2} (g_{3\,11}+g_{3\,22})
\right)\,.
\label{mixing:source:algebraic}
\end{align}
Upon partial integration in $d k^z$, this can be cast into the more appealing form
\begin{align}
k^z\frac{\partial}{\partial z}g_{0\,11}
=\frac{1}{4{k^z}^2}\frac{m_1 m_2  v_a^\prime v_b^\prime \sin\varphi}{m_1^2-m_2^2}
(g_{3\,11}-g_{3\,22})\,.
\end{align}
When comparing with the semiclassical force~(\ref{semiclassical:force}) for a single
fermion flavour, we observe here two orders in derivatives as well. However,
the denominator $m_1^2-m_2^2$ leads to a resonant enhancement within the validity of
the present approximation, i.e. $|m_1-m_2|$ must no be smaller than $1/\ell_{\rm w}$.
We also note that the resulting total current is axial [cf. Eq.~(\ref{currents:densities:spodd})] because the
equilibrium form of $g_{3ii}^s$ is odd in the spin $s$, cf. Eq.~(\ref{g3:moment}).

Since the present approximation treats the off-diagonal elements of the
mass matrix as small perturbations, we also compare with an apparently related
approach taken
in Refs.~\cite{Riotto:1998zb,Lee:2004we}. There, the Kadanoff--Baym equations are extended to include
a collision term as on the right-hand side of Eq.~(\ref{KBE:Wigner:Fermions}).
The latter is approximated by considering mass insertions as vertices. It turns
out that this does not readily lead to a $CP$-violating source but only in case
the propagators ${\rm i}S^{<,>}$ that appear explicitly and implicitly through $\slashed\Sigma^{<,>}$
are replaced with finite-width expressions that provide extra $CP$-even phases.
As a result, the source for the axial current is of the schematic form
\begin{align}
\label{mixing:source:vev:coll}
\sim\frac{(v_a v_b^\prime - v_b v_a^\prime)(m_1-m_2)(\Gamma_1+\Gamma_2)}{[(m_1-m_2)^2+(\Gamma_1+\Gamma_2)^2]^2}\sin\varphi\,,
\end{align}
where $\Gamma_{1,2}$ are the values for the widths of the mixing particles.
In future work, it may be of interest to investigate whether the finite-width
terms crucial for the source~(\ref{mixing:source:vev:coll}) can be reproduced
in a systematic extension of the present approach based on the gradient expansion.
To this end, we note  that the source~(\ref{mixing:source:algebraic})
leads to $CP$-violating effects
without relying on a finite width, while the result~(\ref{mixing:source:vev:coll})
vanishes in the zero-width limit. It therefore appears that both
sources are distinct and that the source~(\ref{mixing:source:vev:coll}) may
possibly be reproduced when including a collision term
in Eqs.~(\ref{KBE:fermi:decmop}) and then solving these
equations algebraically for the axial current, neglecting the
derivative terms, as it has been done for the collision-free equations
in this section. Further, it may then be relevant to carefully distinguish
between finite-width effects that directly suppress the off-diagonal correlations
and those that preserve these, as it has been investigated for correlations
among doublet leptons in the context of flavoured leptogenesis.
In that example, Yukawa interactions directly
damp flavour correlations whereas gauge interactions preserve these to leading order~\cite{Beneke:2010dz}. Note however that for electroweak baryogenesis,
due to electroweak symmetry breaking, typically
degrees of freedom with different gauge charges are allowed to mix such that
there may indeed be source terms that are proportional to
the finite-width effects mediated by gauge interactions.

We may conclude to this end, that in approximations relying
on the local insertion of spacetime-dependent mass terms
and neglecting the spatial gradients acting on the distribution functions,
there is an extra source term~(\ref{mixing:source:algebraic}) in addition
to the one~(\ref{mixing:source:vev:coll}) reported previously. Phenomenologically,
the source~(\ref{mixing:source:algebraic}) is suppressed by an extra order in derivatives whereas 
Eq.~(\ref{mixing:source:vev:coll}) is suppressed by the finite
width of the mixing particles. In addition, there are extra factors of $v_{a,b}$
in the term~(\ref{mixing:source:vev:coll}) that may lead to an important
relative suppression because the asymmetry may most effectively be produced
close to the symmetric phase where $v_{a,b}$ are small because of the rapid
damping of the chiral asymmetries in the broken phase for large values of $v_{a,b}$.

Before drawing phenomenological conclusions, one should of course
spell out that the approximations~(\ref{mixing:source:algebraic}) and~(\ref{mixing:source:vev:coll}), besides neglecting spatial gradients on the distribution
functions, rely on using the particle propagators of the symmetric phase
even inside the wall. The advantage of numerical solutions is that such
a far-reaching assumption is not required~\cite{Konstandin:2005cd}. However,
a realistic coupling to the background degrees of freedom in the SM appears
challenging and has yet to be accomplished.

First steps toward a reliable numerical calculation in the mixing scenarios
have been taken in Refs.~\cite{Cirigliano:2009yt,Cirigliano:2011di}. In both cases, a toy model of mixing scalar
particles is considered. The kinetic equations that descend from the
Kadanoff--Baym equations are then solved self-consistently, crucially including
the collision term in a background thermal bath. The calculations in
the time-dependent model~\cite{Cirigliano:2009yt} closely resemble the numerical solutions
for resonant leptogenesis in Ref.~\cite{Garbrecht:2014aga}, whereas the model assuming a space-dependent
background~\cite{Cirigliano:2011di} yields insights into electroweak baryogenesis. Since, as discussed
in Section~\ref{sec:reslg}, analytic approximations are available for resonant
leptogenesis in a substantial region of parameter space, including also cases
when the mass difference of the RHNs is smaller than their width, it would be
very interesting to consider whether also for electroweak baryogenesis, more insights
can be gained from a further improvement of both, numerical and analytical approximations.

\subsection{$CP$-conserving transport effects and solving for the asymmetry}
\label{sec:CPconservingprocesses}

In order to complete this overview over the calculational methods for electroweak baryogenesis,
we need to account for some crucial effects SM processes have on the way the axial currents
are processed. First, in absence of extra collision terms, the kinetic equations just describe
convective motion of the particles in a force field. In the presence of SM particles
at finite temperature, the particles will however experience flavour-conserving scatterings
via gauge bosons that drive the system toward kinetic equilibrium as well as flavour-converting scatterings that drive toward chemical equilibration.
In the symmetric electroweak phase, left and right-chiral fermions effectively
act as different species, such that we consider the Yukawa couplings
of the SM as flavour-converting in the present context (different from
the notion of flavour violation in the broken electroweak phase).
Finally, the asymmetry in chiral fermions that has been produced inside the
wall and transported ahead of it has to be converted into a baryon asymmetry,
a process that is carried out by weak sphalerons.

\paragraph{Diffusion}
The flavour-conserving processes can effectively be described by a diffusion law
\begin{align}
\label{Fick:law}
\mathbf j_{i}=-D_i \nabla q_i\,,
\end{align}
where $\mathbf j_{i}$ is the three-current of the particle species $i$ in the plasma frame,
$q_i$ is the charge density and $D_i$ a phenomenological parameter called diffusion constant.

Let us now drop the subscripts labeling the species for simplicity.
When neglecting forces due to spatial gradients and restricting to stationary solutions,
the kinetic equations reduce to
\begin{align}
\label{fl:cons:eq}
\frac{1}{|k^0|}\mathbf k\cdot\nabla f(\mathbf k)={\cal C}[f]\,,
\end{align}
where the right-hand side stands for the portion of the collision term responsible for
flavour-conserving processes. The leading contribution to ${\cal C}$ is from two-by-two scatterings
with gauge bosons. The phase-space integrals over tree-level contributions
with $t$-channel exchange of gauge bosons or fermions can be logarithmically divergent,
which is physically regulated by Landau damping and Debye screening and thus requires
a resummation of the insertions of the pertinent self energies~\cite{Arnold:2000dr,Arnold:2003zc}.

Now, we decompose the distribution function as $f=f^{(0)}+\delta f$,
where
\begin{align}
f^{(0)}(k)=\frac{1}{{\rm e}^\frac{k^0-\mu(\mathbf x)}{T}\pm1}
\end{align}
as forced by local kinetic equilibrium due to fast gauge interactions. This first approximation
to the solution is isotropic. The first correction then corresponds the leading anisotropic
contributions and takes the form
\begin{align}
\delta f(\mathbf k)=h(\mathbf k) \mathbf k \cdot \nabla\mu\,.
\end{align}
In terms of this anisotropic contribution, the current is given by
\begin{align}
\mathbf j_=\int\frac{d^3 k}{(2\pi)^3}\frac{\mathbf k}{|\mathbf k|}\delta f(\mathbf k)\,.
\end{align}
Since $q\propto \mu,$
we thus see that determining $h(\mathbf k)$ by solving
Eq.~(\ref{fl:cons:eq}) amounts to finding the diffusion constant.
Values for the diffusion constants widely used in phenomenological studies can be found in Ref.~\cite{Joyce:1994zn}.
It may presently be in order to reevaluate the estimates in that work using state-of-the art methods such as
developed in Refs.~\cite{Arnold:2000dr,Arnold:2003zc}.

Taking the divergence of Eq.~(\ref{Fick:law}) yields $\nabla\cdot\mathbf j_i=-D_i \Delta q_i$.
Minding that this relation holds in the plasma frame, we write down the covariant form and
deduce from this the correct expression for the frame moving along with the wall
in negative $z$-direction:
\begin{align}
\label{Diffusion:comoving}
\partial_\mu j_i^\mu=v_{\rm w} \partial_z q_i -D_i\partial_z^2 q_i\,.
\end{align}
In general, also flavour-converting interactions contribute to
the diffusion transport what should be of phenomenological relevance e.g.
for the left and right-handed top quarks and the Higgs bosons. To account for
this, one may promote the diffusion constants to matrices in flavour space in future
work. Note that diffusion corresponds to a random walk, such that the
distance a particle proceeds in a time $t$ is given by $\sqrt{D t}$ for a diffusion
constant $D$, while
the wall proceeds by $v_{\rm w} t$. Therefore, a particle typically remains a time
$t_{\rm diff}\sim D/v_{\rm w}^2$ ahead of the wall before it is caught up by the broken
electroweak phase.

\paragraph{Flavour conversion}
Flavour-converting interactions mediated by Yukawa couplings
are discussed in the context of the CTP framework in Refs.~\cite{Cirigliano:2006wh,Chung:2009qs}. In that work, the thermal effects necessary to account
for the conversion of massless particles in the symmetric phase are estimated through the
inclusion of thermal masses. Applying techniques that have been developed more recently
for leptogenesis~\cite{Anisimov:2010gy,Besak:2012qm,Garbrecht:2013gd,Garbrecht:2013urw,Laine:2013lka,Biondini:2017rpb}, it may be worthwhile to update these calculations such that
they include two-by-two scatterings involving the radiation of one gauge boson, in particular including the $t$-channel enhanced contributions.

At the level of fluid equations, the flavour-violating rates take the effect of
forcing the system toward chemical equilibrium, i.e. they are proportional to sums
of chemical potentials $\mu_i$ that are related to the charge densities $q_i$ via Eq.~(\ref{q:mu}).
Putting this together with the source terms that are e.g. given by the second term in Eq.~(\ref{semiclassical:force}) for the source from the semiclassical
force or the
right-hand side of Eq.~(\ref{mixing:source:algebraic}) for the source from resonantly enhanced
mixing, one obtains the fluid equation
\begin{align}
\label{fluid:eq:EWBG}
\partial_\mu j_i^\mu=-\sum\limits_a \Gamma^{(a)} (\mu_1\pm\cdots\pm\mu_n)+S^{\cancel{CP}}_i\,.
\end{align}
The averaged decay rate for the reaction enumerated by $a$ is given by $\Gamma^{(a)}$, and the signs
are positive for an incoming particle and negative for an incoming antiparticle
(negative for an outgoing particle and positive for an outgoing antiparticle).
When substituting the diffusion relation~(\ref{Diffusion:comoving}) for the left
hand side of Eq.~(\ref{fluid:eq:EWBG}), this equation can be readily solved for
the charge distributions for all relevant particle species with vanishing
boundary conditions far ahead and behind the wall.

In phenomenological calculations, the reaction rates $\Gamma^{(a)}$ also account for
strong sphalerons as well as the damping of chiral asymmetries
in the broken phase due to the nonvanishing expectation value of the Higgs field. The latter effect
has again been computed from the insertion of expectation values of the Higgs field~\cite{Lee:2004we}. For the future, it would be desirable to derive this in a framework where one takes account of the locally correct dispersion relation for the quasiparticles as their mass changes through the wall.

\paragraph{Baryon number violation}
As the final step of the calculation, we sum the charge densities of
all left handed SM fermions from the solution to Eq.~(\ref{fluid:eq:EWBG}) to obtain $q_{\rm left}$ (recall that these count one isodoublet component each in the 
convention of the present work).
Ahead of the bubble wall, also weak sphalerons are active, turning
$q_{\rm left}$ into a baryon number density $q_B$ but at the same time
wash out $q_B$. These processes are described by the equation~\cite{Huet:1995sh,Cline:2000nw,Carena:2002ss,Lee:2004we}
\begin{align}
-v_{\rm w} \frac{d}{dz}q_B(z)+\frac{15}{4}\Gamma_{\rm ws}q_B(z)=3 \Gamma_{\rm ws}2 q_{\rm left}(z)\,,\quad \textnormal{for $z<0$}
\end{align}
where the weak sphaleron rate in the symmetric phase has been found to be
$\Gamma_{\rm ws}\approx 6\kappa \alpha_w^5 T$, where
$\alpha_w=g^2/(4/\pi)$, $g$ is the coupling constant of the ${\rm SU}(2)_{\rm L}$
gauge interactions and $\kappa\sim 20$~\cite{Bodeker:1999gx,Moore:1999fs,Moore:2000mx}.
One eventually one finds the solution for the baryon charge density:
\begin{equation}
\label{nB:sphal}
q_{B}(z=0)=-3\frac{\Gamma_{\rm ws}}{v_{\rm w}}
\int\limits_{-\infty}^0 dz\; 2 q_{\rm left}(z)
{\rm e}^{\frac{15}{4} \frac{\Gamma_{\rm ws}}{v_{\rm w}}z}\,.
\end{equation}
Provided the sphaleron rate inside the bubble is smaller than the Hubble rate,
i.e. in the case of a so-called strong first order phase transition, the baryon-to-entropy ratio $Y_B=q_B/s$ is then conserved and may correspond to the value observed today~(\ref{BAU:CMB}). In fact, the solution~\eqref{nB:sphal} relies on modeling the spaleron rate as constant for $z<0$ and zero for $z>0$.
To first approximation this can be justified because
in the SM, $v_{\rm w} t_{\rm diff}\sim D/v_{\rm w}\gg l_{\rm w}$, such that the
chiral asymmetries spread over a large region compared to
the size of the bubble wall, wherein  $\Gamma_{\rm ws}$ smoothly changes. As an additional
consequence of this relation, only a small fraction of the chiral asymmetry typically ends up being converted into baryons.

\section{Conclusions}

While the idea that particle physics processes  are accountable for the creation of the
baryon asymmetry of the Universe~(\ref{BAU:CMB}) is very compelling, 
it is yet unknown which
mechanism precisely is at work and even
in which sector of a complete theory it resides. Sakharov's nonequilibrium condition
may well be realized in a somewhat exotic manner, e.g. through the out-of-equilibrium decay
of condensates relying on sometimes random or arbitrary initial conditions.
Nonetheless, it is very plausible that baryogenesis is after
all realized based on simple extensions of the SM, e.g. through a limited number of new particles such
as RHNs or an electroweak sector featuring a first order phase transition.
In that case, calculations of baryogenesis rely on statistical physics, i.e.
on QFT at finite temperature in conjunction with an appropriate description
of the deviations from thermal equilibrium from the expansion of the Universe.
Methods addressing these tasks are given by the CTP formalism, that
is reviewed in the present work.

Even for generic scenarios, such as baryogenesis from out-of-equilibrium decays or at phase boundaries,
there are quite a few questions that are not covered in the present work,
most importantly those concerning the dynamics of the electroweak phase transition,
i.e. whether it is of first order or a crossover as well as how large the rate for
$B+L$-violating sphaleron-transitions is. These aspects must be considered in combination with the matters we have been focusing
on in this article, 
i.e. $CP$ violation in the early Universe, fluid equations and the calculational
foundation of these using CTP techniques. So far, baryogenesis has thus been
a challenge leading to fascinating insights into the real-time evolution
of systems governed by QFT and statistical physics. It is very plausible,
but it remains to be seen, that these methods and results are applicable to the mechanism that
is responsible for the baryon asymmetry of the Universe and that is yet to be discovered.

\begin{appendix}
\renewcommand{\theequation}{\Alph{section}\arabic{equation}}
\setcounter{equation}{0}

\section{Discrete symmetries}
\label{app:discrete}

A thorough and comprehensive discussion on this topic is given in the monograph~\cite{Branco:1999fs}.
Here, we quote the identities that are most useful in the present context.

For a given representation of Dirac matrices, there are nonsingular $4\times 4$
matrices $A$ and $C$ such that
\begin{align}
\label{A:C}
A \gamma_\mu=\gamma_\mu^\dagger A\,,\qquad
\gamma_\mu C=-C\gamma_\mu^t\,,
\end{align}
where $t$ denotes transposition. In the Weyl representation, $A=\gamma_0$ and $C={\rm i}\gamma_2\gamma_0$.
These satisfy the useful identities
\begin{align}
\label{ccidentities}
\begin{array}{rclrcl}
A^\dagger&=&A\,,    &   A\gamma_5&=&-\gamma_5^\dagger A\,,\\
C^t&=&-C\,,\quad   &    \gamma_5 C &=& C \gamma_5^t\,,\\
CA^*C^*A&=&1\,, & A\sigma_{\mu\nu}&=& \sigma_{\mu\nu}^\dagger A\,,\\
&&&  \sigma_{\mu\nu} C &=& -C \sigma_{\mu\nu}^t\,,
\end{array}
\end{align}
where $\sigma_{\mu\nu}=\frac{\rm i}{2}[\gamma_\mu,\gamma_\nu]$. For a spinor $\psi$, its conjugate is given
by $\bar\psi=\psi^\dagger A$. The charge-conjugate spinor is
\begin{align}
\psi^C={\rm e}^{{\rm i}\alpha^{C}}C{\bar\psi}^t\,,
\end{align}
and with the help of one of the identities~(\ref{ccidentities}), we see that
$\bar{\psi^C}=-\psi^t C^{-1}$. Including a parity reflection in addition,
the $CP$ conjugate spinor is given by
\begin{align}
\psi^{CP}={\rm e}^{{\rm i}\alpha^{CP}}\gamma^0C{\bar\psi}^t\,.
\end{align}
For a scalar field $\phi$, of course, $C$ and $CP$ conjugation take the same
effect (up to possible arbitrary phases), namely $\phi\to\phi^{C,CP}= {\rm e}^{{\rm i}\beta^{C,CP}}\phi^*$. The phases $\alpha^{C,CP}$ and $\beta^{C,CP}$ are arbitrary, i.e. physical
effects of $C$ and $CP$ violation do not depend on their choice.

The action of $CP$-conjugation in terms of a unitary operator ${\cal CP}$
on a spinor and on a complex scalar field $\phi$ is
\begin{subequations}
\begin{align}
({\cal CP})\psi({\cal CP})^\dagger=\psi^{CP}\,,\\
({\cal CP})\phi({\cal CP})^\dagger= \phi^{CP}\,,
\end{align}
\end{subequations}
such that the bilinear terms that appear in Lagrangians or correlation functions
transform as (see Ref.~\cite{Branco:1999fs} for all relevant details as well as
comprehensive tables on discrete symmetries)
\begin{subequations}
\begin{align}
\label{CP:scalar}
({\cal CP})\bar\psi\chi({\cal CP})^\dagger=&{\rm e}^{{\rm i}(\alpha_\chi^{CP}-\alpha_\psi^{CP})}\bar\chi \psi\,,\\
\label{CP:pseudoscalar}
({\cal CP})\bar\psi\gamma^5\chi({\cal CP})^\dagger=&-{\rm e}^{{\rm i}(\alpha_\chi^{CP}-\alpha_\psi^{CP})}\bar\chi \gamma^5\psi\,,\\
({\cal CP})\bar\psi\gamma^\mu\chi({\cal CP})^\dagger=&-{\rm e}^{{\rm i}(\alpha_\chi^{CP}-\alpha_\psi^{CP})}\bar\chi\gamma_\mu \psi\,,\\
({\cal CP})\bar\psi\gamma^\mu\gamma^5\chi({\cal CP})^\dagger=&-{\rm e}^{{\rm i}(\alpha_\chi^{CP}-\alpha_\psi^{CP})}\bar\chi\gamma_\mu \gamma^5\psi\,.
\end{align}
\end{subequations}

\section{Tree-level propagators on the CTP}
\label{App:treeprop}

Useful basic elements of perturbative calculations on the CTP can be the
tree-level propagators. For a scalar field of mass $m$, these are given by
\begin{subequations}
\label{prop:phi:expl}
\begin{align}
{\rm i}\Delta^<(p)&=
2\pi \delta(p^2-m^2)\left[
\vartheta(p_0) f(\mathbf p)
+\vartheta(-p_0) (1+\bar f(-\mathbf p))\right]
\,,
\\
{\rm i}\Delta^>(p)&=
2\pi \delta(p^2-m^2)\left[
\vartheta(p_0) (1+f_\phi(\mathbf p))
+\vartheta(-p_0) \bar f(-\mathbf p)\right]
\,,
\\
{\rm i}\Delta^T(p)&=
\frac{\rm i}{p^2-m^2+{\rm i}\varepsilon}+
2\pi \delta(p^2)\left[
\vartheta(p_0) f(\mathbf p)
+\vartheta(-p_0) \bar f(-\mathbf p)\right]
\,,
\\
{\rm i}\Delta^{\bar T}(p)&=
-\frac{\rm i}{p^2-m^2-{\rm i}\varepsilon}+
2\pi \delta(p^2-m^2)\left[
\vartheta(p_0) f_(\mathbf p)
+\vartheta(-p_0) \bar f(-\mathbf p)\right]
\,,
\end{align}
\end{subequations}
and for a four-component spinor with mass $m$
\begin{subequations}
\label{prop:N:expl}
\begin{align}
{\rm i}S^{<}(p)
&=-2\pi\delta(p^2-m^2)(p\!\!\!/+m)\left[
\vartheta(p_0)f(\mathbf{p})
-\vartheta(-p_0)(1-\bar f(-\mathbf{p}))
\right]\,,\\
{\rm i}S^{>}(p)
&=-2\pi\delta(p^2-m^2)(p\!\!\!/+m)\left[
-\vartheta(p_0)(1-f(\mathbf{p}))
+\vartheta(-p_0)\bar f(-\mathbf{p})
\right]\,,\\
\label{S:T}
{\rm i}S^{T}(p)
&=
\frac{{\rm i}(p\!\!\!/+m)}{p^2-m^2+{\rm i}\varepsilon}
-2\pi\delta(p^2-m^2)(p\!\!\!/+m)\left[
\vartheta(p_0)f(\mathbf{p})
+\vartheta(-p_0)\bar f(-\mathbf{p})
\right]\,,\\
\label{S:Tbar}
{\rm i}S^{\bar T}(p)
&=
-\frac{{\rm i}(p\!\!\!/+m)}{p^2-m^2-{\rm i}\varepsilon}
-2\pi\delta(p^2-m^2)(p\!\!\!/+m)\left[
\vartheta(p_0)f(\mathbf{p})
+\vartheta(-p_0)\bar f(-\mathbf{p})
\right]
\,.
\end{align}
\end{subequations}
The distribution functions $f$ are for particles and $\bar f$ for antiparticles.
These solutions are useful as elements in the computation of
many correlation functions. Nonetheless, there are important situations
where one may want to solve the Schwinger--Dyson equations such that
the resulting propagators readily include effects from finite width
(cf. Refs.~\cite{Garbrecht:2011xw,Garbrecht:2013urw}), flavour mixing (cf. Section~\ref{sec:reslg} and
Ref.~\cite{Garbrecht:2011aw} on resonant leptogenesis, Ref.~\cite{Beneke:2010dz} on flavour
effects for doublet leptons in leptogenesis in the CTP formalism) or gradient effects (cf. Refs.~\cite{Kainulainen:2001cn,Prokopec:2003pj,Prokopec:2004ic,Konstandin:2004gy,Konstandin:2005cd,Konstandin:2013caa}).

Concerning the propagators for the spinor fields, we note that Eqs.~(\ref{prop:N:expl})
describe helicity-symmetric states of Dirac fermions. It is nonetheless applicable to the
examples in the present article, where also (massless) chiral fermions (such as $\ell$)
and nonrelativistic Majorana fermions (i.e. the $N_i$) occur. The reduction to
chiral fermions is achieved by the use of the chiral projection operators $P_{\rm L,R}$.
As for the Majorana fermions, imposing the Majorana condition as in Ref.~\cite{Garbrecht:2011aw} implies that $f(\mathbf p)=\bar f(\mathbf p)$. In the relativistic regime,
also helicity asymmetries of Majorana neutrinos are of importance, and the
necessary constructions within the CTP framework can be found in Ref.~\cite{Drewes:2016gmt}.

\section{Details on the calculation of the $CP$-violating vertex contribution to leptogenesis
in the CTP approach}
\label{app:vertex}

We start from Eq.~(\ref{Sigmavert:CTP}) and apply
Eq.~(\ref{CTP:indices}) in order to take the combination of CTP indices
corresponding to the Wightman function $\slashed\Sigma^{{\rm vert}>}$,
\begin{align}
{\rm i}\slashed\Sigma^{-+}(q)=-Y_1^2{Y_2^*}^2\int\limits_{k k^\prime q^\prime p}\delta_{p-k-k^\prime}\delta_{p-q-q^\prime}\big\{&
P_{\rm R} {\rm i}S_{N2}^{-+}(k-q^\prime)P_{\rm R} {\rm i}{S_\ell^{CP}}^{++}(k)
P_{\rm L}{\rm i}S^{++}_{N1}(p)P_{\rm L}
{\rm i}\Delta^{+-}(-k^\prime){\rm i}\Delta^{++}(q^\prime)\notag\\[-6mm]
-&P_{\rm R} {\rm i}S_{N2}^{--}(k-q^\prime)P_{\rm R} {\rm i}{S_\ell^{CP}}^{-+}(k)
P_{\rm L}{\rm i}S^{++}_{N1}(p)P_{\rm L}
{\rm i}\Delta^{+-}(-k^\prime){\rm i}\Delta^{+-}(q^\prime)\notag\\
-&P_{\rm R} {\rm i}S_{N2}^{-+}(k-q^\prime)P_{\rm R} {\rm i}{S_\ell^{CP}}^{+-}(k)
P_{\rm L}{\rm i}S^{-+}_{N1}(p)P_{\rm L}
{\rm i}\Delta^{--}(-k^\prime){\rm i}\Delta^{++}(q^\prime)\notag\\
+&P_{\rm R} {\rm i}S_{N2}^{--}(k-q^\prime)P_{\rm R} {\rm i}{S_\ell^{CP}}^{--}(k)
P_{\rm L}{\rm i}S^{-+}_{N1}(p)P_{\rm L}
{\rm i}\Delta^{--}(-k^\prime){\rm i}\Delta^{+-}(q^\prime)
\big\}\notag\\
+1\leftrightarrow 2\,.
\end{align}
Next, when restricting to the contributions that do not require an on shell $N_2$
(This is a good approximation in the hierarchical regime $N_1\ll N_2$, when
e.g. the temperature never is high enough to produce $N_2$.
Alternatively,
we can add the contribution from on shell $N_2$ in the same way we calculate
here the asymmetry generated by decays and inverse decays of $N_1$.),
only terms with ${\rm i}S_{N2}^T$ or ${\rm i}S_{N2}^{\bar T}$ are left, where
we can replace ${\rm i}S_{N2}^{\bar T}\to - {\rm i}S_{N2}^T$ because $N_2$ is off shell, cf. Eqs.~(\ref{S:T}) and~(\ref{S:Tbar}).
We thus obtain
\begin{align}
{\rm i}\slashed\Sigma^{{\rm vert}>}(q)=-Y_1^2{Y_2^*}^2\int\limits_{k k^\prime q^\prime p}\delta_{p-k-k^\prime}\delta_{p-q-q^\prime}\big\{&
P_{\rm R} {\rm i}S_{N2}^{T}(k-q^\prime)P_{\rm R} {\rm i}{S_\ell^{CP}}^{>}(k)
P_{\rm L}{\rm i}S^{T}_{N1}(p)P_{\rm L}
{\rm i}\Delta^{<}(-k^\prime){\rm i}\Delta^{<}(q^\prime)\notag\\[-6mm]
-&P_{\rm R} {\rm i}S_{N2}^{T}(k-q^\prime)P_{\rm R} {\rm i}{S_\ell^{CP}}^{\bar T}(k)
P_{\rm L}{\rm i}S^{>}_{N1}(p)P_{\rm L}
{\rm i}\Delta^{\bar T}(-k^\prime){\rm i}\Delta^{<}(q^\prime)\big\}&\notag\\
-{Y_1^*}^2Y_2^2\int\limits_{k k^\prime q^\prime p}\delta_{p-k-k^\prime}\delta_{p-q-q^\prime}\big\{&
P_{\rm R} {\rm i}S_{N1}^{>}(k-q^\prime)P_{\rm R} {\rm i}{S_\ell^{CP}}^{T}(k)
P_{\rm L}{\rm i}S^{T}_{N2}(p)P_{\rm L}
{\rm i}\Delta^{<}(-k^\prime){\rm i}\Delta^{T}(q^\prime)\notag\\[-6mm]
-&P_{\rm R} {\rm i}S_{N1}^{>}(k-q^\prime)P_{\rm R} {\rm i}{S_\ell^{CP}}^{>}(k)
P_{\rm L}{\rm i}S^{T}_{N2}(p)P_{\rm L}
{\rm i}\Delta^{<}(-k^\prime){\rm i}\Delta^{<}(q^\prime)
\big\}\,.
\end{align}
The Wightman self-energy ${\rm i}\slashed\Sigma^{>}(q)$ follows through
$<\leftrightarrow>$ and $T\leftrightarrow \bar T$.
From these self energies, we build the vertex contribution
to the source term~(\ref{coll:vert})
\begin{align}
&\hskip-6.1cm{\cal S}^{\rm vert}=
\int\limits_q{\rm tr}
\left[{\rm i}\slashed \Sigma_\ell^{{\rm vert}>}(q){\rm i}S_\ell^<(q)-{\rm i}\slashed \Sigma_\ell^{{\rm vert}>}(q){\rm i}S_\ell^<(q)\right]
\notag\\
=-Y_1^2{Y_2^*}^2\int\limits_{k k^\prime q q^\prime p}\delta_{p-k-k^\prime}\delta_{p-q-q^\prime}{\rm tr}\Big[&
P_{\rm R} {\rm i}S_{N2}^{T}(k-q^\prime)P_{\rm R}
{\rm i}{S_\ell^{CP}}^{>}(k)
P_{\rm L}{\rm i}S^{T}_{N1}(p)P_{\rm L}
{\rm i}\Delta^{<}(-k^\prime)
{\rm i}\Delta^{<}(q^\prime)
{\rm i}S_\ell^<(q)\notag\\[-6mm]
-&P_{\rm R} {\rm i}S_{N2}^{T}(k-q^\prime)P_{\rm R}
{\rm i}{S_\ell^{CP}}^{\bar T}(k)
P_{\rm L}{\rm i}S^{>}_{N1}(p)P_{\rm L}
{\rm i}\Delta^{\bar T}(-k^\prime)
{\rm i}\Delta^{<}(q^\prime)
{\rm i}S_\ell^<(q)\notag\\
+&P_{\rm R} {\rm i}S_{N2}^{T}(k-q^\prime)P_{\rm R}
{\rm i}{S_\ell^{CP}}^{<}(k)
P_{\rm L}{\rm i}S^{\bar T}_{N1}(p)P_{\rm L}
{\rm i}\Delta^{>}(-k^\prime)
{\rm i}\Delta^{>}(q^\prime)
{\rm i}S_\ell^>(q)\notag\\
-&P_{\rm R} {\rm i}S_{N2}^{T}(k-q^\prime)P_{\rm R}
{\rm i}{S_\ell^{CP}}^{T}(k)
P_{\rm L}{\rm i}S^{<}_{N1}(p)P_{\rm L}
{\rm i}\Delta^{T}(-k^\prime)
{\rm i}\Delta^{>}(q^\prime)
{\rm i}S_\ell^>(q)\Big]\notag\\
-{Y_1^*}^2Y_2^2\int\limits_{k k^\prime q q^\prime p}\delta_{p-k-k^\prime}\delta_{p-q-q^\prime}{\rm tr}\Big[&
P_{\rm R} {\rm i}S_{N1}^{>}(p)P_{\rm R}
{\rm i}{S_\ell^{CP}}^{T}(k)
P_{\rm L}{\rm i}S^{T}_{N2}(k-q^\prime)P_{\rm L}
{\rm i}\Delta^{<}(q^\prime)
{\rm i}\Delta^{T}(-k^\prime)
{\rm i}S_\ell^<(q)\notag\\[-6mm]
-&P_{\rm R} {\rm i}S_{N1}^{\bar T}(p)P_{\rm R}
{\rm i}{S_\ell^{CP}}^{>}(k)
P_{\rm L}{\rm i}S^{T}_{N2}(k-q^\prime)P_{\rm L}
{\rm i}\Delta^{<}(q^\prime)
{\rm i}\Delta^{<}(-k^\prime)
{\rm i}S_\ell^<(q)\notag\\
+&P_{\rm R} {\rm i}S_{N1}^{<}(p)P_{\rm R}
{\rm i}{S_\ell^{CP}}^{\bar T}(k)
P_{\rm L}{\rm i}S^{T}_{N2}(k-q^\prime)P_{\rm L}
{\rm i}\Delta^{>}(q^\prime)
{\rm i}\Delta^{\bar T}(-k^\prime)
{\rm i}S_\ell^>(q)\notag\\
-&P_{\rm R} {\rm i}S_{N1}^{T}(p)P_{\rm R}
{\rm i}{S_\ell^{CP}}^{<}(k)
P_{\rm L}{\rm i}S^{T}_{N2}(k-q^\prime)P_{\rm L}
{\rm i}\Delta^{>}(q^\prime)
{\rm i}\Delta^{>}(-k^\prime)
{\rm i}S_\ell^>(q)\Big]\,,
\end{align}
where, under the second integral, we have replaced
$-k^\prime \leftrightarrow q$ and $p\to k+q-p$.

Next, as it is shown in detail in Ref.~\cite{Garbrecht:2013iga}, we can replace
e.g.
\begin{align}
&\int\limits_k\delta_{p-k-k^\prime}\left(
{\rm i}{S_\ell^{CP}}^{>}(k){\rm i}\Delta^{<}(-k^\prime)
-{\rm i}{S_\ell^{CP}}^{\bar T}(k){\rm i}\Delta^{\bar T}(-k^\prime)\right)g(k)
\notag\\
\to&
\int\limits_k\delta_{p-k-k^\prime}
\frac12\left({\rm i}{S_\ell^{CP}}^{>}(k){\rm i}\Delta^{<}(-k^\prime)
-{\rm i}{S_\ell^{CP}}^{<}(k){\rm i}\Delta^{>}(-k^\prime) \right)g(k)\,,
\end{align}
where  $g(k)$ is an arbitrary function. The proof is straightforward and relies
on substitution of the tree-level propagators~(\ref{prop:phi:expl}), (\ref{prop:N:expl}) and considering all
product terms separately. Making use of this replacement, the source term
reduces to
\begin{align}
{\cal S}^{\rm vert}=
-Y_1^2{Y_2^*}^2\int\limits_{kk^\prime qq^\prime p}\delta_{p-k-k^\prime}\delta_{p-q-q^\prime}\frac12{\rm tr}\Big[&
P_{\rm R} {\rm i}S_{N2}^{T}(k-q^\prime)P_{\rm R}
\left({\rm i}{S_\ell^{CP}}^{>}(k){\rm i}\Delta^{<}(-k^\prime)-{\rm i}{S_\ell^{CP}}^{<}(k){\rm i}\Delta^{>}(-k^\prime)\right)\notag\\[-6mm]
&P_{\rm L}{\rm i}\delta S_{N1}(p)P_{\rm L}\left({\rm i}\Delta^{<}(q^\prime){\rm i}S_\ell^<(q)-\Delta^{>}(q^\prime){\rm i}S_\ell^>(q)\right)\Big]\notag\\
+{Y_1^*}^2Y_2^2\int\limits_{kk^\prime qq^\prime p}\delta_{p-k-k^\prime}\delta_{p-q-q^\prime}\frac12{\rm tr}\Big[&
P_{\rm R} {\rm i}\delta S_{N1}(p)P_{\rm R}
\left({\rm i}{S_\ell^{CP}}^{>}(k){\rm i}\Delta^{<}(-k^\prime)-{\rm i}{S_\ell^{CP}}^{<}(k){\rm i}\Delta^{>}(-k^\prime)\right)\notag\\[-6mm]
&P_{\rm L}{\rm i}S^T_{N2}(k-q^\prime)P_{\rm L}\left({\rm i}\Delta^{<}(q^\prime){\rm i}S_\ell^<(q)-\Delta^{>}(q^\prime){\rm i}S_\ell^>(q)\right)\Big]\,.
\end{align}
Substituting the tree-level propagators~(\ref{prop:N:expl}), taking the trace
and carrying out the integrals over the zero-components of them momenta by making
use of the on-shell $\delta$-functions, we obtain the result~(\ref{coll:vert}).

\end{appendix}


\begin{thebibliography}{99}
\bibliographystyle{unsrt}


\bibitem{Riotto:1999yt}
  A.~Riotto and M.~Trodden,
  ``Recent progress in baryogenesis,''
  Ann.\ Rev.\ Nucl.\ Part.\ Sci.\  {\bf 49} (1999) 35
  doi:10.1146/annurev.nucl.49.1.35
  [hep-ph/9901362].



\bibitem{Cline:2006ts}
  J.~M.~Cline,
  ``Baryogenesis,''
  hep-ph/0609145.



\bibitem{Canetti:2012zc}
  L.~Canetti, M.~Drewes and M.~Shaposhnikov,
  ``Matter and Antimatter in the Universe,''
  New J.\ Phys.\  {\bf 14} (2012) 095012
  doi:10.1088/1367-2630/14/9/095012
  [arXiv:1204.4186 [hep-ph]].



\bibitem{Steigman:2008ap}
  G.~Steigman,
  ``When Clusters Collide: Constraints On Antimatter On The Largest Scales,''
  JCAP {\bf 0810} (2008) 001
  doi:10.1088/1475-7516/2008/10/001
  [arXiv:0808.1122 [astro-ph]].



\bibitem{Branco:1999fs}
  G.~C.~Branco, L.~Lavoura and J.~P.~Silva,
  ``CP Violation,''
  Int.\ Ser.\ Monogr.\ Phys.\  {\bf 103} (1999) 1.



\bibitem{Wu:1957my}
  C.~S.~Wu, E.~Ambler, R.~W.~Hayward, D.~D.~Hoppes and R.~P.~Hudson,
  ``Experimental Test of Parity Conservation in Beta Decay,''
  Phys.\ Rev.\  {\bf 105} (1957) 1413.
  doi:10.1103/PhysRev.105.1413



\bibitem{Christenson:1964fg}
  J.~H.~Christenson, J.~W.~Cronin, V.~L.~Fitch and R.~Turlay,
  ``Evidence for the $2\pi$ Decay of the $K_2^0$ Meson,''
  Phys.\ Rev.\ Lett.\  {\bf 13} (1964) 138.
  doi:10.1103/PhysRevLett.13.138

\bibitem{Kostelecky:2008ts}
  V.~A.~Kostelecky and N.~Russell,
  ``Data Tables for Lorentz and CPT Violation,''
  Rev.\ Mod.\ Phys.\  {\bf 83} (2011) 11
  doi:10.1103/RevModPhys.83.11
  [arXiv:0801.0287 [hep-ph]].


\bibitem{Penzias:1965wn}
  A.~A.~Penzias and R.~W.~Wilson,
  ``A Measurement of excess antenna temperature at 4080-Mc/s,''
  Astrophys.\ J.\  {\bf 142} (1965) 419.
  doi:10.1086/148307
 
 
\bibitem{AlpherHerman1}
R.~A.~Alpher and R.~C.~Herman,
``On the Relative Abundance of the Elements,''
Phys.\ Rev.\  {\bf 74} (1948) 1737.


\bibitem{AlpherHerman2}
R.~A.~Alpher and R.~C.~Herman,
``Evolution of the Universe,''
Nature  {\bf 162} (1948) 774.

\bibitem{Gamov1}
G.~Gamow,
``The Origin of Elements and the Separation of Galaxies,''
Phys.\ Rev.\ {\bf 74} (1948) 4.

\bibitem{Gamov2}
G.~Gamow,
``The evolution of the Universe,''
Nature {\bf 162} (1948) 680.




\bibitem{Sakharov:1967dj}
  A.~D.~Sakharov,
  ``Violation of CP Invariance, C asymmetry, and baryon asymmetry of the universe,''
  Pisma Zh.\ Eksp.\ Teor.\ Fiz.\  {\bf 5} (1967) 32
   [JETP Lett.\  {\bf 5} (1967) 24]
   [Sov.\ Phys.\ Usp.\  {\bf 34} (1991) no.5,  392]
   [Usp.\ Fiz.\ Nauk {\bf 161} (1991) no.5,  61].
  doi:10.1070/PU1991v034n05ABEH002497



\bibitem{tHooft:1976rip}
  G.~'t Hooft,
  ``Symmetry Breaking Through Bell-Jackiw Anomalies,''
  Phys.\ Rev.\ Lett.\  {\bf 37} (1976) 8.
  doi:10.1103/PhysRevLett.37.8



\bibitem{Kuzmin:1985mm}
  V.~A.~Kuzmin, V.~A.~Rubakov and M.~E.~Shaposhnikov,
  ``On the Anomalous Electroweak Baryon Number Nonconservation in the Early Universe,''
  Phys.\ Lett.\  {\bf 155B} (1985) 36.
  doi:10.1016/0370-2693(85)91028-7



\bibitem{Kajantie:1996mn}
  K.~Kajantie, M.~Laine, K.~Rummukainen and M.~E.~Shaposhnikov,
  ``Is there a hot electroweak phase transition at m(H) larger or equal to m(W)?,''
  Phys.\ Rev.\ Lett.\  {\bf 77} (1996) 2887
  doi:10.1103/PhysRevLett.77.2887
  [hep-ph/9605288].



\bibitem{Rummukainen:1998as}
  K.~Rummukainen, M.~Tsypin, K.~Kajantie, M.~Laine and M.~E.~Shaposhnikov,
  ``The Universality class of the electroweak theory,''
  Nucl.\ Phys.\ B {\bf 532} (1998) 283
  doi:10.1016/S0550-3213(98)00494-5
  [hep-lat/9805013].



\bibitem{Jarlskog:1985ht}
  C.~Jarlskog,
  ``Commutator of the Quark Mass Matrices in the Standard Electroweak Model and a Measure of Maximal CP Violation,''
  Phys.\ Rev.\ Lett.\  {\bf 55} (1985) 1039.
  doi:10.1103/PhysRevLett.55.1039



\bibitem{Farrar:1993hn}
  G.~R.~Farrar and M.~E.~Shaposhnikov,
  ``Baryon asymmetry of the universe in the standard electroweak theory,''
  Phys.\ Rev.\ D {\bf 50} (1994) 774
  doi:10.1103/PhysRevD.50.774
  [hep-ph/9305275].



\bibitem{Gavela:1993ts}
  M.~B.~Gavela, P.~Hernandez, J.~Orloff and O.~Pene,
  ``Standard model CP violation and baryon asymmetry,''
  Mod.\ Phys.\ Lett.\ A {\bf 9} (1994) 795
  doi:10.1142/S0217732394000629
  [hep-ph/9312215].



\bibitem{Gavela:1994dt}
  M.~B.~Gavela, P.~Hernandez, J.~Orloff, O.~Pene and C.~Quimbay,
  ``Standard model CP violation and baryon asymmetry. Part 2: Finite temperature,''
  Nucl.\ Phys.\ B {\bf 430} (1994) 382
  doi:10.1016/0550-3213(94)00410-2
  [hep-ph/9406289].



\bibitem{Huet:1994jb}
  P.~Huet and E.~Sather,
  ``Electroweak baryogenesis and standard model CP violation,''
  Phys.\ Rev.\ D {\bf 51} (1995) 379
  doi:10.1103/PhysRevD.51.379
  [hep-ph/9404302].



\bibitem{Tanabashi:2018oca}
  M.~Tanabashi {\it et al.} [Particle Data Group],
  ``Review of Particle Physics,''
  Phys.\ Rev.\ D {\bf 98} (2018) no.3,  030001.
  doi:10.1103/PhysRevD.98.030001



\bibitem{Hinshaw:2012aka}
  G.~Hinshaw {\it et al.} [WMAP Collaboration],
  ``Nine-Year Wilkinson Microwave Anisotropy Probe (WMAP) Observations: Cosmological Parameter Results,''
  Astrophys.\ J.\ Suppl.\  {\bf 208} (2013) 19
  doi:10.1088/0067-0049/208/2/19
  [arXiv:1212.5226 [astro-ph.CO]].



\bibitem{Aghanim:2018eyx}
  N.~Aghanim {\it et al.} [Planck Collaboration],
  ``Planck 2018 results. VI. Cosmological parameters,''
  arXiv:1807.06209 [astro-ph.CO].

\bibitem{Fukugita:1986hr}
  M.~Fukugita and T.~Yanagida,
  ``Baryogenesis Without Grand Unification,''
  Phys.\ Lett.\ B {\bf 174} (1986) 45.
  doi:10.1016/0370-2693(86)91126-3

\bibitem{Shaposhnikov:1986jp}
  M.~E.~Shaposhnikov,
  ``Possible Appearance of the Baryon Asymmetry of the Universe in an Electroweak Theory,''
  JETP Lett.\  {\bf 44} (1986) 465
   [Pisma Zh.\ Eksp.\ Teor.\ Fiz.\  {\bf 44} (1986) 364].

\bibitem{Shaposhnikov:1987tw}
  M.~E.~Shaposhnikov,
  ``Baryon Asymmetry of the Universe in Standard Electroweak Theory,''
  Nucl.\ Phys.\ B {\bf 287} (1987) 757.
  doi:10.1016/0550-3213(87)90127-1


\bibitem{Cohen:1994ss}
  A.~G.~Cohen, D.~B.~Kaplan and A.~E.~Nelson,
  ``Diffusion enhances spontaneous electroweak baryogenesis,''
  Phys.\ Lett.\ B {\bf 336} (1994) 41
  doi:10.1016/0370-2693(94)00935-X
  [hep-ph/9406345].



\bibitem{Schwinger:1960qe}
  J.~S.~Schwinger,
  ``Brownian motion of a quantum oscillator,''
  J.\ Math.\ Phys.\  {\bf 2} (1961) 407.
  doi:10.1063/1.1703727



\bibitem{Keldysh:1964ud}
  L.~V.~Keldysh,
  ``Diagram technique for nonequilibrium processes,''
  Zh.\ Eksp.\ Teor.\ Fiz.\  {\bf 47} (1964) 1515
   [Sov.\ Phys.\ JETP {\bf 20} (1965) 1018].



\bibitem{Calzetta:1986cq}
  E.~Calzetta and B.~L.~Hu,
  ``Nonequilibrium Quantum Fields: Closed Time Path Effective Action, Wigner Function and Boltzmann Equation,''
  Phys.\ Rev.\ D {\bf 37} (1988) 2878.
  doi:10.1103/PhysRevD.37.2878



\bibitem{Dev:2017trv}
  P.~S.~B.~Dev, P.~Di Bari, B.~Garbrecht, S.~Lavignac, P.~Millington and D.~Teresi,
  ``Flavor effects in leptogenesis,''
  Int.\ J.\ Mod.\ Phys.\ A {\bf 33} (2018) 1842001
  doi:10.1142/S0217751X18420010
  [arXiv:1711.02861 [hep-ph]].



\bibitem{Drewes:2017zyw}
  M.~Drewes {\it et al.},
  ``ARS Leptogenesis,''
  Int.\ J.\ Mod.\ Phys.\ A {\bf 33} (2018) no.05n06,  1842002
  doi:10.1142/S0217751X18420022
  [arXiv:1711.02862 [hep-ph]].



\bibitem{Dev:2017wwc}
  B.~Dev, M.~Garny, J.~Klaric, P.~Millington and D.~Teresi,
  ``Resonant enhancement in leptogenesis,''
  Int.\ J.\ Mod.\ Phys.\ A {\bf 33} (2018) 1842003
  doi:10.1142/S0217751X18420034
  [arXiv:1711.02863 [hep-ph]].



\bibitem{Biondini:2017rpb}
  S.~Biondini {\it et al.},
  ``Status of rates and rate equations for thermal leptogenesis,''
  Int.\ J.\ Mod.\ Phys.\ A {\bf 33} (2018) no.05n06,  1842004
  doi:10.1142/S0217751X18420046
  [arXiv:1711.02864 [hep-ph]].



\bibitem{Chun:2017spz}
  E.~J.~Chun {\it et al.},
  ``Probing Leptogenesis,''
  Int.\ J.\ Mod.\ Phys.\ A {\bf 33} (2018) no.05n06,  1842005
  doi:10.1142/S0217751X18420058
  [arXiv:1711.02865 [hep-ph]].



\bibitem{Hagedorn:2017wjy}
  C.~Hagedorn, R.~N.~Mohapatra, E.~Molinaro, C.~C.~Nishi and S.~T.~Petcov,
  ``CP Violation in the Lepton Sector and Implications for Leptogenesis,''
  Int.\ J.\ Mod.\ Phys.\ A {\bf 33} (2018) no.05n06,  1842006
  doi:10.1142/S0217751X1842006X
  [arXiv:1711.02866 [hep-ph]].



\bibitem{Akhmedov:1998qx}
  E.~K.~Akhmedov, V.~A.~Rubakov and A.~Y.~Smirnov,
  ``Baryogenesis via neutrino oscillations,''
  Phys.\ Rev.\ Lett.\  {\bf 81} (1998) 1359
  doi:10.1103/PhysRevLett.81.1359
  [hep-ph/9803255].



\bibitem{Asaka:2005pn}
  T.~Asaka and M.~Shaposhnikov,
  ``The nuMSM, dark matter and baryon asymmetry of the universe,''
  Phys.\ Lett.\ B {\bf 620} (2005) 17
  doi:10.1016/j.physletb.2005.06.020
  [hep-ph/0505013].



\bibitem{Abada:2015rta}
  A.~Abada, G.~Arcadi, V.~Domcke and M.~Lucente,
  ``Lepton number violation as a key to low-scale leptogenesis,''
  JCAP {\bf 1511} (2015) no.11,  041
  doi:10.1088/1475-7516/2015/11/041
  [arXiv:1507.06215 [hep-ph]].



\bibitem{Hernandez:2015wna}
  P.~Hernandez, M.~Kekic, J.~Lopez-Pavon, J.~Racker and N.~Rius,
  ``Leptogenesis in GeV scale seesaw models,''
  JHEP {\bf 1510} (2015) 067
  doi:10.1007/JHEP10(2015)067
  [arXiv:1508.03676 [hep-ph]].



\bibitem{Drewes:2016gmt}
  M.~Drewes, B.~Garbrecht, D.~Gueter and J.~Klaric,
  ``Leptogenesis from Oscillations of Heavy Neutrinos with Large Mixing Angles,''
  JHEP {\bf 1612} (2016) 150
  doi:10.1007/JHEP12(2016)150
  [arXiv:1606.06690 [hep-ph]].

\bibitem{Antusch:2017pkq}
  S.~Antusch, E.~Cazzato, M.~Drewes, O.~Fischer, B.~Garbrecht, D.~Gueter and J.~Klaric,
  ``Probing Leptogenesis at Future Colliders,''
  JHEP {\bf 1809} (2018) 124
  doi:10.1007/JHEP09(2018)124
  [arXiv:1710.03744 [hep-ph]].


\bibitem{Ghiglieri:2018wbs}
  J.~Ghiglieri and M.~Laine,
  ``Precision study of GeV-scale resonant leptogenesis,''
  arXiv:1811.01971 [hep-ph].



\bibitem{Endoh:2003mz}
  T.~Endoh, T.~Morozumi and Z.~h.~Xiong,
  ``Primordial lepton family asymmetries in seesaw model,''
  Prog.\ Theor.\ Phys.\  {\bf 111} (2004) 123
  doi:10.1143/PTP.111.123
  [hep-ph/0308276].



\bibitem{Pilaftsis:2005rv}
  A.~Pilaftsis and T.~E.~J.~Underwood,
  ``Electroweak-scale resonant leptogenesis,''
  Phys.\ Rev.\ D {\bf 72} (2005) 113001
  doi:10.1103/PhysRevD.72.113001
  [hep-ph/0506107].



\bibitem{Abada:2006fw}
  A.~Abada, S.~Davidson, F.~X.~Josse-Michaux, M.~Losada and A.~Riotto,
  ``Flavor issues in leptogenesis,''
  JCAP {\bf 0604} (2006) 004
  doi:10.1088/1475-7516/2006/04/004
  [hep-ph/0601083].



\bibitem{Nardi:2006fx}
  E.~Nardi, Y.~Nir, E.~Roulet and J.~Racker,
  ``The Importance of flavor in leptogenesis,''
  JHEP {\bf 0601} (2006) 164
  doi:10.1088/1126-6708/2006/01/164
  [hep-ph/0601084].



\bibitem{Beneke:2010dz}
  M.~Beneke, B.~Garbrecht, C.~Fidler, M.~Herranen and P.~Schwaller,
  ``Flavoured Leptogenesis in the CTP Formalism,''
  Nucl.\ Phys.\ B {\bf 843} (2011) 177
  doi:10.1016/j.nuclphysb.2010.10.001
  [arXiv:1007.4783 [hep-ph]].



\bibitem{Blanchet:2011xq}
  S.~Blanchet, P.~Di Bari, D.~A.~Jones and L.~Marzola,
  ``Leptogenesis with heavy neutrino flavours: from density matrix to Boltzmann equations,''
  JCAP {\bf 1301} (2013) 041
  doi:10.1088/1475-7516/2013/01/041
  [arXiv:1112.4528 [hep-ph]].



\bibitem{Anisimov:2010gy}
  A.~Anisimov, D.~Besak and D.~Bodeker,
  ``Thermal production of relativistic Majorana neutrinos: Strong enhancement by multiple soft scattering,''
  JCAP {\bf 1103} (2011) 042
  doi:10.1088/1475-7516/2011/03/042
  [arXiv:1012.3784 [hep-ph]].



\bibitem{Laine:2011pq}
  M.~Laine and Y.~Schroder,
  ``Thermal right-handed neutrino production rate in the non-relativistic regime,''
  JHEP {\bf 1202} (2012) 068
  doi:10.1007/JHEP02(2012)068
  [arXiv:1112.1205 [hep-ph]].



\bibitem{Besak:2012qm}
  D.~Besak and D.~Bodeker,
  ``Thermal production of ultrarelativistic right-handed neutrinos: Complete leading-order results,''
  JCAP {\bf 1203} (2012) 029
  doi:10.1088/1475-7516/2012/03/029
  [arXiv:1202.1288 [hep-ph]].



\bibitem{Garbrecht:2013gd}
  B.~Garbrecht, F.~Glowna and M.~Herranen,
  ``Right-Handed Neutrino Production at Finite Temperature: Radiative Corrections, Soft and Collinear Divergences,''
  JHEP {\bf 1304} (2013) 099
  doi:10.1007/JHEP04(2013)099
  [arXiv:1302.0743 [hep-ph]].



\bibitem{Garbrecht:2013urw}
  B.~Garbrecht, F.~Glowna and P.~Schwaller,
  ``Scattering Rates For Leptogenesis: Damping of Lepton Flavour Coherence and Production of Singlet Neutrinos,''
  Nucl.\ Phys.\ B {\bf 877} (2013) 1
  doi:10.1016/j.nuclphysb.2013.08.020
  [arXiv:1303.5498 [hep-ph]].



\bibitem{Laine:2013lka}
  M.~Laine,
  ``Thermal right-handed neutrino production rate in the relativistic regime,''
  JHEP {\bf 1308} (2013) 138
  doi:10.1007/JHEP08(2013)138
  [arXiv:1307.4909 [hep-ph]].



\bibitem{Biondini:2013xua}
  S.~Biondini, N.~Brambilla, M.~A.~Escobedo and A.~Vairo,
  ``An effective field theory for non-relativistic Majorana neutrinos,''
  JHEP {\bf 1312} (2013) 028
  doi:10.1007/JHEP12(2013)028
  [arXiv:1307.7680 [hep-ph]].



\bibitem{Prokopec:2003pj}
  T.~Prokopec, M.~G.~Schmidt and S.~Weinstock,
  ``Transport equations for chiral fermions to order h bar and electroweak baryogenesis. Part 1,''
  Annals Phys.\  {\bf 314} (2004) 208
  doi:10.1016/j.aop.2004.06.002
  [hep-ph/0312110].



\bibitem{Prokopec:2004ic}
  T.~Prokopec, M.~G.~Schmidt and S.~Weinstock,
  ``Transport equations for chiral fermions to order h-bar and electroweak baryogenesis. Part II,''
  Annals Phys.\  {\bf 314} (2004) 267
  doi:10.1016/j.aop.2004.06.001
  [hep-ph/0406140].



\bibitem{Berges:2004yj}
  J.~Berges,
  ``Introduction to nonequilibrium quantum field theory,''
  AIP Conf.\ Proc.\  {\bf 739} (2005) 3
  doi:10.1063/1.1843591
  [hep-ph/0409233].



\bibitem{Groenewold:1946kp}
  H.~J.~Groenewold,
  ``On the Principles of elementary quantum mechanics,''
  Physica {\bf 12} (1946) 405.
  doi:10.1016/S0031-8914(46)80059-4



\bibitem{Moyal:1949sk}
  J.~E.~Moyal,
  ``Quantum mechanics as a statistical theory,''
  Proc.\ Cambridge Phil.\ Soc.\  {\bf 45} (1949) 99.
  doi:10.1017/S0305004100000487



\bibitem{Greiner:1998vd}
  C.~Greiner and S.~Leupold,
  ``Stochastic interpretation of Kadanoff-Baym equations and their relation to Langevin processes,''
  Annals Phys.\  {\bf 270} (1998) 328
  doi:10.1006/aphy.1998.5849
  [hep-ph/9802312].



\bibitem{Kubo:1957mj}
  R.~Kubo,
  ``Statistical mechanical theory of irreversible processes. 1. General theory and simple applications in magnetic and conduction problems,''
  J.\ Phys.\ Soc.\ Jap.\  {\bf 12} (1957) 570.
  doi:10.1143/JPSJ.12.570



\bibitem{Martin:1959jp}
  P.~C.~Martin and J.~S.~Schwinger,
  ``Theory of many particle systems. 1.,''
  Phys.\ Rev.\  {\bf 115} (1959) 1342.
  doi:10.1103/PhysRev.115.1342




\bibitem{Yanagida:1980gf}
  T.~Yanagida and M.~Yoshimura,
  ``Heavy Majorana Leptons and Cosmological Baryon Excess,''
  Phys.\ Rev.\ D {\bf 23} (1981) 2048.
  doi:10.1103/PhysRevD.23.2048



\bibitem{Masiero:1981vp}
  A.~Masiero and T.~Yanagida,
  ``Cosmological Baryon Production Versus Intermediate Mass Scales?,''
  Phys.\ Lett.\  {\bf 112B} (1982) 336.
  doi:10.1016/0370-2693(82)91063-2



\bibitem{Covi:1996wh}
  L.~Covi, E.~Roulet and F.~Vissani,
  ``CP violating decays in leptogenesis scenarios,''
  Phys.\ Lett.\ B {\bf 384} (1996) 169
  doi:10.1016/0370-2693(96)00817-9
  [hep-ph/9605319].



\bibitem{Pilaftsis:1997jf}
  A.~Pilaftsis,
  ``CP violation and baryogenesis due to heavy Majorana neutrinos,''
  Phys.\ Rev.\ D {\bf 56} (1997) 5431
  doi:10.1103/PhysRevD.56.5431
  [hep-ph/9707235].



\bibitem{Pilaftsis:2003gt}
  A.~Pilaftsis and T.~E.~J.~Underwood,
  ``Resonant leptogenesis,''
  Nucl.\ Phys.\ B {\bf 692} (2004) 303
  doi:10.1016/j.nuclphysb.2004.05.029
  [hep-ph/0309342].

\bibitem{Sigl:1992fn}
  G.~Sigl and G.~Raffelt,
  ``General kinetic description of relativistic mixed neutrinos,''
  Nucl.\ Phys.\ B {\bf 406} (1993) 423.
  doi:10.1016/0550-3213(93)90175-O


\bibitem{Flanz:1994yx}
  M.~Flanz, E.~A.~Paschos and U.~Sarkar,
  ``Baryogenesis from a lepton asymmetric universe,''
  Phys.\ Lett.\ B {\bf 345} (1995) 248
   Erratum: [Phys.\ Lett.\ B {\bf 384} (1996) 487]
   Erratum: [Phys.\ Lett.\ B {\bf 382} (1996) 447]
  doi:10.1016/0370-2693(96)00866-0, 10.1016/0370-2693(96)00842-8, 10.1016/0370-2693(94)01555-Q
  [hep-ph/9411366].



\bibitem{Flanz:1996fb}
  M.~Flanz, E.~A.~Paschos, U.~Sarkar and J.~Weiss,
  ``Baryogenesis through mixing of heavy Majorana neutrinos,''
  Phys.\ Lett.\ B {\bf 389} (1996) 693
  doi:10.1016/S0370-2693(96)01337-8, 10.1016/S0370-2693(96)80011-6
  [hep-ph/9607310].



\bibitem{Garbrecht:2011aw}
  B.~Garbrecht and M.~Herranen,
  ``Effective Theory of Resonant Leptogenesis in the Closed-Time-Path Approach,''
  Nucl.\ Phys.\ B {\bf 861} (2012) 17
  doi:10.1016/j.nuclphysb.2012.03.009
  [arXiv:1112.5954 [hep-ph]].



\bibitem{Anisimov:2005hr}
  A.~Anisimov, A.~Broncano and M.~Plumacher,
  ``The CP-asymmetry in resonant leptogenesis,''
  Nucl.\ Phys.\ B {\bf 737} (2006) 176
  doi:10.1016/j.nuclphysb.2006.01.003
  [hep-ph/0511248].



\bibitem{Garny:2011hg}
  M.~Garny, A.~Kartavtsev and A.~Hohenegger,
  ``Leptogenesis from first principles in the resonant regime,''
  Annals Phys.\  {\bf 328} (2013) 26
  doi:10.1016/j.aop.2012.10.007
  [arXiv:1112.6428 [hep-ph]].



\bibitem{Iso:2013lba}
  S.~Iso, K.~Shimada and M.~Yamanaka,
  ``Kadanoff-Baym approach to the thermal resonant leptogenesis,''
  JHEP {\bf 1404} (2014) 062
  doi:10.1007/JHEP04(2014)062
  [arXiv:1312.7680 [hep-ph]].



\bibitem{Iso:2014afa}
  S.~Iso and K.~Shimada,
  ``Coherent Flavour Oscillation and CP Violating Parameter in Thermal Resonant Leptogenesis,''
  JHEP {\bf 1408} (2014) 043
  doi:10.1007/JHEP08(2014)043
  [arXiv:1404.4816 [hep-ph]].



\bibitem{Garbrecht:2014aga}
  B.~Garbrecht, F.~Gautier and J.~Klaric,
  ``Strong Washout Approximation to Resonant Leptogenesis,''
  JCAP {\bf 1409} (2014) no.09,  033
  doi:10.1088/1475-7516/2014/09/033
  [arXiv:1406.4190 [hep-ph]].



\bibitem{Dev:2014laa}
  P.~S.~Bhupal Dev, P.~Millington, A.~Pilaftsis and D.~Teresi,
  ``Flavour Covariant Transport Equations: an Application to Resonant Leptogenesis,''
  Nucl.\ Phys.\ B {\bf 886} (2014) 569
  doi:10.1016/j.nuclphysb.2014.06.020
  [arXiv:1404.1003 [hep-ph]].



\bibitem{DiBari:2005st}
  P.~Di Bari,
  ``Seesaw geometry and leptogenesis,''
  Nucl.\ Phys.\ B {\bf 727} (2005) 318
  doi:10.1016/j.nuclphysb.2005.08.032
  [hep-ph/0502082].


\bibitem{Engelhard:2006yg}
  G.~Engelhard, Y.~Grossman, E.~Nardi and Y.~Nir,
  ``The Importance of $N_2$ leptogenesis,''
  Phys.\ Rev.\ Lett.\  {\bf 99} (2007) 081802
  doi:10.1103/PhysRevLett.99.081802
  [hep-ph/0612187].


\bibitem{Buchmuller:2004nz}
  W.~Buchmuller, P.~Di Bari and M.~Plumacher,
  ``Leptogenesis for pedestrians,''
  Annals Phys.\  {\bf 315} (2005) 305
  doi:10.1016/j.aop.2004.02.003
  [hep-ph/0401240].

\bibitem{Garbrecht:2019zaa}
  B.~Garbrecht, P.~Klose and C.~Tamarit,
  ``Relativistic and spectator effects in leptogenesis with heavy sterile neutrinos,''
  arXiv:1904.09956 [hep-ph].

\bibitem{Kolb:1979qa}
  E.~W.~Kolb and S.~Wolfram,
  ``Baryon Number Generation in the Early Universe,''
  Nucl.\ Phys.\ B {\bf 172} (1980) 224
   Erratum: [Nucl.\ Phys.\ B {\bf 195} (1982) 542].
  doi:10.1016/0550-3213(80)90167-4, 10.1016/0550-3213(82)90012-8

  
  
\bibitem{Buchmuller:2002rq}
  W.~Buchmuller, P.~Di Bari and M.~Plumacher,
  ``Cosmic microwave background, matter - antimatter asymmetry and neutrino masses,''
  Nucl.\ Phys.\ B {\bf 643} (2002) 367
   Erratum: [Nucl.\ Phys.\ B {\bf 793} (2008) 362]
  doi:10.1016/S0550-3213(02)00737-X, 10.1016/j.nuclphysb.2007.11.030
  [hep-ph/0205349].
  
\bibitem{Buchmuller:2002jk}
  W.~Buchmuller, P.~Di Bari and M.~Plumacher,
  ``A Bound on neutrino masses from baryogenesis,''
  Phys.\ Lett.\ B {\bf 547} (2002) 128
  doi:10.1016/S0370-2693(02)02758-2
  [hep-ph/0209301].
  
\bibitem{Buchmuller:2003gz}
  W.~Buchmuller, P.~Di Bari and M.~Plumacher,
  ``The Neutrino mass window for baryogenesis,''
  Nucl.\ Phys.\ B {\bf 665} (2003) 445
  doi:10.1016/S0550-3213(03)00449-8
  [hep-ph/0302092].

\bibitem{Blanchet:2009kk}
  S.~Blanchet, T.~Hambye and F.~X.~Josse-Michaux,
  ``Reconciling leptogenesis with observable $\mu\to e\gamma$ rates,''
  JHEP {\bf 1004} (2010) 023
  doi:10.1007/JHEP04(2010)023
  [arXiv:0912.3153 [hep-ph]].

\bibitem{Buchmuller:2000nd}
  W.~Buchmuller and S.~Fredenhagen,
  ``Quantum mechanics of baryogenesis,''
  Phys.\ Lett.\ B {\bf 483} (2000) 217
  doi:10.1016/S0370-2693(00)00573-6
  [hep-ph/0004145].

\bibitem{DeSimone:2007gkc}
  A.~De Simone and A.~Riotto,
  ``Quantum Boltzmann Equations and Leptogenesis,''
  JCAP {\bf 0708} (2007) 002
  doi:10.1088/1475-7516/2007/08/002
  [hep-ph/0703175].

\bibitem{Garny:2009rv}
  M.~Garny, A.~Hohenegger, A.~Kartavtsev and M.~Lindner,
  ``Systematic approach to leptogenesis in nonequilibrium QFT: Vertex contribution to the CP-violating parameter,''
  Phys.\ Rev.\ D {\bf 80} (2009) 125027
  doi:10.1103/PhysRevD.80.125027
  [arXiv:0909.1559 [hep-ph]].



\bibitem{Garny:2009qn}
  M.~Garny, A.~Hohenegger, A.~Kartavtsev and M.~Lindner,
  ``Systematic approach to leptogenesis in nonequilibrium QFT: Self-energy contribution to the CP-violating parameter,''
  Phys.\ Rev.\ D {\bf 81} (2010) 085027
  doi:10.1103/PhysRevD.81.085027
  [arXiv:0911.4122 [hep-ph]].



\bibitem{Anisimov:2010aq}
  A.~Anisimov, W.~Buchmuller, M.~Drewes and S.~Mendizabal,
  ``Leptogenesis from Quantum Interference in a Thermal Bath,''
  Phys.\ Rev.\ Lett.\  {\bf 104} (2010) 121102
  doi:10.1103/PhysRevLett.104.121102
  [arXiv:1001.3856 [hep-ph]].



\bibitem{Garny:2010nj}
  M.~Garny, A.~Hohenegger and A.~Kartavtsev,
  ``Medium corrections to the CP-violating parameter in leptogenesis,''
  Phys.\ Rev.\ D {\bf 81} (2010) 085028
  doi:10.1103/PhysRevD.81.085028
  [arXiv:1002.0331 [hep-ph]].



\bibitem{Beneke:2010wd}
  M.~Beneke, B.~Garbrecht, M.~Herranen and P.~Schwaller,
  ``Finite Number Density Corrections to Leptogenesis,''
  Nucl.\ Phys.\ B {\bf 838} (2010) 1
  doi:10.1016/j.nuclphysb.2010.05.003
  [arXiv:1002.1326 [hep-ph]].



\bibitem{Garny:2010nz}
  M.~Garny, A.~Hohenegger and A.~Kartavtsev,
  ``Quantum corrections to leptogenesis from the gradient expansion,''
  arXiv:1005.5385 [hep-ph].



\bibitem{Anisimov:2010dk}
  A.~Anisimov, W.~Buchmüller, M.~Drewes and S.~Mendizabal,
  ``Quantum Leptogenesis I,''
  Annals Phys.\  {\bf 326} (2011) 1998
   Erratum: [Annals Phys.\  {\bf 338} (2011) 376]
  doi:10.1016/j.aop.2011.02.002, 10.1016/j.aop.2013.05.00
  [arXiv:1012.5821 [hep-ph]].



\bibitem{Barbieri:1999ma}
  R.~Barbieri, P.~Creminelli, A.~Strumia and N.~Tetradis,
  ``Baryogenesis through leptogenesis,''
  Nucl.\ Phys.\ B {\bf 575} (2000) 61
  doi:10.1016/S0550-3213(00)00011-0
  [hep-ph/9911315].



\bibitem{Buchmuller:2001sr}
  W.~Buchmuller and M.~Plumacher,
  ``Spectator processes and baryogenesis,''
  Phys.\ Lett.\ B {\bf 511} (2001) 74
  doi:10.1016/S0370-2693(01)00614-1
  [hep-ph/0104189].



\bibitem{Garbrecht:2014kda}
  B.~Garbrecht and P.~Schwaller,
  ``Spectator Effects during Leptogenesis in the Strong Washout Regime,''
  JCAP {\bf 1410} (2014) no.10,  012
  doi:10.1088/1475-7516/2014/10/012
  [arXiv:1404.2915 [hep-ph]].



\bibitem{Garbrecht:2011xw}
  B.~Garbrecht and M.~Garny,
  ``Finite Width in out-of-Equilibrium Propagators and Kinetic Theory,''
  Annals Phys.\  {\bf 327} (2012) 914
  doi:10.1016/j.aop.2011.10.005
  [arXiv:1108.3688 [hep-ph]].

\bibitem{Shaposhnikov:2008pf}
  M.~Shaposhnikov,
  ``The nuMSM, leptonic asymmetries, and properties of singlet fermions,''
  JHEP {\bf 0808} (2008) 008
  doi:10.1088/1126-6708/2008/08/008
  [arXiv:0804.4542 [hep-ph]].

\bibitem{Eijima:2017anv}
  S.~Eijima and M.~Shaposhnikov,
  ``Fermion number violating effects in low scale leptogenesis,''
  Phys.\ Lett.\ B {\bf 771} (2017) 288
  doi:10.1016/j.physletb.2017.05.068
  [arXiv:1703.06085 [hep-ph]].

\bibitem{Ghiglieri:2017gjz}
  J.~Ghiglieri and M.~Laine,
  ``GeV-scale hot sterile neutrino oscillations: a derivation of evolution equations,''
  JHEP {\bf 1705} (2017) 132
  doi:10.1007/JHEP05(2017)132
  [arXiv:1703.06087 [hep-ph]].

\bibitem{Eijima:2018qke}
  S.~Eijima, M.~Shaposhnikov and I.~Timiryasov,
  ``Parameter space of baryogenesis in the $\nu$MSM,''
  JHEP {\bf 1907} (2019) 077
  doi:10.1007/JHEP07(2019)077
  [arXiv:1808.10833 [hep-ph]].


\bibitem{Joyce:1999uf}
  M.~Joyce, K.~Kainulainen and T.~Prokopec,
  ``Quantum transport equations for a scalar field,''
  Phys.\ Lett.\ B {\bf 474} (2000) 402
  doi:10.1016/S0370-2693(00)00041-1
  [hep-ph/9910535].



\bibitem{Kainulainen:2001cn}
  K.~Kainulainen, T.~Prokopec, M.~G.~Schmidt and S.~Weinstock,
  ``First principle derivation of semiclassical force for electroweak baryogenesis,''
  JHEP {\bf 0106} (2001) 031
  doi:10.1088/1126-6708/2001/06/031
  [hep-ph/0105295].



\bibitem{Garbrecht:2013iga}
  B.~Garbrecht and M.~J.~Ramsey-Musolf,
  ``Cuts, Cancellations and the Closed Time Path: The Soft Leptogenesis Example,''
  Nucl.\ Phys.\ B {\bf 882} (2014) 145
  doi:10.1016/j.nuclphysb.2014.02.012
  [arXiv:1307.0524 [hep-ph]].



\bibitem{Cirigliano:2009yt}
  V.~Cirigliano, C.~Lee, M.~J.~Ramsey-Musolf and S.~Tulin,
  ``Flavored Quantum Boltzmann Equations,''
  Phys.\ Rev.\ D {\bf 81} (2010) 103503
  doi:10.1103/PhysRevD.81.103503
  [arXiv:0912.3523 [hep-ph]].



\bibitem{Dev:2014wsa}
  P.~S.~Bhupal Dev, P.~Millington, A.~Pilaftsis and D.~Teresi,
  ``Kadanoff–Baym approach to flavour mixing and oscillations in resonant leptogenesis,''
  Nucl.\ Phys.\ B {\bf 891} (2015) 128
  doi:10.1016/j.nuclphysb.2014.12.003
  [arXiv:1410.6434 [hep-ph]].



\bibitem{Kartavtsev:2015vto}
  A.~Kartavtsev, P.~Millington and H.~Vogel,
  ``Lepton asymmetry from mixing and oscillations,''
  JHEP {\bf 1606} (2016) 066
  doi:10.1007/JHEP06(2016)066
  [arXiv:1601.03086 [hep-ph]].



\bibitem{Kolb:1983ni}
  E.~W.~Kolb and M.~S.~Turner,
  ``Grand Unified Theories and the Origin of the Baryon Asymmetry,''
  Ann.\ Rev.\ Nucl.\ Part.\ Sci.\  {\bf 33} (1983) 645.
  doi:10.1146/annurev.ns.33.120183.003241



\bibitem{Khlebnikov:1996vj}
  S.~Y.~Khlebnikov and M.~E.~Shaposhnikov,
  ``Melting of the Higgs vacuum: Conserved numbers at high temperature,''
  Phys.\ Lett.\ B {\bf 387} (1996) 817
  doi:10.1016/0370-2693(96)01116-1
  [hep-ph/9607386].



\bibitem{Laine:1999wv}
  M.~Laine and M.~E.~Shaposhnikov,
  ``A Remark on sphaleron erasure of baryon asymmetry,''
  Phys.\ Rev.\ D {\bf 61} (2000) 117302
  doi:10.1103/PhysRevD.61.117302
  [hep-ph/9911473].



\bibitem{DOnofrio:2014rug}
  M.~D'Onofrio, K.~Rummukainen and A.~Tranberg,
  ``Sphaleron Rate in the Minimal Standard Model,''
  Phys.\ Rev.\ Lett.\  {\bf 113} (2014) no.14,  141602
  doi:10.1103/PhysRevLett.113.141602
  [arXiv:1404.3565 [hep-ph]].



\bibitem{Harvey:1990qw}
  J.~A.~Harvey and M.~S.~Turner,
  ``Cosmological baryon and lepton number in the presence of electroweak fermion number violation,''
  Phys.\ Rev.\ D {\bf 42} (1990) 3344.
  doi:10.1103/PhysRevD.42.3344


\bibitem{Garbrecht:2012qv}
  B.~Garbrecht,
  ``Leptogenesis from Additional Higgs Doublets,''
  Phys.\ Rev.\ D {\bf 85} (2012) 123509
  doi:10.1103/PhysRevD.85.123509
  [arXiv:1201.5126 [hep-ph]].

\bibitem{Garbrecht:2012pq}
  B.~Garbrecht,
  ``Baryogenesis from Mixing of Lepton Doublets,''
  Nucl.\ Phys.\ B {\bf 868} (2013) 557
  doi:10.1016/j.nuclphysb.2012.11.021
  [arXiv:1210.0553 [hep-ph]].

\bibitem{Garbrecht:2014iia}
  B.~Garbrecht and I.~Izaguirre,
  ``Phenomenology of Baryogenesis from Lepton-Doublet Mixing,''
  Nucl.\ Phys.\ B {\bf 896} (2015) 412
  doi:10.1016/j.nuclphysb.2015.04.017
  [arXiv:1411.2834 [hep-ph]].

\bibitem{Salvio:2011sf}
  A.~Salvio, P.~Lodone and A.~Strumia,
  ``Towards leptogenesis at NLO: the right-handed neutrino interaction rate,''
  JHEP {\bf 1108} (2011) 116
  doi:10.1007/JHEP08(2011)116
  [arXiv:1106.2814 [hep-ph]].



\bibitem{Biondini:2015gyw}
  S.~Biondini, N.~Brambilla, M.~A.~Escobedo and A.~Vairo,
  ``CP asymmetry in heavy Majorana neutrino decays at finite temperature: the nearly degenerate case,''
  JHEP {\bf 1603} (2016) 191
   Erratum: [JHEP {\bf 1608} (2016) 072]
  doi:10.1007/JHEP03(2016)191, 10.1007/JHEP08(2016)072
  [arXiv:1511.02803 [hep-ph]].

\bibitem{Biondini:2016arl}
  S.~Biondini, N.~Brambilla and A.~Vairo,
  ``CP asymmetry in heavy Majorana neutrino decays at finite temperature: the hierarchical case,''
  JHEP {\bf 1609} (2016) 126
  doi:10.1007/JHEP09(2016)126
  [arXiv:1608.01979 [hep-ph]].

\bibitem{Ghisoiu:2014ena}
  I.~Ghisoiu and M.~Laine,
  ``Right-handed neutrino production rate at $T > 160{\rm GeV}$,''
  JCAP {\bf 1412} (2014) no.12,  032
  doi:10.1088/1475-7516/2014/12/032
  [arXiv:1411.1765 [hep-ph]].

\bibitem{Hambye:2016sby}
  T.~Hambye and D.~Teresi,
  ``Higgs doublet decay as the origin of the baryon asymmetry,''
  Phys.\ Rev.\ Lett.\  {\bf 117} (2016) no.9,  091801
  doi:10.1103/PhysRevLett.117.091801
  [arXiv:1606.00017 [hep-ph]].


\bibitem{Konstandin:2013caa}
  T.~Konstandin,
  ``Quantum Transport and Electroweak Baryogenesis,''
  Phys.\ Usp.\  {\bf 56} (2013) 747
   [Usp.\ Fiz.\ Nauk {\bf 183} (2013) 785]
  doi:10.3367/UFNe.0183.201308a.0785
  [arXiv:1302.6713 [hep-ph]].



\bibitem{Trodden:1998ym}
  M.~Trodden,
  ``Electroweak baryogenesis,''
  Rev.\ Mod.\ Phys.\  {\bf 71} (1999) 1463
  doi:10.1103/RevModPhys.71.1463
  [hep-ph/9803479].



\bibitem{Morrissey:2012db}
  D.~E.~Morrissey and M.~J.~Ramsey-Musolf,
  ``Electroweak baryogenesis,''
  New J.\ Phys.\  {\bf 14} (2012) 125003
  doi:10.1088/1367-2630/14/12/125003
  [arXiv:1206.2942 [hep-ph]].



\bibitem{White:2016nbo}
  G.~A.~White,
  ``A Pedagogical Introduction to Electroweak Baryogenesis,''
  doi:10.1088/978-1-6817-4457-5



\bibitem{Konstandin:2004gy}
  T.~Konstandin, T.~Prokopec and M.~G.~Schmidt,
  ``Kinetic description of fermion flavor mixing and CP-violating sources for baryogenesis,''
  Nucl.\ Phys.\ B {\bf 716} (2005) 373
  doi:10.1016/j.nuclphysb.2005.03.013
  [hep-ph/0410135].



\bibitem{Konstandin:2005cd}
  T.~Konstandin, T.~Prokopec, M.~G.~Schmidt and M.~Seco,
  ``MSSM electroweak baryogenesis and flavor mixing in transport equations,''
  Nucl.\ Phys.\ B {\bf 738} (2006) 1
  doi:10.1016/j.nuclphysb.2005.11.028
  [hep-ph/0505103].



\bibitem{Riotto:1998zb}
  A.~Riotto,
  ``The More relaxed supersymmetric electroweak baryogenesis,''
  Phys.\ Rev.\ D {\bf 58} (1998) 095009
  doi:10.1103/PhysRevD.58.095009
  [hep-ph/9803357].



\bibitem{Lee:2004we}
  C.~Lee, V.~Cirigliano and M.~J.~Ramsey-Musolf,
  ``Resonant relaxation in electroweak baryogenesis,''
  Phys.\ Rev.\ D {\bf 71} (2005) 075010
  doi:10.1103/PhysRevD.71.075010
  [hep-ph/0412354].



\bibitem{Joyce:1994fu}
  M.~Joyce, T.~Prokopec and N.~Turok,
  ``Electroweak baryogenesis from a classical force,''
  Phys.\ Rev.\ Lett.\  {\bf 75} (1995) 1695
   Erratum: [Phys.\ Rev.\ Lett.\  {\bf 75} (1995) 3375]
  doi:10.1103/PhysRevLett.75.3375, 10.1103/PhysRevLett.75.1695
  [hep-ph/9408339].



\bibitem{Joyce:1994zn}
  M.~Joyce, T.~Prokopec and N.~Turok,
  ``Nonlocal electroweak baryogenesis. Part 1: Thin wall regime,''
  Phys.\ Rev.\ D {\bf 53} (1996) 2930
  doi:10.1103/PhysRevD.53.2930
  [hep-ph/9410281].



\bibitem{Kainulainen:2002th}
  K.~Kainulainen, T.~Prokopec, M.~G.~Schmidt and S.~Weinstock,
  ``Semiclassical force for electroweak baryogenesis: Three-dimensional derivation,''
  Phys.\ Rev.\ D {\bf 66} (2002) 043502
  doi:10.1103/PhysRevD.66.043502
  [hep-ph/0202177].



\bibitem{Cline:2000nw}
  J.~M.~Cline, M.~Joyce and K.~Kainulainen,
  ``Supersymmetric electroweak baryogenesis,''
  JHEP {\bf 0007} (2000) 018
  doi:10.1088/1126-6708/2000/07/018
  [hep-ph/0006119].



\bibitem{Carena:2002ss}
  M.~Carena, M.~Quiros, M.~Seco and C.~E.~M.~Wagner,
  ``Improved Results in Supersymmetric Electroweak Baryogenesis,''
  Nucl.\ Phys.\ B {\bf 650} (2003) 24
  doi:10.1016/S0550-3213(02)01065-9
  [hep-ph/0208043].



\bibitem{Cirigliano:2006wh}
  V.~Cirigliano, M.~J.~Ramsey-Musolf, S.~Tulin and C.~Lee,
  ``Yukawa and tri-scalar processes in electroweak baryogenesis,''
  Phys.\ Rev.\ D {\bf 73} (2006) 115009
  doi:10.1103/PhysRevD.73.115009
  [hep-ph/0603058].



\bibitem{Cirigliano:2011di}
  V.~Cirigliano, C.~Lee and S.~Tulin,
  ``Resonant Flavor Oscillations in Electroweak Baryogenesis,''
  Phys.\ Rev.\ D {\bf 84} (2011) 056006
  doi:10.1103/PhysRevD.84.056006
  [arXiv:1106.0747 [hep-ph]].



\bibitem{Arnold:2000dr}
  P.~B.~Arnold, G.~D.~Moore and L.~G.~Yaffe,
  ``Transport coefficients in high temperature gauge theories. 1. Leading log results,''
  JHEP {\bf 0011} (2000) 001
  doi:10.1088/1126-6708/2000/11/001
  [hep-ph/0010177].



\bibitem{Arnold:2003zc}
  P.~B.~Arnold, G.~D.~Moore and L.~G.~Yaffe,
  ``Transport coefficients in high temperature gauge theories. 2. Beyond leading log,''
  JHEP {\bf 0305} (2003) 051
  doi:10.1088/1126-6708/2003/05/051
  [hep-ph/0302165].



\bibitem{Chung:2009qs}
  D.~J.~H.~Chung, B.~Garbrecht, M.~J.~Ramsey-Musolf and S.~Tulin,
  ``Supergauge interactions and electroweak baryogenesis,''
  JHEP {\bf 0912} (2009) 067
  doi:10.1088/1126-6708/2009/12/067
  [arXiv:0908.2187 [hep-ph]].



\bibitem{Huet:1995sh}
  P.~Huet and A.~E.~Nelson,
  ``Electroweak baryogenesis in supersymmetric models,''
  Phys.\ Rev.\ D {\bf 53} (1996) 4578
  doi:10.1103/PhysRevD.53.4578
  [hep-ph/9506477].



\bibitem{Bodeker:1999gx}
  D.~Bodeker, G.~D.~Moore and K.~Rummukainen,
  ``Chern-Simons number diffusion and hard thermal loops on the lattice,''
  Phys.\ Rev.\ D {\bf 61} (2000) 056003
  doi:10.1103/PhysRevD.61.056003
  [hep-ph/9907545].



\bibitem{Moore:1999fs}
  G.~D.~Moore and K.~Rummukainen,
  ``Classical sphaleron rate on fine lattices,''
  Phys.\ Rev.\ D {\bf 61} (2000) 105008
  doi:10.1103/PhysRevD.61.105008
  [hep-ph/9906259].



\bibitem{Moore:2000mx}
  G.~D.~Moore,
  ``Sphaleron rate in the symmetric electroweak phase,''
  Phys.\ Rev.\ D {\bf 62} (2000) 085011
  doi:10.1103/PhysRevD.62.085011
  [hep-ph/0001216].
  

\end{thebibliography}
\end{document}